\documentclass[aps,pre,twocolumn]{revtex4}
\usepackage{amssymb,macros2erev4,graphicx,amsmath}
\usepackage{times}

\newcommand{\p}{\partial}
\newcommand{\tr}{\mbox{tr}}
 
\newcommand{\fig}[2]{\includegraphics[width=#1]{./figures/#2}}
\newcommand{\Fig}[1]{\includegraphics[width=\columnwidth]{./figures/#1}}
\newlength{\bilderlength}
\newcommand{\bilderscale}{0.35}
\newcommand{\storebilderscale}{\bilderscale}
\newcommand{\bilderskip}{\hspace*{0.8ex}}
\newcommand{\textdiagram}[1]{%
\renewcommand{\bilderscale}{0.25}%
\diagram{#1}\renewcommand{\bilderscale}{\storebilderscale}}
\newcommand{\diagram}[1]{%
\settowidth{\bilderlength}{\bilderskip%
\includegraphics[scale=\bilderscale]{./figures/#1}\bilderskip}%
\parbox{\bilderlength}{\bilderskip%
\includegraphics[scale=\bilderscale]{./figures/#1}\bilderskip}}

\newcommand{\rme}{{\mathrm{e}}}
\newcommand{\rmd}{{\mathrm{d}}}
\newcommand{\nn}{\nonumber}

\newcommand{\E}{\epsilon}
\newcommand{\Tr}{{\mathrm{\bf Tr}}}

\begin{document}

\def\beginincorrect{

\noindent%
{\unitlength1mm
\begin{picture}(100,5)
\put(0,0){\line(0,1){5}}
\put(0,5){\line(1,0){179}}
\put(179,0){\line(0,1){5}}
\put(78,2){\mbox{\normalsize*****incorrect*****}}
\end{picture}}\newline
\scriptsize}
\def\endincorrect{

\noindent%
{\unitlength1mm
\begin{picture}(100,5)
\put(0,0){\line(0,1){5}}
\put(0,0){\line(1,0){179}}
\put(179,0){\line(0,1){5}}
\put(74,1){\mbox{\normalsize*****end incorrect*****}}
\end{picture}}

\normalsize}

%

\title{\sffamily\Large\bfseries Functional Renormalization Group at
Large $N$
for Disordered Elastic Systems, and Relation to Replica Symmetry
Breaking} \author{\sffamily\bfseries\normalsize Pierre Le
Doussal{$^1$} and Kay J\"org Wiese{$^2$} \vspace*{3mm}}
\affiliation{{$^1$} CNRS-Laboratoire de Physique Th{\'e}orique de
l'Ecole Normale Sup{\'e}rieure,
24 rue Lhomond, 75005 Paris, France\\
{$^2$} KITP, University of
California at
Santa Barbara, Santa Barbara, CA 93106-4030, USA\medskip }
\date{\small\today}
\begin{abstract}
We study the replica field theory which describes the pinning of
elastic manifolds of arbitrary internal dimension $d$ in a random
potential, with the aim of bridging the gap between mean field and
renormalization theory. The full effective action is computed exactly
in the limit of large embedding space dimension $N$.  The second
cumulant of the renormalized disorder obeys a closed self-consistent
equation. It is used to derive a Functional Renormalization Group
(FRG) equation valid in any dimension $d$, which correctly matches the
Balents Fisher result to first order in $\epsilon=4-d$. We analyze in
detail the solutions of the large-$N$ FRG for both long-range and
short-range disorder, at zero and finite temperature.  We find
consistent agreement with the results of Mezard Parisi (MP) from the
Gaussian variational method (GVM) in the case where full replica
symmetry breaking (RSB) holds there. We prove that the cusplike
non-analyticity in the large $N$ FRG appears at a finite scale,
corresponding to the instability of the replica symmetric solution of
MP. We show that the FRG exactly reproduces, for any disorder
correlator and with no need to invoke Parisi's spontaneous RSB, the
non-trivial result of the GVM for small overlap. A formula is found
yielding the complete RSB solution for all overlaps. Since our
saddle-point equations for the effective action contain both the MP
equations and the FRG, it can be used to describe the crossover from
FRG to RSB. A qualitative analysis of this crossover is given, as well
as a comparison with previous attempts to relate FRG to GVM. Finally,
we discuss applications to other problems and new perspectives.
\end{abstract}
\maketitle

\section{Introduction} Elastic objects pinned by a quenched random
potential are a relevant model for many experimental systems.  It
describes interfaces in magnets
\cite{NattermannBookYoung,LemerleFerreChappertMathetGiamarchiLeDoussal1998}
which experience either short-range disorder (random bond), or long
range (random field) disorder, the contact line of a liquid wetting a
rough substrate \cite{PrevostRolleyGuthmann2002,ErtasKardar1994b},
vortex lines in superconductors
\cite{BlatterFeigelmanGeshkenbeinLarkinVinokur1994,
GiamarchiLeDoussal1995,GiamarchiBookYoung,NattermannScheidl2000}. It
also provides powerful analogies, via mode coupling theory, to complex
systems such as structural glasses \cite{BookYoung}. One important
observable is the roughness exponent $\zeta$ of the pinned manifold.

From the theoretical side, this problem still offers considerable
challenges.  It is the simplest example of a class of disordered
systems, including random field magnets, where the so called
dimensional reduction
\cite{EfetovLarkin1977,AharonyImryShangkeng1976,Grinstein1976,ParisiSourlas1979,Cardy1983,NattermannBookYoung}
renders conventional perturbation theory trivial and useless at zero
temperature.  The elastic object is usually parameterized by a $N$
component vector $\vec u(x)$ in the embedding space $\mathbb{R}^{N}$,
and $x \in \mathbb{R}^d$ is the coordinate in the internal space.
Apart from the case of the directed polymer (DP) in $1+1$ dimensions
($d=1$, $N=1$), where some exact results were obtained
\cite{Kardar1987,BrunetDerrida2000,BrunetDerrida2000a,Johansson1999,PraehoferSpohn2000},
analytical results are scarce. One important challenge is to
understand the DP for any $N$, due to its exact relation to the
Kardar-Parisi-Zhang growth equation whose upper critical dimension is
at present not known, and even its very existence is debated
\cite{Laessig1995,Wiese1998a,LassigKinzelbach1997,MarinariPagnaniParisi2000}.

Two main analytical approaches have been devised so far.  Each
succeeds in evading dimensional reduction, providing an interesting
physical picture, but comes with its limitations. The first one is the
mean field theory, the replica gaussian variational method (GVM)
\cite{MezardParisi1991} in the statics and the off equilibrium
dynamical version
\cite{CugliandoloKurchanLeDoussal1996a,CugliandoloLeDoussal1996b}.
The GVM approximates the replica measure by a replica symmetry broken
(RSB) gaussian, equivalently, the Gibbs measure for $u$ as a random
superposition of gaussians \cite{MezardParisi1991}, and is argued to
be exact for $N=\infty$.  It yields Flory values for the exponent
$\zeta$.  As for spin glasses, computing the next order corrections
(i.e.\ in $1/N$) at the RSB saddle point is very arduous
\cite{CarlucciDeDominicisTemesvari1996,DeDominicisEtAlBookYoung,%
Goldschmidt1993}. One may question whether it is the most promising
route, since it is as yet unclear whether the huge degeneracy of
states encoded in the Parisi RSB is relevant to describe finite
$N$. There seems to be some agreement that this type of RSB does not
occur for low $d$ and $N$. Certainly, in the simpler but still-non
trivial $d=0$ limit, Parisi type RSB found in the GVM should exist
only at $N=\infty$, apart from the interesting so-called marginal case
of logarithmic correlations \cite{CarpentierLeDoussal2001}. For the
DP, another exactly solvable mean field limit is the Cayley tree and
there too it is not  clear how to meaningfully expand around that
limit \cite{DerridaSpohn1988,CookDerrida1989,CookDerrida1989a}.

The second main analytical method is the functional renormalization
group (FRG) which performs a dimensional expansion around $d=4$ and
was originally developped only to one loop, within a Wilson scheme
\cite{Fisher1985b,DSFisher1986,BalentsDSFisher1993,GiamarchiLeDoussal1995}. Its
aim is to include fluctuations, neglected in the mean-field
approaches. There too, the dynamics
\cite{NattermanStepanowTangLeschhorn1992%
,LeschhornNattermannStepanow1996,NarayanDSFisher1992a,NarayanDSFisher1992b,%
ChauveGiamarchiLeDoussal2000} has been investigated.  The FRG follows
the second cumulant of the random potential $R(u)$ under coarse
graining, a full function since the field is dimensionless in
$d=4$. It was found that $R(u)$ becomes non-analytic already in the
1-loop equation at $T=0$ after a finite renormalization, at the Larkin
scale.

Both methods circumvent dimensional reduction by providing a mechanism
which is non-perturbative in the bare disorder.  The GVM evades DR
thanks to the RSB saddle point. The FRG escapes via the generation of
a cusp-like non-analyticity in $R''(u)$ at $u=0$. Indeed, while the
bare disorder correlator is an analytic function, FRG fixed points for
the renormalized $R(u)$, perturbative in $\epsilon=4-d$, are found
only in the space of non-analytic functions, and subject to the
condition that the resulting exponent $\zeta$ is 
non-trivial. Both methods are disconcertingly different in spirit and
it is an outstanding question in the theory of disordered systems 
how to compare and reconcile them. Comparisons were made between some
predictions of the 1-loop FRG and of the GVM
\cite{BalentsDSFisher1993,GiamarchiLeDoussal1995}. Balents and Fisher
obtained the 1-loop FRG equation for any $N$ restricted to
$O(\epsilon)$, and found that its solution reproduces the Flory value
of $\zeta$ for LR disorder, but yields subtle corrections for SR
disorder, exponential in $N$.

Physically both methods capture the metastable states beyond the
Larkin scale $L_c$ and it is tempting to compare how they describe
them. In \cite{BalentsBouchaudMezard1996} a coarse grained random
potential was defined and it was found within the GVM that its
correlator mimics the one in the FRG, exhibiting some non-analyticity
which was interpreted in terms of shock-like singularities in the
coarse grained disorder. Unfortunately, this analogy was demonstrated
only around the Larkin scale, while a quantitative and more general
connection able to reach perturbatively the true large scale
behaviour, as is achieved in the field theoretic FRG, is still
missing.

The need for a study of the FRG at large $N$ is all the more pressing
since we have developped systematic higher loop approaches within the
$\epsilon$-expansion
\cite{ChauveLeDoussalWiese2000a,LeDoussalWieseChauve2002,ChauveLeDoussal2001}.
Within these studies, we have found that higher loop FRG equations for
$R(u)$ at $u \neq 0$ contain non-trivial, potentially ambiguous
``anomalous terms'' involving the non-analytic structure of $R(u)$ at
$u = 0$. We have proposed a solution to lift these ambiguities in the
statics at two loops
\cite{ChauveLeDoussalWiese2000a,LeDoussalWieseChauve2002,ChauveLeDoussal2001}.
Since the large-$N$ limit allows in principle to handle higher-loop
corrections (i.e.\ to treat any $\epsilon$) it should be useful to
understand the many-loop structure of the field theory.  Stated
differently, we want to understand which physical quantity precisely
does the FRG computes?  Finally, developping a systematic $1/N$
expansion within the FRG for any $d$ should provide a novel handle to
attack problems such as KPZ, maybe avoiding the need for spontaneous
RSB altogether if it proves to be non-essential.

The aim of this paper is to study the FRG at large $N$.  For this
purpose we first perform an exact calculation of the effective action
$\Gamma[u]$ of the replicated field theory at large $N$. Its value for
a uniform mode and further expansion in cumulants yields a definition
of the renormalized disorder consistent with field theoretic
approaches. The second disorder cumulant is found to obey a closed
self-consistent equation. All higher cumulants can be constructed
recursively from the lower ones.  It can be easily inverted below the
Larkin scale and there the solution is analytic and corresponds to the
replica symmetric solution of MP. Varying with respect to an infrared
scale, here the mass, we obtain the FRG $\beta$-function in any $d$ at
dominant order, $N=\infty$. The continuation beyond the Larkin scale
is remarkably easier to perform on the resulting FRG equation.  Its
solution reveals that the FRG {\it exactly} reproduces the non-trivial
result of the GVM with full RSB for small overlap.  We also give a
formula which yields the complete RSB solution for all overlaps. At no
point in our derivation Parisi-RSB is invoked, as replica symmetry is
broken explicitly here. Since our saddle point equations for the
effective action contain both the MP equations and the FRG, it can be
used to describe the crossover from FRG to RSB. A qualitative analysis
of this crossover is given, as well as a comparison with previous
attempts to relate the FRG to the GVM
\cite{BalentsBouchaudMezard1996}. Finally, applications to other
problems and new perspectives are discussed.  A short version of this
work has appeared in \cite{LeDoussalWiese2001}.  In a related paper
\cite{LeDoussalWiesePREPg}, we give all details of the calculation
of the $O(1/N)$ corrections, with the aim of understanding finite but
large $N$.

The outline of the paper is as follows. In Section \ref{sec:Model and
program} we define the model, the effective action and its physical
interpretation. In Section \ref{sec:calculgamma} we compute the
effective action at large $N$, using the saddle point method and
perform a cumulant expansion (Section \ref{sec:Self-consistent
equation for the renormalized disorder}). A graphical interpretation
is given in Section \ref{s:graphinterprete}. In Section
\ref{sec:Functional renormalization group equations} we establish the
FRG equation at large $N$ (the $\beta$-function of the theory). Then
in Section \ref{detailed} we perfom a detailed analysis of the FRG
equation for a specific class of disorder correlators, both below and
above the Larkin scale. In Section \ref{sec:ComparisonRSBFRG} we
compare the FRG with the MP solution using RSB. First we recall the MP
approach and find agreement with the predictions of the FRG
calculation. Next we extend these results to an arbitrary disorder
correlator for which the GVM gives full RSB. Finally we discuss the
physical interpretation and compare our approach with the one of
Ref.~\cite{BalentsBouchaudMezard1996}.  Section \ref{sec:Discussion
and conclusion} presents the conclusion. The appendices contain
several generalizations, the calculation of the third and fourth
disorder cumulant, finite temperature fixed points, and an analysis
and comparison with the effective action in more conventional field
theories.

\section{Model and program}
\label{sec:Model and program}
\subsection{Model and large-$N$ limit} We consider the general model
for an elastic manifold of internal dimension $d$ embedded in a space
of dimension $N$. The position of the manifold in the embedding space
is described by a single valued displacement field $u(x)$, where $x$
belongs to the internal space and $u$ is a $N$ component vector which
belongs to the embedding space.  (Its components $u^i$, $i=1,\dotsc,
N$, are specified below only when strictly necessary.) A well studied
example is that of an interface (e.g.\ a domain wall in a magnet)
where $d=2$ and $N=1$. There $u(x)$ denotes the height of the
interface. Other examples are the directed polymer ($d=1$) in a $N$
dimensional space, which can be mapped to the $N$-dimensional Burgers
and Kardar-Parisi- Zhang (KPZ) equations \cite{KPZ}, or a vortex
lattice in the absence of dislocations described by $d=3$, $N=2$,
where $u(x)$ is there the deformation from the ideal crystal
\cite{BlatterFeigelmanGeshkenbeinLarkinVinokur1994,GiamarchiLeDoussal1995}.

We will study here the equilibrium statistical mechanics of such
an elastic manifold in presence of quenched disorder, modeled by
a random potential $V(x,u(x))$. It is described, in a given
realization of the random potential, by the partition function
\begin{equation}\label{lf1}
{\cal Z}_V=\int {\cal D}[u]\, \rme^{- {\cal H}_V[u]/T}
\ ,
\end{equation}
where
\begin{equation}\label{lf2}
\frac{{\cal H}_V[u]}{T} = \int_q ~ \frac{1}{2} ~ C(q)^{-1} ~ u(-q) \cdot u(q) + \frac{1}{T}
\int_x V(x,u(x)) 
\end{equation}
consists out of an elastic energy (expressed here in Fourier space and
taken to be isotropic), and of a pinning energy due to
disorder. Here and below we denote 
\begin{equation}\label{lf3}
 \int_q := \int \frac{\rmd^d q}{(2 \pi)^d} \quad , \quad 
\int_x := \int \rmd^d x 
\end{equation}
and $u \cdot v = \sum_{i=1}^N u^i v^i$. Throughout, square brackets as
e.g.\ in $A[u]$ denote a functional, here $A$ of the field $u_a(x)$,
while parenthesis as in $A(u)$ denote functions.

A convenient form for the inverse bare propagator, used below, is:
\begin{equation}\label{choice}
C(q)^{-1} = \frac{q^2 + m^2}{T}
\ ,
\end{equation}
where $T$ is the temperature and the elastic constant is set to unity
by a choice of units. The role of the additional mass term $m$ will be
discussed below. An additional small scale (ultraviolet, UV) cutoff
$\Lambda$ is implied here and will be made explicit when needed.

This model is highly non-trivial and, apart from the cases of $N=1$
and $d=0,1$, very few exact results are known
\cite{Kardar1987,BrunetDerrida2000,BrunetDerrida2000a,Johansson1999,%
PraehoferSpohn2000,LeDoussalMonthus2003}.  To obtain exact results for
large embedding space $N \to  \infty$, we need to consider a fully
isotropic version of the model with $O(N)$ symmetry such that the
model remains non-trivial in that limit. As in standard large-$N$
treatment (as for instance of the $\phi^4$ $O(N)$ model) one defines the
rescaled field
\begin{equation}\label{lf4}
v(x) = \frac{u(x)}{\sqrt{N}} \ .
\end{equation}
We will freely switch from one to the other in the following.  One
also chooses the distribution of the random potential to be $O(N)$
rotationally invariant. It can be parameterized by its set of
connected cumulants, of  the form
\begin{align}
&\overline{V(x,u) V(x',u')} = R(|u-u'|) \delta^d(x-x') \nonumber \\
& \qquad = N
B((v-v')^2) \delta^d(x-x') \qquad \label{correlator}\\
 &\overline{V(x_1,u_1)\dots V(x_p,u_p)}^{\mathrm{con}} \nonumber \\
& \qquad = N
\delta_{x_1,\dots ,x_p} (-1)^p S^{(p)}(v_1,\dots ,v_p)\ ,  \quad p \geq
3  ~~~ \\
& \delta_{x_{1},\dotsc ,x_{p}} :=
\prod_{i=2}^{p}\delta^{d} (x_{1}-x_{i})
\end{align}
This adequately models the case of uncorrelated (or short-range
correlated) disorder in the internal space, studied here. The second
cumulant, which plays the central role, is thus defined in terms of a
function $B(z)$. The higher cumulants are not strictly necessary in
the bare model, but they appear, as we will see, under coarse
graining. The distribution of disorder being translationally
invariant, these functions satisfy $S^{(p)}(v_1+v,\dots ,v_p+v) =
S^{(p)}(v_1,\dots ,v_p)$ for any $v$. The model studied here is thus a
slight generalization of the model studied by Mezard and Parisi
\cite{MezardParisi1991}, henceforth also referred to as MP, in the
same limit.

Although we will consider the general case, it is useful, as in MP
\cite{MezardParisi1991} to define two sets of simple models for which
more specific results will be given. These are, respectively, the
gaussian, short-range (SR) disorder, correlator
\begin{equation}\label{corrsr}
B(z)= g e^{- z}
\end{equation}
and the power-law correlations
\begin{eqnarray}\label{corrlr}
B(z) =  \frac{g}{(\gamma - 1)}  (a^2 + z)^{1 - \gamma} 
\ ,
\end{eqnarray}
which, for infinite $N$ always corresponds to long-range (LR)
disorder, a different universality class, as we will see below. For
finite $N$, the long-range disorder corresponds, at the bare level, to
$\gamma < 1 + N/2$; but this is modified at the renormalized level, and
the true frontier LR-SR for finite $N$ is non-trivial.

\subsection{Program} Having defined the model, and before turning to
calculations, let us first outline what we aim at.  All the
considerations in the present section are valid for any $N$, but,
since in the next section we will consider the large $N$ limit
explicitly, we already make apparent the rescalings.

The model defined above has already been studied in MP
\cite{MezardParisi1991}.  One of the aims of this study was to compute
the roughness exponent of the manifold, defined from the 2-point
function as
\begin{equation}\label{lf5}
\overline{\left<(u(x) - u(x'))^2 \right>} \sim A |x-x'|^{2 \zeta} \ .
\end{equation}
Besides the roughness exponent $\zeta$, the amplitude $A$ is also of
interest whenever it is universal, as it is the case e.g.\ for long
range disorder.  To this aim the model was replicated ($ u \to u_a$),
averaged over disorder and self-consistent saddle point equations
where derived for the 2-point function
\begin{equation}\label{lf6}
G_{ab}(q) \equiv \left<v_a(q) v_b(-q)\right> 
\ .
\end{equation}
This can always be done in a large-$N$ limit, and is then solved via a
RSB ansatz.

Our goal is in a sense broader. We want to understand the full
structure of the field theory, i.e.\ all correlation functions and not
only the 2-point one. We will thus instead study the generating
function of correlations as well as the effective action functional
which yields the renormalized vertices. This program, defined here,
will be carried out in the following sections explicitly for large
$N$. In this article we will restrict ourselves to dominant order, but
the aim is to understand large but finite $N$, including calculating
of $1/N$ corrections. This is deferred to \cite{LeDoussalWiesePREPg}.

\subsubsection{Effective action and field theory} All physical
observables for any $N$ can be obtained from the replicated action in
presence of a source, i.e.\ an external force $J_a(x)$ acting on each
replica:
\begin{equation}\label{repzj}
 {\cal Z}[J] = \overline{ \int \prod_a {\cal D} [u_a] e^{- \sum_a
H_V[u_a]/T +
\int_x \sum_a J_a(x)\cdot u_{a}(x) } } 
\ ,
\end{equation}
where $u_a(x)$, $a=1,\dots ,n$ are the replicated fields (each one
being an $N$ component vector $u^i_a(x)$).  Differentiating with
respect to the replicated source $J_a(x)$ in the limit $n \to 0$
yields all correlation functions. The finite-$n$ information is also
interesting. For instance from
\begin{equation}\label{lf7}
 {\cal Z} := {\cal Z}[J=0] = \overline{\exp(- n {\cal F}_V/T)}
\end{equation}
one can retrieve the sample to sample distribution of the free energy
${\cal F}_V = - T \ln {\cal Z}_V$, as was done e.g.\ in a finite size
system for $d=1$ \cite{BrunetDerrida2000a,GorokhovBlatter1999}.  Thus,
unless specified we will keep $n$ arbitrary.

One can explicitly perform the disorder average in (\ref{repzj}):
\begin{eqnarray}\label{lf8}
 {\cal Z}[J]&=& \int \prod_a {\cal D} [u_a] \rme^{ - N {\cal
S}[u,j] } \\
 {\cal S}[u,j] &=&  \frac{1}{2} \int_q  C(q)^{-1} v_a(-q) \cdot v_a(q) 
\nonumber \\ 
&& + \int_x [ U(\chi(x)) - j_a(x) \cdot
v_a(x) ]\qquad \label{lf9}\ ,
\end{eqnarray}
where here $\chi_{ab}(x) = v_{a}(x) \cdot v_{b}(x)$ and here and below
summations over repeated replica indices are implicit. We have
rescaled the source in a manner complementary to the field:
\begin{equation}\label{lf10a}
 J_a(x) = \sqrt{N} j_a(x)
\ .
\end{equation}
We have also introduced the bare interaction potential
\begin{equation}\label{U}
 U(\chi) = \frac{- 1}{2 T^2} \sum_{ab} B(\tilde{\chi}_{ab}) -
\frac{1}{3! T^3} \sum_{abc} S(\tilde{\chi}_{ab},
\tilde{\chi}_{bc}, \tilde{\chi}_{ca}) + \dots 
\ ,
\end{equation}
which is a function of a $n$ by $n$ replica matrix $\chi_{ab}$ and has
a cumulant expansion in terms of sums with higher numbers of replicas.
Because of translational symmetry and $O(N)$ invariance it depends
only on the matrix
\begin{equation}\label{lf11}
 \tilde{\chi}_{ab} := \chi_{aa} + \chi_{bb} - \chi_{ab} -
\chi_{ba} 
\end{equation}
and the form of each cumulant is restricted. For instance one has
$S^{(3)}(v_1,v_2,v_3) = S((v_1-v_2)^2,(v_2-v_3)^2,(v_3-v_1)^2)$
etc..  The matrix potential $U(\chi)$ can thus be considered as a
convenient way to parameterize the disorder (here the bare disorder).

The physical object which contains the information about the field
theory at large scale is the effective action. It is the generating
function of the 1-particle irreducible diagrams and in conventional
field theories its formal expansion in powers of the field yields the
renormalized vertices. All correlation functions are then obtained
simply as tree diagrams from these renormalized vertices.  In
particular it is known that within a $d=4-\epsilon$ expansion at zero
temperature to at least 2-loop order the theory can be renormalized
(i.e.\ rendered UV finite and yielding universal results) by
considering counter-terms only to the second cumulant. The latter is a
function $R(u)$, and can be viewed as the set of all coupling
constants which simultaneously become marginal in $d=4$.  To probe
renormalizability to any number of loops, we want to compute the
effective action from first principles.

The effective action functional is defined as a Legendre transform:
\begin{eqnarray}
 \Gamma[u] + {\cal W}[J] &=& \int_x \sum_a J_a(x) \cdot u_a(x) \\
 {\cal W} [J] &=& \ln {\cal Z}[J] 
\ .
\end{eqnarray}
Strictly speaking the definition is the convex envelope
$\Gamma[u]=\min_J (\int_x \sum_a J_a(x) \cdot u_a(x) - {\cal
W}[J])$. Here we apply the definition to the replicated action, and
will content ourselves with the differential definition
\begin{eqnarray}
 \frac{\delta \Gamma[u]}{\delta u_a(x)} &=& J_a(x) \\
 \frac{\delta {\cal W}[J]}{\delta J_a(x)} &=& u_a(x)
\ ,
\end{eqnarray}
which relates a pair of values $(J,u)$, later also denoted by
$(J[u],u)$.  Since $\Gamma[u]$ defines the renormalized vertices, its
zero momentum limit defines the {\it renormalized disorder}.  Thus in
order to compute the renormalized disorder, we only need to compute
$\Gamma[u]$ (per unit volume) for a {\it uniform} configuration of the
replica field $u_a(x)=u_a=\sqrt{N} v_a$ (a so-called fixed background
configuration). Because of the statistical tilt symmetry
\cite{SchulzVillainBrezinOrland1988,HwaFisher1994b,stsproof}, i.e.
invariance of disorder term in the replicated action (\ref{lf9}) under
the translation $v_a(x) \to v_a(x) + w(x)$, and of the $O(N)$
invariance one can argue, and this is what we find below, that for the
model (\ref{choice}) the scaled effective action per unit volume
(which for a uniform mode is simply a function of $u_a$) should have
the following form
\begin{equation}\label{lf12}
\hat{\Gamma}(v) := \frac{1}{L^d N} \Gamma(u) = \frac{1}{2 T} m^2
v_a^2 + \tilde{U}(v v)
\ ,
\end{equation}
where $L^d$ is the volume of the system, and here and below we use the
notation:
\begin{equation}\label{lf13}
v v := v_a \cdot v_b
\end{equation}
for the $n$ by $n$ replica matrix. This defines the renormalized
disorder. Furthermore, whenever $\tilde{U}(v v)$ can be expanded, up
to a constant, in the form:
\begin{equation}\label{cumulants}
\tilde{U}(v v)  = \frac{- 1}{2 T^2} \sum_{ab} \tilde{B}(v_{ab}^2)
- \frac{1}{3! T^3} \sum_{abc}
\tilde{S}(v_{ab}^2,v_{bc}^2,v_{ca}^2) + \dots 
\ ,
\end{equation}
where here and in the following we denote
\begin{equation}\label{lf14}
v_{ab} := v_a - v_b\ ,
\end{equation}
then (\ref{cumulants}) defines the {\it renormalized cumulant}
functions $\tilde{B}(z)$ etc.. As we will see below this is correct up
to some very subtle behavior at coinciding replica vectors (i.e.\
$v_{ab}=0$ for some pair $a,b$).  Also note that the constant part
$\tilde{U}(v.v=0)$ is the free energy.

The main result of the following sections will be the exact
calculation of the uniform part of the effective action, i.e.\ of the
function $\tilde{U}(v v)$. This will be performed within a large $N$
expansion:
\begin{equation}\label{lf15}
\tilde{U}(v v) = \tilde{U}_0(v v) + \frac{1}{N} \tilde{U}_1(v v) + \dotsb 
\end{equation}
and here we will obtain the dominant order $\tilde{U}_0(v v)$; the
corrections $\tilde{U}_1(v v)$ are calculated in
\cite{LeDoussalWiesePREPg}. It will be a function of a scale
parameter. We choose to add a mass-term $m$ which provides such a
scale. It is a convenient choice since for $m=\infty$ one has
$\tilde{U}=U$: Fluctuations are totally suppressed and the effective
action equals the action. One can then progressively lower the mass
down to zero, starting from this initial condition, since ultimately
one is interested in the massless limit. Another choice is to change
the UV cutoff, as will be discussed again below.

It is now useful to give a more direct physical interpretation of
this quantity, in addition to the above field theoretic
interpretation.

\subsubsection{Effective action as the distribution of the order
parameter} 
\label{orderparam}

The effective action for a uniform background is also known
to be related to the distribution of the order parameter. Let us
recall the relation for a simple pure ferromagnet. The unnormalized
probability distribution of the order parameter $\Phi = \frac{1}{L^d}
\int_x \phi(x)$ where $\phi(x)$ is the local magnetization is by
definition
\begin{equation}\label{lf16}
 Z(\Phi) = \int {\cal D}[\phi]\, \delta\left(\Phi - \frac{1}{L^d} \int_x
\phi(x)\right) \rme^{ - S[\phi] } \ ,
\end{equation}
where $S[\phi]$ is the action which describes the ferromagnet
(e.g.\ a $\phi^4$ theory or a Landau Ginsburg model). The
functional $W(J)$ evaluated for a uniform $J$ reads:
\begin{eqnarray}
W(J) &=& \int \rmd \Phi\, Z(\Phi) \, \rme^{L^d J \Phi} \nonumber \\
 &=& \int \rmd \Phi\,
\rme^{L^d ( \frac{1}{L^d} \ln Z(\Phi) + J \Phi )}  
\ .
\end{eqnarray}
In the large-volume limit, the saddle point can be taken and since the
Legendre transform is involutive, this yields the relation between
the effective action at $q=0$ per unit volume and the probability
distribution of the order parameter as:
\begin{equation}\label{lf17}
-\hat{\Gamma}(\Phi) =\lim_{L \to \infty}\frac{1}{L^d} \ln Z(\Phi) \ .
\end{equation}
In the thermodynamic limit the effective action per unit volume can
very well be a non-analytic function. This is the case e.g.\ in the
ferromagnetic phase where its left and right second derivatives at
$\Phi=M$ do not coincide ($M$ is the spontaneous magnetization per
unit volume).  While the right derivative at $\Phi=M$ is related to
the inverse susceptibility, the left one is zero, mathematically due
to the prescription to take the convex envelope, and physically
because one can always lower the magnetization at no cost in free
energy per unit volume by introducing a domain wall. The above
property (\ref{lf17}) can be extended to a given $q$ mode. Finally,
note that in $d=0$ the above does not hold since there is no large
factor $L^{d}$, and the probability distribution is directly given by
the action $S(\Phi=\phi)$.

What is then the physical meaning of the quantity that we will be
computing in the next sections? Let us in analogy to the magnetization
for a ferromagnet define the center of mass of an interface:
\begin{equation}\label{lf18}
w =  \frac{1}{N^{1/2} L^d} \int_x u(x)
\ .
\end{equation}
Since we have added a mass in the elastic energy (\ref{choice}), which
acts as an extra quadratic well bounding the fluctuations of the
interface, the disorder-induced fluctuations of the center of mass are
always finite. One expects that they diverge typically as $w \sim
m^{-\zeta}$ as $m \to 0$, thus their behavior as a function of $m$ is
of high interest and yields e.g.\ the information about the roughness
exponent.

One can then define the probability distribution $P_V(w)$ of the
center of mass of the interface in a given realization of the random
potential $V$ (and in presence of the quadratic well induced by the
mass). One can see that by definition the generating function for a
uniform $j$ is the Laplace transform of the probability distribution of
$w$, namely
\begin{equation}\label{lf19}
 Z(j) = \int \rmd w_1 \ldots \rmd w_n \overline{P_V(w_1) \dots
P_V(w_n)} \,\rme^{ - N L^d \sum_a j_a w_a}\ ,
\end{equation}
then by the same saddle point argument as for the ferromagnet one
expects, at least naively, that
\begin{eqnarray}
\label{prob1}
\hat{\Gamma}[\{w_a \}] = - \lim_{L \to \infty} \frac{1}{N L^d}
\ln \overline{P_V(w_1) \dots P_V(w_n)} \ .
\end{eqnarray}
Symbolically one can write:
\begin{eqnarray}
\label{prob2}
\overline{P_V(w_1) \dots P_V(w_n)} \approx e^{ - L^d N
\hat{\Gamma}[\{w_a \}] } \ ,
\end{eqnarray}
provided this is taken with a grain of salt. Thus one can also think of the
renormalized disorder $N \tilde{U}(v\cdot v)$ as parameterizing the set
of correlations of an effective equivalent toy model ($d=0$) which
has the same set of correlations as the center of mass variable
in the original model.

The $p$-th connected moment of the center of mass is identical, up
to a volume factor, to the zero momentum limit of the connected m
point correlator of the $u$ field, e.g.
\begin{equation}\label{lf20}
\left< w_{a_1}\dots w_{a_p} \right>_c = \frac{1}{L^d} \left<
v_{a_1}(q_1)\dots v_{a_p}(q_p)\right>_c |_{q_i=0}
\end{equation}
and, once the effective action is known, both can thus be obtained
in principle as the sum of all tree graphs made from
$\hat{\Gamma}[v]$ vertices. For instance the 2-point function
should be obtainable from:
\begin{equation}\label{lf21}
 \left< w_a w_b \right> = \frac{1}{L^d} G_{ab}(q=0) 
= \frac{1}{L^d} [\hat{\Gamma}''[v=0]]^{-1}_{ab} 
\end{equation}
and the connected 4-point function from:
\begin{eqnarray} \label{4pt}
 \left< w_a w_b w_c w_d\right>^{\mathrm{con}} &=& \frac{1}{L^{d}} 
G_{abcd}(q_i=0) \nonumber \\
& = & \frac{1}{N L^d} \sum_{efgh} [\hat{\Gamma}''[v=0]]^{-1}_{ae}
[\hat{\Gamma}''[v=0]]^{-1}_{bf}
 \nonumber \\
&& \times
[\hat{\Gamma}''[v=0]]^{-1}_{cg}
[\hat{\Gamma}''[v=0]]^{-1}_{dh} \nonumber \\
&& \times \hat{\Gamma}''''[v=0]_{efgh}
\end{eqnarray}
this however assumes analyticity, which as we will see below, does not
always hold. Another integral relation holds
\begin{eqnarray}
\left< w_a w_b \right> & \equiv& \int \rmd w_1 \dotsb \rmd w_n\, w_1 w_2
\overline{P_V(w_1)
\dots P_V(w_n)} \nonumber \\
& \equiv& W''_{12}(j=0)  \nonumber \\
& \approx& \int \rmd w_1 \dotsb \rmd w_n\, w_1 w_2 e^{ - L^d N
\tilde{\Gamma}[\{w_a \}] } \ .
\end{eqnarray}

\section{Calculation of the effective action}
\label{sec:calculgamma}

Let us now consider explicitly the large $N$ limit. One can
rewrite for any $N$ the starting generating function 
(\ref{lf8}-\ref{lf9}) as:
\begin{align}\label{lf84}
& {\cal Z}[J] = \int {\cal D} [u] {\cal D}[ \chi] {\cal D}[\lambda]
\rme^{ - N {\cal S}[u,\chi,\lambda,J] } \\
& {\cal S}[u,\chi,\lambda,j] = \frac{1}{2}\! \int_q C(q)^{-1} v_a(-q)
\cdot v_a(q) -\! \int_x j_{a}(x) \cdot
v_{a}(x) \qquad \nonumber \\
& + \int_x U(\chi(x)) - \frac{1}{2} i \lambda_{ab}(x)
\left[\chi_{ab}(x) - v_{a}(x) \cdot v_{b}(x)\right] \ ,\hspace{-1 cm}
\end{align}
where the replica matrix field $\chi(x) \equiv \chi_{ab}(x)$ has been
introduced through a Lagrange multiplier matrix
$\lambda_{ab}(x)$. Here and below summations over repeated replica
indices are implicit. One can then explicitly perform the functional
integration over the field $u(x)$ and obtain:
\begin{eqnarray}\label{lf85}
{\cal Z}[J] &=& \int D \chi D\lambda e^{ - N S[\chi,\lambda,j] } \\
 S[\chi,\lambda,j] &=& \frac{1}{2} \Tr \ln ( C^{-1} + i
\lambda ) \nonumber \\
&& + \int_x  U(\chi(x)) - \frac{i}{2} \lambda^{ab}(x) \chi^{ab}(x) \nonumber \\
&& - \frac{1}{2} \int_{x,x'} j_a(x) (C^{-1} + i \lambda)^{-1}_{a x, b
x'} j_b(x') \ , \qquad 
\end{eqnarray}
where the inversion and trace are performed in both replica space and
spatial coordinate space.

It has now the standard form for a saddle point evaluation 
of the functional ${\cal W}(J) = \ln {\cal Z}(J)$ except
that the saddle point is not, in general, uniform in space. It is 
useful to define the scaled functional $\tilde{W}(j)$ through
\begin{equation}\label{lf22}
 {\cal W}[J] = N \tilde{W}[j = J/\sqrt{N}]
\ ,
\end{equation}
which has a well defined large-$N$ limit and
can be expanded in $1/N$ as:
\begin{equation}\label{lf23}
 \tilde{W}[j] = W^0[j] + \frac{1}{N} W^1[j] +\dots 
\ .
\end{equation}
Deferring the calculation of the corrections to a future publication
\cite{LeDoussalWiesePREPg}, we obtain here the dominant order in $1/N$
as:
\begin{equation}\label{W0}
 W^0[j] = - S[\chi_{j},\lambda_{j},j] 
\ ,
\end{equation}
where $\chi_j$ and $\lambda_j$ depend on $j(x)$ and are the solutions
of the saddle point equations obtained respectively by setting to zero
the functional derivatives (at fixed $j(x)$):
\begin{eqnarray}\label{spgenWb}
 \frac{\delta S[\chi,\lambda,j]}
{\delta \lambda_{ab}(x)}\bigg| _{\chi=\chi_j, \lambda = \lambda_{j}} &=& 0  \\
 \frac{\delta S[\chi,\lambda,j]}
{\delta \chi_{ab}(x)}\bigg |_{\chi=\chi_j, \lambda = \lambda_{j}}  &=&  0
\label{spgenW}
\ .
\end{eqnarray}
The result is
\begin{eqnarray}
  \chi_j^{ab}(x) &=& (G_j)_{ax,bx} + (G_j :j)_{a x} \cdot (G_j :j)_{b
x}\qquad  \\ 
 i \lambda_j^{ab}(x) &=& 2 \partial_{ab} U (\chi_j(x)) \\
 G_j^{-1} &=& C^{-1} + i \lambda_j \label{spW} \ ,
\end{eqnarray}
where $G_j$ is a matrix with both replica indices and spatial
coordinates and inversion is carried out for both. Here and below,
replica indices are raised whenever explicit dependency is given,
e.g.\ $\chi_{ab} \equiv \chi_j^{ab}$. The notation for the
$N$-component vector $(G:j)^i_{b x} = \sum_c \int_y G_{b x,c y}
j^i_c(y)$ is a shorthand for a matrix product, and everywhere we
denote by
\begin{equation}\label{lf24}
 \partial_{ab} U(\phi) := \partial_{\phi_{ab}} U(\phi)
\end{equation}
the simple derivative of the function $U(\phi)$ with respect
to its matrix argument $\phi_{ab}$. Of course, if, for a
given $j(x)$ there are several solutions to these equations, then
one must sum over all saddle points, to the same order
\begin{equation}\label{lf25}
  {\cal W}[j] \approx \ln \Big( \sum_{\mathrm{sp}} \rme^{ - N
S[\chi_{\mathrm{sp}}(j),\lambda_{\mathrm{sp}}(j),j] } \Big)\ .
\end{equation}
This case will be discussed below, for now we ignore this possible
complication, as well as issues of stability of the saddle point.

Now we want to take the Legendre transform and trade the variable $j$
for the variable $v$ to obtain the effective action $\Gamma[u]$. One
also defines the scaled functional, and its $1/N$ expansion through:
\begin{eqnarray}
 \Gamma[u] &=&  N \tilde \Gamma[v=u/\sqrt{N}] \\
 \tilde{\Gamma}[v] &=& \Gamma_0[v] + \frac{1}{N} \Gamma_1[v] + \dots 
\ .
\end{eqnarray}
Then ($\tilde \Gamma[v]$,$\tilde{W}[j]$) and ($\tilde
\Gamma_0[v]$,$W_0[j]$) are also two pairs of Legendre transforms. Thus
the dominant order of the effective action functional in the large-$N$
limit is given by
\begin{equation}
 \Gamma_0[v] = \int_x v_a(x) \cdot j^a_v(x) - W_0[j_v] 
\label{legendre}
\end{equation}
with $W_0[j]$ given by (\ref{W0}), (\ref{spW}) and
where $j_v(x)$ is the $v(x)$-dependent source solution of:
\begin{equation}\label{v}
 \frac{\delta W_0[j_v]}{\delta j^a_v(x)} = v_a(x)
\ .
\end{equation}
One can now derive a self-consistent functional saddle point equation
for $\Gamma_0[v]$. First we establish the relation between $v$ and
$j_v$, namely
\begin{equation}
v_a(x) = (G_v :j_v)_{a x} \quad \Leftrightarrow \quad j_v^a(x) =
(G_v^{-1}:v)_{a x} \label{vj} \ ,
\end{equation} 
where from now on we define
\begin{equation}\label{lf26}
 G_v := G_{j_v} \ .
\end{equation}
Eq.~(\ref{vj}) is obtained noting that
\begin{eqnarray}
 v_a(x) &=& \frac{\delta W_0[j_v]}{\delta j^a_v(x)}
= - \frac{\rmd }{\rmd j_a(x)} S[\chi_j,\lambda_j,j]\bigg|_{j=j_v} \nonumber \\
& =& - \int_y \left[ \frac{\partial \chi_{j}(y) } {\p j_a(x)} \frac{\p
S}{\p \chi_j(y)} + \frac{\p \lambda_j(y)}{\p j_a(x)} \frac{\p S}{\p
\lambda_j(y)} \right]\bigg|_{j=j_v}
\nonumber \\
&& - \frac{\p S}{\p j_a(x)}\bigg|_{j=j_v} = 
- \frac{\p S}{\p j_a(x)}\bigg|_{j=j_v} = (G_{j_v}:j_v)_{a x} \nonumber
\ ,\\
\end{eqnarray}
where we have used the saddle-point equations (\ref{spgenWb}), (\ref{spgenW}).

We can now use (\ref{vj}) in the saddle point equations
(\ref{spgenWb}), (\ref{spgenW}) and defining
\begin{equation}\label{lf27}
\chi_v := \chi_{j_v} \ , \qquad \lambda_v := \lambda_{j_v} \nonumber\
,
\end{equation}
this yields a self-consistent equation for $\chi_v(x)$
\begin{eqnarray}
 \chi^{ab}_v(x) &=& v_a(x) \cdot v_b(x) + (G_v)_{ax,bx} \label{scchi} \\
 (G_v^{-1})_{ax,by} &=& (C^{-1})_{x,y} \delta_{ab}
+ 2 \partial_{ab} U(\chi_v(x)) \delta^d(x-y)\ , \nonumber \\
&& \label{scG}
\end{eqnarray}
which is also a self-consistent equation for $G_v$. Since the Legendre
transform is involutive, one can also write:
\begin{equation}\label{lf28}
 \frac{\delta \Gamma_0[v]}{\delta v_a(x)} = j_a(x) = (G_v^{-1} : v)_{ax} 
\ ,
\end{equation}
which determines the derivative of $\Gamma_0[v]$ once (\ref{scG}) is solved.

One can however do better. Using (\ref{vj}) in (\ref{legendre}) one
obtains the effective action for a spatially varying field $v(x)$:
\begin{equation}\label{lf29}
 \Gamma_0[v] = v:(G_v^{-1}):v + S[\chi_v,\lambda_v,j_v] \ ,
\end{equation}
which gives
\begin{eqnarray}
\Gamma_0[v] &=& \frac{1}{2}\nonumber
\int_{xy} C^{-1}_{a x, b y} v_a(x) v_b(y) \\
&& + \frac{1}{2} \Tr \ln( C^{-1} + 2 \partial U(\chi_v) )  
+ \int_x U(\chi_v(x)) \nonumber \\ 
&& {+}\! \int_x\!  v_a(x) \partial_{ab} U(\chi_v(x)) v_b(x)
{-} \chi^{ab}_v(x) \partial_{ab} U(\chi_v(x)) \ . \nonumber \\
&&
\end{eqnarray}
It is interesting to rewrite it with the help of (\ref{scG}) as a
functional of $G_v$ and $v$ only:
\begin{eqnarray}  
\Gamma_0[G_v,v] &: =& \Gamma_0[v] =
- \frac{1}{2} \Tr \ln G_v \nonumber \\
&& + \frac{1}{2} \int_{xy} C^{-1}_{a x, b y} \left[v_a(x) v_b(y) +
(G_v)_{a x, b y} \right] \nonumber \\
&& + \int_x U(vv(x) + (G_v)_{x,x}) \label{var}\ .
\end{eqnarray}
We have dropped a constant $\sim n$. (\ref{var}) has the property
\begin{equation}
 \partial_G \Gamma_0[G,v] |_{G=G_v} = 0 \label{saddleG}
\ ,
\end{equation}
where the derivative acts only on $G$, leaving fixed all $v$, since
this coincides with the saddle point equation (\ref{scG}). This makes
apparent that it can also be obtained from a {\it variational method}
where the average of the field is fixed, as we detail in Appendix
\ref{sec:var}. Since the explicit non-trivial $v$-dependence in
(\ref{var}) using (\ref{scG}) is purely in terms of the bilinears
$v_a(x) \cdot v_b(x)$ at the same space points, it also shows that one
can write:
\begin{equation}\label{form}
\Gamma_0[v] = \frac{1}{2}\, v:C^{-1}:v + \tilde U_0[v\cdot v] 
\ ,
\end{equation}
where the interaction (i.e.\ disorder) part satisfies
\begin{equation}\label{lf31}
 \frac{\delta \tilde U_0[v\cdot v]}{\delta (v_a(x) \cdot v_b(y)) } = 0\ ,
\qquad  x \neq y
\end{equation}
and is the solution of a self-consistent functional equation:
\begin{eqnarray}
&&\!\!\! \frac{\delta \tilde U_0[v\cdot v]}{\delta (v_a(x) \cdot v_b(x)) }
= \partial_{ab} U(vv(x) + (G_v)_{x,x}) \label{scfunctional} \\
&&\!\!\! (G_v^{-1})_{ax,by} = (C^{-1})_{x,y} \delta_{ab} +
\frac{\delta \tilde U_0[v\cdot v]}{\delta (v_a(x) \cdot v_b(x)) }
\, \delta^{d}(x-y) \nonumber \ .
\end{eqnarray}
A generalization of this equation is presented in Appendix
\ref{sec:appB}.

\section{Self-consistent equation for the renormalized disorder}
\label{sec:Self-consistent equation for the renormalized disorder}
\subsection{Uniform configuration and saddle point equation} 
\label{sec:sp} 
Let us now consider the simpler problem of computing the effective
action for a uniform field configuration, which can be solved
self-consistently. To be more specific we will focus on the form
(\ref{choice}) for the elastic energy. Also, to simplify notations and
since we will restrict ourselves to dominant order in $1/N$, we drop
the index $0$, so we set:
\begin{equation}\label{lf32}
 \Gamma_0 \to \tilde \Gamma \quad , \quad \tilde{U}_0 \to \tilde{U}
\end{equation}
and so on. 

For a uniform field $v_a(x)=v_a$ the effective action (\ref{form}) per
unit volume takes the form:
\begin{equation}\label{lf33}
\hat{\Gamma}(v) := \frac{1}{L^d} \tilde \Gamma(v) = \frac{1}{2 T} m^2
v_a^2 + \tilde{U}(v v) \ .
\end{equation}
Note that these are now simply functions (not functionals) of a
$N\times n$-component vector, and $\tilde{U}(v v)$ is a function of
the $n$ by $n$ matrix $v_a v_b$.

Eq.~(\ref{var}) yields also a formula for $\tilde U_{0}(v v)$ (up to a
constant):
\begin{eqnarray}
 \tilde U(v v) &=&  U(\chi_v) 
+ \frac{1}{2} \int_q  \tr  \Big\{  \ln\!\big((q^2 + m^2) \delta 
+ 2 T \partial U(\chi_v)\big) \nonumber \\
&& + (q^2 + m^2) \left[ (q^2 + m^2) \delta + 2 T \partial
U(\chi_v)\right]^{-1} \Big\}\ .\qquad \label{freeenergy}
\end{eqnarray}
The trace acts in replica space, and the log is a function of a
matrix, to be defined as usual.  Since (\ref{freeenergy}) contains the
derivative $\partial_{ab} \tilde{U}^0(v v)$ we must first determine
the latter.  One finds that analogous to (\ref{scfunctional})
\begin{eqnarray}
 \partial_{ab} \tilde U(v v) &=& \partial_{ab} U(\chi_v) \\
 \chi_v^{ab} &=& v_a v_b + T \int_{q} [ (q^2 + m^2) \delta + 2 T
\partial U(\chi_v) ]_{ab}^{-1} \qquad \label{saddle} \ .
\end{eqnarray}
Since one can replace the matrix $\partial U(\chi_v)$ by $\partial
\tilde U(v v)$ in the denominator of (\ref{saddle}), this is also a
self-consistent equation, which involves only $\partial_{ab}
\tilde{U}(v v)$. Here inversion is simple $n$ by $n$ matrix inversion
and $\delta$ is the Kronecker $n$ by $n$ identity matrix
$\delta_{ab}$. One must be careful that
\begin{equation}\label{lf34}
\partial_{ab} \tilde U(v v) = \frac{\partial \tilde U(v
v)}{\partial(v_a \cdot v_b)}
\end{equation}
is a {\it first derivative} of $\tilde U(v v)$ with respect to the matrix
element $v_a \cdot v_b$.  One can also check that taking the
derivative of (\ref{freeenergy}) with respect to $v_a \cdot v_b$
correctly reproduces (\ref{saddle}). A direct derivation uses
$\partial\chi\partial (vv)$ from Eq.~(\ref{saddle}). A more clever way
is to remember that because of (\ref{saddleG}), one is allowed to
differentiate only with respect to the explicit $vv$ in $\chi_v$ in
the first term, and that the remaining terms can be written as a
function of $G_{v}$ only, and using again (\ref{saddleG}). 

This self-consistent equation for $\partial \tilde{U}(v v)$, i.e.\ for
the uniform part of the effective action is one of our main results
and the remainder of this paper is devoted to analyze it. It contains
a huge amount of information, since it encodes the {\it full
distribution} (i.e.\ all cumulants) of the renormalized disorder, and
is thus quite non-trivial to analyze. It includes both the Gaussian
variational Method (GVM) of Mezard-Parisi\cite{MezardParisi1991} and
the functional renormalization group (FRG). For simplicity, we now
consider the bare disorder to be gaussian and set all bare cumulants
except the second cumulant $B(z)$ to zero.

The GVM is recovered upon setting $v=0$ which is one limit in which
the equation ``simplifies''. One sees that (\ref{saddle}) then
reproduces the Mezard Parisi equations, the self energy $\sigma_{ab}$ and two
point function $G_{ab}(k)$ in Ref.\ \cite{MezardParisi1991} being:
\begin{eqnarray}
 \sigma_{ab} &=& 2 T \partial_{ab} U(\chi_{v=0}) \\
 G_{ab}(k) &=& G_{v=0}^{ab}(k) \\
 (\chi_{v=0})_{ab}&=& \int_k G_{ab}(k) \ .
\end{eqnarray}
In the glass phases, these exhibit spontaneous replica symmetry breaking
(RSB) with multiple solutions corresponding to saddle points
obtained via replica permutations, and the above equations are
solved by a hierarchical Parisi
ansatz for $\chi(v{=}0)_{ab} = \chi(v{=}0)({\sf u})$ where $0 \leq
{\sf u} \leq 1$ is the overlap between replicas $a$ and $b$.  We will
give more details about this correspondence in the following.

For now we will study the opposite limit of ``strong'' explicit
symmetry breaking field (all $v_{ab}\equiv v_a - v_b \neq 0$). Then we
expect that the renormalized disorder $\tilde{U}(v v)$ is given by a
single saddle point and can be expanded in replica sums in terms of
unambiguous renormalized cumulants, i.e.\ up to a constant
\begin{equation}\label{cumulexp}
\tilde{U}(v v)  =  \frac{- 1}{2 T^2} \sum_{ab}
\tilde{B}(v_{ab}^2) - \frac{1}{3! T^3} \sum_{abc}
\tilde{S}(v_{ab}^2,v_{bc}^2,v_{ca}^2) + \dots 
\ .
\end{equation}
This is the limit solved here, which we will show below is the natural
limit in the FRG, and amounts, as we will discuss, to forcing the
manifold in distant states within the RSB picture. The rich crossover
to RSB contained in (\ref{saddle}), when some of the $v_{ab}$ are set
to zero will be discussed below.

\subsection{Cumulant expansion} We now transform equation
(\ref{saddle}) for the formal function $\tilde U(vv)$ in a set of
equations for the second, third, fourth, a.s.o.\ cumulants. This is
performed through an expansion in sums over an increasing number of
free replica indices, and {\em is not an approximation}. The such
obtained equations are as exact as (\ref{saddle}), i.e.\ exact to
dominant order at large $N$, albeit more explicit.  In fact, they allow
a recursive exact calculation of {\it all cumulants}. Their increasing
complexity will illustrate the wealth of information summarized in
(\ref{saddle}).

Let us first rewrite (\ref{saddle}) using an infinite series:
\begin{eqnarray}
 \partial_{ab} \tilde{U}(v v) &=& \partial_{ab} U(\chi_v)  \label{series} \\
 \chi_v &=& v v + T I_1 \delta + 
T \sum_{n=1}^{\infty} I_{n+1} (-2 T \partial \tilde{U}(v v))^{n} \nonumber \\
 I_{n} &: =& \int_k \frac{1}{(k^2 + m^2)^n} 
\ , \label{chidef}
\end{eqnarray}
where the $n$-th power here denotes the matrix product. 

Since we consider a gaussian bare model (\ref{U}) where only the
second cumulant is non-zero one has:
\begin{equation}
 - 2 T \partial_{ab} U(\chi_v) = \frac{2}{T} \left[ -
B'(\tilde{\chi}^{ab}_v) + \delta_{ab} \sum_c B'(\tilde{\chi}^{ac}_v)
\right] \label{chipower1}
\end{equation}
using that $\partial_{ab} \tilde{\chi}_{ab} = 2 (\delta_{ab} - 1)$
\cite{Note1}.  The same quantity for the renormalized disorder reads:
\begin{eqnarray}
 - 2 T \partial_{ab} \tilde U(vv) &=& \frac{2}{T} ( - \tilde B'_{ab} +
\delta_{ab} \sum_c
\tilde B'_{ac} ) \nonumber \\
&& + \frac{2}{T^2} \left[- \sum_g \tilde{S}'_{1,abg} + \delta_{ab} \sum_{c
g} \tilde{S}'_{1,acg} \right] \nonumber \\
&&+ \cdots 
\ , \label{power1}
\end{eqnarray} 
where we denote $B'_{ab}=B'(v^2_{ab})$,
$\tilde{S}_{abc}=\tilde{S}(v^2_{ab},v^2_{bc},v^2_{ac})$ and
$\tilde{S}'_{1,abc}$ denotes a derivative with respect to the first
argument of the function $\tilde{S}$ ($S$ has the symmetries implied
by replica permutation symmetry). All matrices we will encounter can
be parameterized as:
\begin{eqnarray}
X_{ab} &=& x_{ab} + \delta_{ab} x_a \\
x_{ab} &=& x^{(0)}_{ab} + x^{(1)}_{ab} + x^{(2)}_{ab} + \dots  \\
 x_a &=& x^{(0)}_a + x^{(1)}_{a} + x^{(2)}_{a} + \dots 
\ ,
\end{eqnarray}
where $x_{ab}$ do not contain any explicit Kronecker $\delta_{ab}$,
the upper index denotes the number of free replica sums,
e.g.\ $x^{(1)}_{ab} = \sum_f x_{ab;f}$, $x^{(2)}_{ab} = \sum_{fg}
x_{ab;fg}$.  Since under matrix product $(X^p)_{ab}$ or Hadamar
product $(X_{ab})^p$ the number of sums can only increase, one gets
only a finite number of terms in projecting out on terms with a given
number of free replica sums.

If we parameterize in the same way:
\begin{eqnarray} \label{defoff}
 \chi^{ab}_v &=& \chi_{ab} + \delta_{ab} \chi_{a} \\
 \tilde{\chi}^{ab}_v &=& \tilde{\chi}_{ab} + \delta_{ab} \tilde{\chi}_{a} 
\end{eqnarray}
then one easily sees that:
\begin{eqnarray}
&& B'(\tilde \chi^{ab}_v) = \delta_{ab} [ B'(\tilde{\chi}_{aa} +
\tilde{\chi}_{a}) - B'(\tilde{\chi}_{aa}) ] +
B'(\tilde{\chi}_{ab}) \nonumber \\
&& B'(\tilde \chi^{ab}_v ) -  \delta_{ab} \sum_c B'(\tilde \chi^{ac}_{v}) =
B'(\tilde{\chi}_{ab}) -  \delta_{ab} \sum_c B'(\tilde{\chi}_{ac})
\nonumber\ . \\ 
&& \label{simpli}
\end{eqnarray}
We can now expand in number of sums:
\begin{equation}\label{lf35}
 B'(\tilde{\chi}_{ab}) = B'(\tilde{\chi}^{(0)}_{ab}) +
B''(\tilde{\chi}^{(0)}_{ab}) \tilde{\chi}^{(1)}_{ab} + \dots 
\end{equation}
and the equivalence of (\ref{chipower1}) and (\ref{power1}) using
(\ref{simpli}), implies:
\begin{eqnarray}
 \tilde B_{ab}' &=& B'(\tilde{\chi}^{(0)}_{ab}) \\
 \frac{1}{T} \sum_g \tilde{S}'_{1,abg} &=&
B''(\tilde{\chi}^{(0)}_{ab}) \tilde{\chi}^{(1)}_{ab} \label{eqcum}
\end{eqnarray}
and so on for higher cumulants.
Thus to obtain the second renormalized cumulant we only need to
compute the part $\tilde{\chi}^{(0)}_{ab}$ of $\tilde \chi^{ab}_v$
which contains zero sum and no explicit $\delta_{ab}$. One has in
general:
\begin{equation}\label{p}
 \tilde{\chi}^{(p)}_{ab} = \chi^{(p)}_{aa} + \chi^{(p)}_{bb} +
\chi^{(p)}_{a} + \chi^{(p)}_{b} - 2 \chi^{(p)}_{ab} \ .
\end{equation}
Thus for the second cumulant we need only $\chi^0_{ab}$ and $\chi^0_a$. Since one has, 
to be explicit:
\begin{widetext}
\begin{eqnarray}\label{power2}
[(- 2 T \partial \tilde{U})^2]_{ab} &=&
\frac{4}{T^2} \!\left[ \delta_{ab} \sum_{ef} \tilde{B}'_{ae} \tilde{B}'_{af} 
- \tilde{B}'_{ab} \sum_f (\tilde{B}'_{af} + \tilde{B}'_{bf}) 
+ \sum_c \tilde{B}'_{ac} \tilde{B}'_{cb} \right] \nonumber  \\
&& + \frac{4}{T^3} \!\left[ 2 \delta_{ab} \sum_{egh} \tilde{B}'_{ae}
\tilde{S}'_{agh} - \tilde{B}'_{ab} \sum_{gh} ( \tilde{S}'_{bgh} {+}
\tilde{S}'_{agh} ) - \sum_{eh} (\tilde{B}'_{ae} \tilde{S}'_{abh} +
\tilde{B}'_{be} \tilde{S}'_{abh})
+ \sum_{h c} (\tilde{B}'_{ac} \tilde{S}'_{cbh} {+} \tilde{B}'_{bc}
\tilde{S}'_{cah})\right]\!\!\nonumber \\ 
&& + \cdots
\end{eqnarray}
\end{widetext}
where all terms not written have at least three free replica sums
(this is the case for $O(S^2)$ as well as terms involving the fourth
cumulant and higher). Similarly $[(- 2 T \partial_\chi
\tilde{U})^p]_{ab}$ has at least $p-1$ free replica sums (from the
$O(B^p)$ term). This is much more what we need, which comes only from
(\ref{power1}) and, using (\ref{series}):
\begin{eqnarray}
 \chi^{(0)}_{ab} &=& v_a v_b - 2 I_2 \tilde B'_{ab} \\
 \chi^{(0)}_a &=& T I_1 
\ .
\end{eqnarray}
This yields:
\begin{equation}\label{lf36}
\tilde{\chi}^{(0)}_{ab} = (v_a - v_b)^2 + 2 T I_1 + 4 I_2 (\tilde
B'_{ab} - \tilde B'_{aa}) \ .
\end{equation}
Thus we find that the renormalized second cumulant satisfies a closed
equation at any $T$:
\begin{equation}
\label{saddlepointforB} \tilde{B}'(v_{ab}^2) = B'(v_{ab}^2 + 2 T I_1 +
4 I_2 (\tilde{B}'(v_{ab}^2) - \tilde{B}'(0)))
\end{equation}
with no other contributions from higher cumulants at any $T$.
Appendix \ref{scnonloc} contains a non-local extension of this
formula.
Eq.~(\ref{saddlepointforB}) can be integrated with the result
\begin{eqnarray}\label{tildeBmagic}
\tilde B (v^{2}) &=& B \!\left(v^{2}+2 T I_{1}+ 4 I_{2}[\tilde B'
(v^{2})-\tilde B' (0)]\right)\nonumber \\
&& - 2 I_{2} \left\{ B' \!\left(v^{2}+2 T I_{1}+ 4 I_{2}[\tilde B'
(v^{2})-\tilde B' (0)]\right)\right\}^{2}\ .\nonumber \\
&&
\end{eqnarray}
A direct derivation from  (\ref{lf85}) is also possible.

\subsection{Higher cumulants}
\begin{figure}[b]
\fig{0.5\columnwidth}{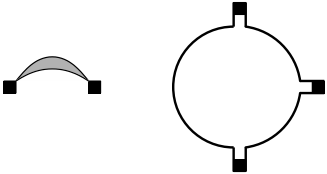} \caption{Graphical representation
of the third cumulant. The notation is explained in
\protect\cite{LeDoussalWiesePREPg}.  The first diagram yields the
terms proportinal to $I_{2}$, the second diagram the terms
proportional to $I_{3}$ in Eq.~(\ref{3cum}).}  \label{f:3rdcumulant}
\end{figure}Higher cumulants of the renormalized disorder can be
obtained by the same method using (\ref{eqcum}) and its
extensions. They can also be obtained by the graphical method. For
simplicity here we give only the expression of the third cumulant. The
complete expression for the fourth cumulant, together with all
calculational details and an introduction to the graphical method, can
be found in Appendix \ref{sec:highercum}.

The third cumulant is found to be:
\begin{eqnarray}
 \tilde S(x,y,z) &=& \frac{ 6 T I_2 }{1 + 4 I_2 \tilde{B}''(0)} 
\text{sym}_{x,y,z} \Big[\tilde{B}'(x) \tilde{B}'(y)\Big] \nonumber \\
&& + 24 I_3 \text{sym}_{x,y,z} \Big[
 (\tilde{B}'(x) - \tilde{B}'(0)) \tilde{B}'(x) \tilde{B}'(y) \Big] \nonumber \\
&& - 8 I_3 \tilde{B}'(y) \tilde{B}'(z) \tilde{B}'(x) \label{3cum} \ ,
\end{eqnarray}
where $\text{sym}_{x,y,z}$ is $1/6$ times the sum of all permutations
of $x,y,z$. Note that this relation is exact for all values of the
mass $m$, and not just a fixed point form. The only input in the
derivation is the absence of a third cumulant for the bare model
($m=\infty$). It would be interesting to include an additional bare
third cumulant. The fourth cumulant is derived in appendix
\ref{sec:highercum}, where also details for the graphical method are
given.

\section{Graphical Interpretation}\label{s:graphinterprete}
\begin{figure}[tb]
\centerline{\fig{0.75\columnwidth}{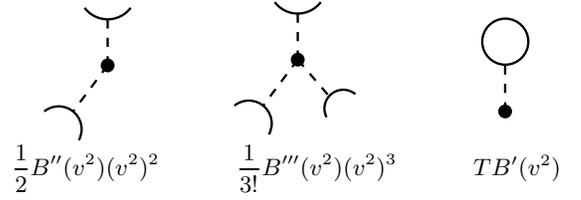}}
\centerline{$\displaystyle \frac{1}{2}B''(v^2) (v^{2})^{2} $
\hspace{0.1\columnwidth} $\displaystyle \frac{1}{3!}B'''(v^2)
(v^{2})^{3} $ \hspace{0.1\columnwidth} $TB' (v^{2}) $}
\caption{Examples for vertices and the 1-loop tadpole diagram which is
dominant at large $N$.}  \label{fig:vert+tadpole}
\end{figure}
\begin{figure}[tb] 
\centerline{\Fig{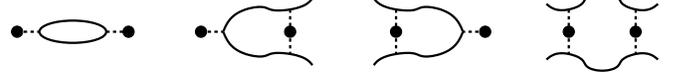}} \caption{The four 1-loop diagrams
correcting the disorder. A fat dot represents a vertex $B$, a solid
line the field $u$, and its correlator. A dashed line attaches two
fields $u$ to a vertex $B$.  We do not draw replica-indices.}
\label{fig:1loopNcomp}
\end{figure}
\begin{figure}[tb]
\Fig{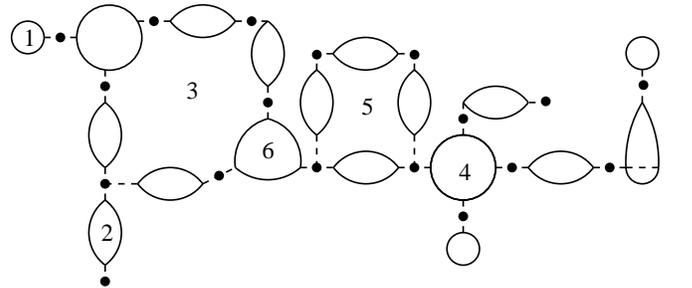}
\caption{Loops which give additional factors of $1/N$, as explained in
the main text.}
\label{fig:badloops}
\end{figure}
\begin{figure}[tb]
\Fig{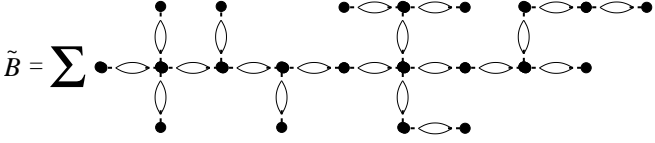}
\caption{Tree-configurations which contribute to $\tilde B (v^{2})$.}
\label{fig:tree1}
\end{figure}
\begin{figure}[tb]
\centerline{ \Fig{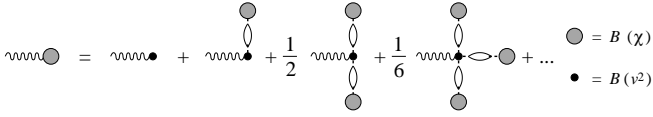} }
\caption{Self-consistent equation at leading order for  $\tilde
B' (v_{ab}^{2})=B' (\chi_{ab} )$. The wiggly line denotes
a derivative, and is combinatorially equivalent to choosing one
$B$. At finite $T$ one can attach an additional arbitrary number
of tadpoles to any $B$.} \label{fig1}
\end{figure}
In this section, we sketch how the central results at large $N$ can be
obtained graphically, first the saddle-point equation
(\ref{saddlepointforB}), which gives the effective disorder $\tilde B$
as a function of the bare disorder $B$, and second the
$\beta$-function (\ref{betafinal}).

The graphical rules for the perturbation theory of the replicated
model have been described in detail in \cite{LeDoussalWieseChauve2003}
for $N=1$ and we refer the reader to this work for elementary
details. Here there are in addition $N$ components of the field
$u_a^i$, the propagator being diagonal in all indices.  For the
present purpose we are mostly interested in the counting in $N$, and
since it is difficult to represent graphically both vector- and
replica-indices, we work with unsplitted vertices (see
\cite{LeDoussalWieseChauve2003}) and specify the replica content only
when needed.  Disorder vertices may contain arbitrary number of
derivatives and some examples are represented on Fig.\
\ref{fig:vert+tadpole}.  As usual there is a factor of $1/N$ per
derivative (i.e.\ per dashed line) at each vertex (see e.g.\ Fig.\
\ref{fig:vert+tadpole}, using $v^2=u^2/N$), $N$ per vertex, and $N$ per
loop.

We consider the effective action, i.e.\ the sum of all 1-particle
irreducible diagrams (1PI), and later focus on its 2-replica part.
We start our analysis at $T=0$ with the three possible 1-loop
diagrams, as presented on figure \ref{fig:1loopNcomp}. They are
obtained from contracting
\begin{equation}\label{iosdu}
\frac{N}{2}\sum_{ab} B\big(\textstyle \frac{(
u^{a}_{x}-u^{b}_{x})^{2}}N \big)\ \displaystyle \frac{N}{2}\sum_{cd}
B\big(\textstyle \frac{(u^{c}_{y}-u^{d}_{y})^{2}}N \big) \ .
\end{equation}
In order to simplfy the calculation we omit the terms taken at
coinciding replicas (e.g.~$B' (0)$), they can be added at the
end. Contracting (\ref{iosdu}) twice between points $x$ and $y$ gives
\begin{eqnarray}\label{sl;dfj}
\frac{N}{2} C_{xy}^{2}\sum_{ab} \!&\bigg[&\! B'\big(\textstyle \frac{(
u^{a}_{x}-u^{b}_{x})^{2}}N \big) B'\big(\textstyle \frac{(
u^{a}_{y}-u^{b}_{y})^{2}}N \big) \nonumber \\
&&\hspace{-1cm}+\displaystyle \frac{2}{N} B'\big(\textstyle \frac{(
u^{a}_{x}-u^{b}_{x})^{2}}N \big) B''\big(\textstyle \frac{(
u^{a}_{y}-u^{b}_{y})^{2}}N \big)\frac{( u^{a}_{y}-u^{b}_{y})^{2}}N\nonumber \\
&&\hspace{-1cm} +\displaystyle \frac{2}{N}  B''\big(\textstyle \frac{( 
u^{a}_{x}-u^{b}_{x})^{2}}N \big)\frac{( u^{a}_{x}-u^{b}_{x})^{2}}N
B'\big(\textstyle \frac{( 
u^{a}_{y}-u^{b}_{y})^{2}}N \big)\nonumber \\
&&\hspace{-1cm} +\frac{4}{N}B''\big(\textstyle \frac{( 
u^{a}_{x}-u^{b}_{x})^{2}}N \big)\frac{( u^{a}_{x}-u^{b}_{x})^{2}}N
B''\big(\textstyle \frac{(
u^{a}_{y}-u^{b}_{y})^{2}}N \big)\frac{(
u^{a}_{y}-u^{b}_{y})^{2}}N\bigg]\nonumber \\
&&\hspace{-1.8cm} +\,\mbox{higher replica terms}
\end{eqnarray}
This is graphically depicted on figure \ref{fig:1loopNcomp}.  The
important observation is that only the first diagram, with a closed
$u$-loop is contributing in the limit of large $N$. This analysis can
be repeated to higher loop order. Again, only diagrams as the first
one on figure \ref{fig:1loopNcomp} contribute. Especially, there are
no loops with three propagators or more, as loops 4 or 6 on figure
\ref{fig:badloops}. Also, there are no ``meta-loops'', i.e.\ loops
formed by loops, as loop 5 on figure \ref{fig:badloops}. Finally, only
diagrams as those on figure \ref{fig:tree1} survive, which as building
block have only the elementary 1-loop diagram with a closed loop
contributing a factor of $N$, as the first diagram on figure
\ref{fig:1loopNcomp}.  These are tree-like diagrams, where the nodes
are made out of $B$ and the links out of the before-mentioned 1-loop
diagram (two parallel replica lines in the splitted diagrammatics
\cite{LeDoussalWieseChauve2003} which produce the desired 2-replica
term). At junction points the replica lines branch also in
parallel. These are of course not tree diagrams, i.e.\ they are 1PI
and contribute to the effective action. Note on Fig.\ \ref{fig:tree1}
that  since there is a $1/(2 T^2)$ factor per vertex, but that
each vertex (except one) comes with two propagators (factor $T^2$) the
counting in temperature is right to produce a 2-replica term with the
expected $1/(2 T^2)$ global factor (the 3-replica terms proportional
to $T$ have been discarded etc.).

We are now in a position to derive the self-consistent equation
(\ref{saddlepointforB}). The key-observation is that deriving $\tilde
B (v^{2})$ once with respect to its argument, amounts in the graphical
intepreation of figure \ref{fig:tree1} to choose one of the bare
vertices $B (v^{2})$, and deriving it. $\tilde B' (v^{2})$ thus is $B'
(v^{2})$, with as many branches attached as one wants. Every branch
consists out of a 1-loop integral $\textdiagram{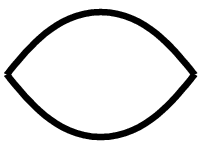}=I_{2}$ times
another tree; the latter is again $\tilde B (v^{2})$, given that one
of the bare vertices is chosen, i.e.\ again $\tilde B' (v^{2})$. Since
attaching loops to $B' (v^{2})$ amounts to deriving $B' (v^{2})$ once
for every loop, we arrive at
\begin{eqnarray}\label{lf86}
\tilde B' (v^{2}) &=& \sum_{\ell=0}^{\infty} \frac{B^{(n+1)} (v^{2})
\left[4 I_{2} (\tilde B' (v^{2})-\tilde B' (0))\right]^{\ell }}{\ell
!} \nonumber \\
&=& \label{resumtildeB} B'\! \left(v^{2} + 4 I_{2} [\tilde B'
(v^{2})-\tilde B' (0)] \right)\ .
\end{eqnarray}
Note that we have added the term with coinciding replica-indices,
dropped previously. The combinatorial factor comes from the expansion
of the exponential function in $\rme^{-\cal S}$. That it indeed resums
to $B'$ with a shifted argument is natural: For a function $f(x)$
taking the expectation value $\left< f(x) \right>$ in a theory with
only a first moment $\left< x \right>$ is equivalent to calculating
$f(\left< x \right>)$. Taylor-expanding the latter leads to the above
combinatorics.

By the same arguments the full effective action can be written as the
sum over tree-like (but not tree) diagrams represented in
Fig.\ \ref{fig:tree1} where, in addition, each vertex can be dressed by
an arbitrary number of tadpoles (see Fig.\ref{fig:vert+tadpole}).
Each tadpole brings an additional factor of $T$, thus tadpoles
contribute to the two replica term only at $T>0$. At finite
temperature, any of the $v^{2}$'s could be contracted, leading to the
relacement
\begin{equation}\label{treplace}
(v^{a}-v^{b})^{2} \to (v^{a}-v^{b})^{2} + 2 T I_{1}\ .
\end{equation}
(This offers another possibility to verify the combinatorics in
\ref{lf86}.)  Thus the final result is
\begin{equation}\label{tildeBfinal}
\tilde B' (v^{2}) = B' \!\left(v^{2}+2 T I_{1}+ 4 I_{2}[\tilde B'
(v^{2})-\tilde B' (0)]\right)
\end{equation}

We now illustrate how to recover the $\beta$-function. Applying
$-m\partial /\partial m$ to $\tilde B$ implies to derive each integral
w.r.t.\ $m$ appearing in each loop of Fig \ref{fig:tree1}.
Diagrammatically this amounts to choosing in the tree of Fig
\ref{fig:tree1} one of the bonds (loop $I_{2}$) which connects two
$B$'s. Summing over all trees, it gives a term
\begin{equation}\label{sdfjkl}
2 \left( -m\frac{\partial}{\partial m}I_{2} \right) \left[ \tilde B'
(v^{2})^{2} -2 \tilde B' (v^{2}) \tilde B' (0)\right]
\end{equation}
since the two trees attached to the loop $I_{2}$ are nothing but $ B
(v^{2})$, derived once, and again itself with things attached,
i.e.~$\tilde B' (v^{2})$ as given in (\ref{tildeBfinal}). This
reproduces the $T=0$ term in (\ref{betafinal}).

The second contribution comes from deriving $TI_{1}$. The graphical
derivation is complicated, and we refer the interested reader to
\cite{LeDoussalWiesePREPg} where a more complete, but much more
involved, diagrammatic method is presented.

\section{Functional renormalization group equations}
\label{sec:Functional renormalization group equations}
\subsection{From self-consistent to FRG equation}
We will now study the self-consistent equation, exact for $N=\infty$,
for the second cumulant correlator of the random potential that we
have derived in the previous Section:
\begin{equation}\label{starting}
 \tilde{B}'(x) = B' \Big( x + 2 T I_1 + 4 I_2 (\tilde{B}'(x) - \tilde{B}'(0)) \Big) 
\ ,
\end{equation}
which involves only the two one loop integrals:
\begin{eqnarray}
  I_1 &=& \int_0^\Lambda \frac{d^d k}{(2 \pi)^d} \frac{1}{k^2 + m^2} \\
  I_2 &=& \int_0^\Lambda \frac{d^d k}{(2 \pi)^d} \frac{1}{(k^2 + m^2)^2}
\ ,
\end{eqnarray}
where we have indicated symbolically that a short scale UV cutoff is
needed for $I_2$ to be finite if $d \geq 4$ and for $I_1$ for $d \geq
2$.

There is a simple way to obtain directly the solutions of
(\ref{starting}) which we will detail below. It is also interesting to
turn this equation into a FRG equation for the function $\tilde{B}(x)$
as a function of the scale parameter $m$. Indeed this yields the
$\beta$-function of the field theory in the limit of infinite $N$,
which is our main goal. Let us show first how one does it.

Let us first take a derivative of (\ref{starting}) with respect to
$x$. One obtains:
\begin{equation}\label{secder}
 \frac{\tilde{B}''(x)}{1 + 4 I_2 \tilde{B}''(x)} = 
B''\Big( x + 2 T I_1 + 4 I_2 (\tilde{B}'(x) - \tilde{B}'(0)) \Big)
\ .
\end{equation}
Taking the derivative $m \partial_m$ of (\ref{starting}) and using
(\ref{secder}) gives:
\begin{eqnarray}
 m \partial_m \tilde{B}'(x) &=& \frac{\tilde{B}''(x)}{1 + 4 I_2
\tilde{B}''(x)}
\Big[2 m \partial_m T I_1 \nonumber \\
&& + 4 (m \partial_m I_2) (\tilde{B}'(x) - \tilde{B}'(0)) \nonumber \\
&& +
4 I_2 m \partial_m \tilde{B}'(x)  - 4 I_2 m \partial_m \tilde{B}'(0)
\Big] \ .\qquad  \label{5.5}
\end{eqnarray}
Regrouping the terms one obtains:
\begin{eqnarray}
 m \partial_m \tilde{B}'(x) &=& \tilde{B}''(x) \Big[ 2 m \partial_m T
I_1 - 4 I_2 m
\partial_m \tilde{B}'(0) \nonumber
\\
&& \hphantom{\tilde{B}''(x) \Big[}+ 4 (m \partial_m I_2) 
(\tilde{B}'(x) - \tilde{B}'(0))  \Big] \ . \qquad \label{5.7}
\end{eqnarray}
From (\ref{5.5}) one also has
\begin{equation}\label{lf37}
 m \partial_m \tilde{B}'(0) =  \frac{\tilde{B}''(0)}{1 + 4 I_2 \tilde{B}''(0)}
2 m \partial_m (T I_1)
\ .
\end{equation}
Inserting (\ref{lf37}) into (\ref{5.7} ) finally yields
\begin{eqnarray} 
m \partial_m \tilde B'(x) &=& \tilde B''(x)
\Big[ 2 (m  \partial_m T I_1) \frac1{1+ 4 I_2 \tilde B''(0)} \nonumber
\\
&& \hphantom{\tilde B''(x) \Big[}+ 4 (m \partial_m I_2)(\tilde
B'(x)-\tilde B'(0)) \Big] \label{frg1} \ .\qquad
\end{eqnarray}
This equation is valid for any space dimension $d$.  It can be
integrated once w.r.t.~$x$ to obtain the final result
\begin{eqnarray}
m \partial_m \tilde B(x) &=& \frac{2 (m \partial_m T I_1)}{1+ 4 I_2
\tilde B''(0)} \tilde B' (x) \nonumber \\
\label{betafinal}&& + 2 (m \partial_m I_2)\left[\tilde
B'(x)^{2}-2\tilde B'(0)\tilde B' (x)\right]\ ,\qquad
\end{eqnarray}
where we have dropped an $m$-dependent integration constant.

A general method to study (and solve) the FRG equation (\ref{frg1}) is
then to start from $m= \infty$ where the initial condition is
$\tilde{B}(x) = B(x)$ in the presence of a UV momentum cutoff
$\Lambda$, or a lattice with lattice constant $a=1/\Lambda$. Then one
studies how $\tilde{B}(x)$ evolves as $m$ is slowly decreased.

There are thus two possible paths to solve the problem, namely the
direct inversion of the self-consistent equation and the solution of
(\ref{frg1}) with the above initial condition.  Both are studied
below. These two methods are clearly equivalent when the solution
$\tilde{B}(x)$ is analytic at $x=0$. Indeed, in the above derivation,
we have assumed that $\tilde{B}''(0)$ exists.  This will not always
hold, as we now discuss.  What the proper ensuing modifications are is
a subtle point which will be examined later.

\subsection{General features: Analytic vs non-analytic solution}

Before solving this equation let us first 
find the conditions under which there exists 
an analytic solution. This will give us insight in the
phases of the model. One notes from (\ref{secder})
that:
\begin{equation}\label{secder0}
 \frac{1}{\tilde{B}''(0)} = \frac{1}{B''(2 T I_1)} - 4 I_2
\ .
\end{equation}
For $m=\infty$ the starting value is $\tilde{B}''(0) = B''(0) > 0$, in
any dimension $d$. (The force correlator decays for small distances.)
As $m$ is decreased several things can happen.

Let us start with $T=0$. Then for $d < 4$, since $I_2$ diverges for small $m$,
one sees from (\ref{secder0}) that $\tilde{B}''(0)$ becomes infinite as
$m \to m_c^+$, where the Larkin mass $m_c$
is the solution of:
\begin{equation}\label{lf38}
 4 S_d \int_0^\Lambda \rmd q \frac{q^{d-1}}{(q^2 + m_c^2)^2} =
\frac{1}{B''(0)}
\end{equation}
with $S_D=1/(2^{d-1} \pi^{d/2} \Gamma[d/2])$ and has the standard
dependence $m_c \sim B''(0)^{1/\epsilon}$ of the inverse Larkin length
on the bare disorder (a Larkin length $L_c =1/m_c$ can be defined).
Since $\tilde B''(0)$ is like $\tilde R''''(0)$ positive, this
divergence is the usual one of the FRG, as also found in 1- and 2-loop
studies \cite{DSFisher1986,BalentsDSFisher1993,%
BucheliWagnerGeshkenbeinLarkinBlatter1998,%
ChauveLeDoussalWiese2000a,LeDoussalWieseChauve2002,ChauveLeDoussal2001},
where it signals that the function $\tilde R(u)$ becomes non-analytic
and that a cusp singularity forms at $u=0$ in the second derivative $-
\tilde R''(u)$, i.e.\ in the correlator of the pinning force. This is
usually interpreted as a glass phase with many metastable states
beyond the Larkin length. Thus for $d<4$ the function always becomes
non-analytic at large scale (small mass), and there is a single glass
phase. For $d>4$, since $I_2$ is convergent, the cusp occurs only if
the bare disorder is sufficiently large.

At non-zero temperature $T>0$ (\ref{secder0}) shows that for $2 < d <
4$ thermal fluctuations do not change the scenario. Since $I_1$
remains finite, temperature only slightly renormalizes the value of
$m_c$ downward, as
\begin{equation}\label{lf39}
 4 S_d \int_0^\Lambda \rmd q \frac{q^{d-1}}{(q^2 + m_c^2)^2} \approx
\frac{1}{B''(2 T S_d \Lambda^{d-2}/ (d-2))}
\end{equation}
for $\Lambda \gg m_c$. For $d<2$ the effect of thermal fluctuations is
more important. For definiteness let us consider the set of models
with power law correlations (\ref{corrlr}). Then (\ref{secder0})
becomes:
\begin{equation}\label{b2}
 \frac{1}{\tilde{B}''(0)} = \frac{1}{g \gamma} (a^2 + 2 T I_1)^{1 +
\gamma} - 4 I_2 \ .
\end{equation}
Since both integrals diverge for small mass as $I_1 \sim 1/m^{2-d}$,
$I_2 \sim 1/m^{4-d}$, one can distinguish three cases:

\begin{itemize}
\item [(i)] If disorder correlations decay fast enough $\gamma >
\gamma_c(d) = 2/(2-d)$ then the $I_1$ term wins and as $m \to 0$ one
has $\tilde{B}''(0) \to 0$, indicating that disorder is subdominant,
resulting in a high-temperature phase. In that case the solution is
analytic as $m \to 0$. There is however a more complicated behavior
for intermediate values of $m$ (see Appendix \ref{app:sr}).
\item [(ii)] If disorder correlations decay slower, i.e.\ $\gamma <
\gamma_c(d)$, the term proportional to $I_2$ wins and the solution
always becomes non-analytic at some Larkin mass.
\item [(iii)] In the marginal case, $\gamma = \gamma_c(d)$ there is a
transition at some critical temperature $T_c$ between a high
temperature phase and a glass phase.
\end{itemize}
These features are very general and each of these cases will be
studied in more details below. 

One can immediately see that the existence of an analytic solution for
$\tilde{B}(u)$ is in one to one correspondence to the existence of a
locally stable replica symmetric solution of the MP equations. Indeed
the condition for the stability of the RS saddle point is precisely
that the replicon eigenvalue be positive, namely that
\cite{MezardParisi1991}:
\begin{eqnarray}\label{lf40}
 \lambda_{\mathrm{rep}}(p) &=& 1 - 4 I_2(p) B''(2 T I_1 ) \\
 I_2(p) &=& \int_k (k^2 + m^2)^{-1}((k+p)^2 + m^2)^{-1} \qquad 
\end{eqnarray}
be positive for all $p$.  The RSB instability occurs when the lowest
eigenvalue, which corresponds to $p=0$, vanishes. The condition
$\lambda_{\mathrm{rep}}(p=0)=0$ is equivalent to the vanishing of
(\ref{secder0}), i.e.\ of the divergence of $\tilde B''(0)$ and the
emergence of non-analytic behavior. Thus the generation of a cusp in
the FRG coincides at large $N$ exactly with the instability of the RS
solution.

It is easy to see that an analytic solution $\tilde{B}(x)$ of
(\ref{starting}) and (\ref{frg1}) cannot describe the glass phase at
$T=0$.  Indeed when $\tilde{B}(x)$ is analytic, Eq.~(\ref{3cum}) and
similar results for higher cumulants indicate that the full effective
action is analytic. It is then immediate to obtain correlations from
its derivatives.  For instance, from (\ref{lf20}) the 2-point function
at $q=0$ is simply:
\begin{equation}\label{lf41}
 \frac{1}{N} \left<u_a(q) \cdot u_b(q)\right>|_{q=0} = 
\frac{T}{m^2} \delta_{ab} - 2 \frac{\tilde{B}'(0)}{m^4} 
\end{equation}
On the other hand, setting $x=0$ in (\ref{starting}) one finds:
\begin{equation}\label{dr}
 \tilde{B}'(0) = B'(2 T I_1)
\ .
\end{equation}
Thus at $T=0$ one recovers the dimensional reduction (DR) result
$\overline{u^2} \sim m^{-d-2 \zeta}$ with $\zeta=\zeta_{\mathrm{DR}}=
(4-d)/2$ instead of a non-trivial value for $\zeta$ expected in the
glass phase. Furthermore since the effective action is analytic, all
higher connected cumulants will trivially vanish at $T=0$ (or be equal
to the bare ones if the bare model contains such higher cumulants)
from the DR property.  Clearly, in the glass phase, the DR scaling is
expected to be incorrect and a non-analytic solution should be found,
as well as a way to escape (\ref{dr}). Below we find how such a
mechanism occurs within the FRG.

It will emerge from our study that for the case where disorder is
relevant in the large scale limit (i.e.\ the long range case $\gamma <
\gamma_c(d)$ mentioned above) the non-analytic solution of the FRG
equation will correspond to the full replica symmetry breaking
solution of MP.  The situation for the short range case is more
delicate. Both are discussed below.

\subsection{FRG equation for rescaled disorder, $d<4$} 
The equation (\ref{frg1}) is valid (for $N=\infty$) in any spatial 
dimension $d$. Since one has the exact relation:
\begin{equation}\label{rel}
 - \frac{1}{2} m \partial_m I_1 = m^2 I_2 
\end{equation}
one sees that the FRG equation (\ref{frg1}) has a well defined limit
$\Lambda \to \infty$ for $d<4$. It makes formulae somewhat simpler so
we will start by considering this case; the case $d \geq 4$ will be
studied later. Note that although the equation has a well-defined
limit, its solution may require a UV cutoff (e.g.\ as is manifestly
the case in integrating (\ref{rel}) above).

Thus from now on we study $d < 4$ and consider the infinite UV cutoff
limit. Then one has
\begin{equation}\label{lf42}
I_2 = A_d \frac{m^{-\epsilon}}{\epsilon} \ ,\qquad A_d = \frac{2}{(4
\pi)^{d/2}} \Gamma ( 3 - \frac{d}{2})
\end{equation}
with $\epsilon=4-d$. It is convenient to define 
the rescaled dimensionless function:
\begin{equation}\label{lf43}
b(x) =  4 A_d m^{4 \zeta - \epsilon} \tilde{B}(x m^{-2 \zeta})
\ ,
\end{equation}
where $\zeta$ is a fixed number, but for now arbitrary.  Note that
whether one works with $\tilde{B}$ or the rescaled $b(x)$ does not
make any difference for the possibility of a non-analyticity or a
divergence of the second derivative.

Then $b(x)$ satisfies {\it the FRG equation in the infinite-$N$ limit}:
\begin{eqnarray}
 - m \partial_m b(x) &=& \beta[b] \nonumber \\
&=& (\epsilon - 4 \zeta) b(x) + 2 \zeta x b'(x)\nonumber  \label{rgscaled}  \\
&& + \frac{1}{2} b'(x)^2 - b'(x) b'(0) + T_m \frac{b'(x)}{ 1 +
\frac{b''(0)}{\epsilon}} + c_m \ .
 \nonumber \\
&&  \label{frg2}
\end{eqnarray}
The rescaled temperature, and the energy exponent $\theta$ are 
defined as
\begin{eqnarray}
 T_m&=& T \frac{4 A_d}{\epsilon} m^{\theta} \\
 \theta &=& d - 2 + 2 \zeta
\ .
\end{eqnarray}
To obtain (\ref{frg2}) we have also integrated (\ref{frg1}) once, so
there is a priori a $m$-dependent integration constant.

We emphasize that this FRG equation (\ref{frg2}) that we have derived
is valid, to dominant order in $1/N$, {\it in any dimension} $d<4$ and
at any temperature $T$. In a previous study \cite{BalentsDSFisher1993}
Balents and Fisher studied another limit: arbitrary $N$ but only to
first order in $\epsilon=4-d$ and $T=0$. If we consider the dominant
order in $N$ of their equation, we find that it is identical to the
$T=0$ part of (\ref{frg2}) (up to some changes in notation). Equation
(\ref{frg2}) however is valid to {\it all orders} in $\epsilon$, an
important point which the method used in \cite{BalentsDSFisher1993}
could not address. Comparison of (\ref{frg2}) to our recent 2-loop,
i.e.\ $O(\epsilon^2)$ studies requires to expand to next order in
$1/N$ and is performed in \cite{LeDoussalWiesePREPg}.

Furthermore (\ref{frg2}) includes the effect of temperature to all
orders in $\epsilon$. Expanding the term proportional to $T$ to lowest
order in disorder $b$, one finds the term $T_m b'(x)$. This is the
large-$N$ limit of the tadpole term obtained in the 1-loop FRG at
$T>0$ \cite{Balents1993,ChauveGiamarchiLeDoussal1998,%
ChauveGiamarchiLeDoussal2000,MullerGorokhovBlatter2001}:
\begin{eqnarray}
\partial_l \tilde R(u) &=& T \sum_{i=1}^N \partial_i^2 \tilde R(u) \nonumber \\
\to \partial_l B(v^2) &=& T \tilde B'(v^2) + \frac{T}{N} v^2 \tilde
B''(v^2) \ ,
\end{eqnarray}
where for infinite $N$ the last term drops out. (It appears however to
next order in $1/N$ \cite{LeDoussalWiesePREPg}.)

The form and the effect of the temperature term in (\ref{frg2}) to all
orders in $\epsilon$ is radically different from its 1-loop
truncation. Indeed, in the 1-loop FRG the temperature is known to
smoothen the cusp and render the function $\tilde R(u)$ analytic in a
boundary layer $u \sim \tilde T_m$ (e.g.\ for $N=1$
\cite{Balents1993,ChauveGiamarchiLeDoussal2000,BalentsLeDoussal2002a})
with $\tilde R''''(0) \sim 1/\tilde T_m$.  Here however, as further
analysis confirms below, for $\theta>0$ the divergence of
$\tilde{b}''(0)$ is self-reinforcing since it kills the term
proportional to $T_m$. We find that it usually occurs at a finite
(Larkin) scale.  In the marginal case $\theta=0$, we will find
non-trivial analytic finite-temperature fixed points.

\section{Detailed analysis of the FRG equations}
\label{detailed}

\subsection{Inversion of self-consistent equation} Let us now show how
one can invert the self-consistent equation (\ref{starting}). We first
rewrite it in terms of the rescaled correlator
\begin{eqnarray}
\label{self}
 b'(x) &=& 4 A_d m^{2 \zeta - \epsilon} \nonumber\\
&& \times B' \Big( m^{- 2 \zeta} ( x + \frac{1}{\epsilon} 
(b'(x) - b'(0))  + 2 T I_1  m^{2 \zeta}) \Big) 
\ ,\nonumber \\
\end{eqnarray} 
where in the term proportional to temperature, for $d>2$ we mean $\lim_{\Lambda
\to \infty} T I_1$ choosing a bare temperature $T \sim
\Lambda^{(2-d)}$ (this choice is known to be necessary to give a
universal and finite $\beta$ function, see e.g.\ the discussion in
Ref.\ \cite{LeDoussalWiesePREPg}). One can of course keep an explicit
$\Lambda$ dependence everywhere, but that leads to needless
complications without changing the result.

The above equation (\ref{self}) is easily inverted into
\begin{equation}
x = m^{2 \zeta} \Phi\Big[ \frac{y}{4 A_d m^{2 \zeta - \epsilon}} \Big]
+ \frac1{\epsilon} ( y - y_0 ) - \tilde{T}_m \label{self2}
\ ,
\end{equation}
where we define
\begin{eqnarray}
 y &=& y(x) = -b'(x) \label{new} \\
 y_0 &=& - b'(0) = - 4 A_d m^{2 \zeta - \epsilon} \tilde B'(0) \\
 \tilde{T}_m &=& 2 T I_1 m^{2 \zeta} 
\end{eqnarray}
with $\tilde{T}_m=T_m/(2-d)$ for $d<2$,
and $\Phi$ is the inverse function of $-B' (x)$ i.e.\ 
\begin{equation}
(-B')(\Phi(y))=y \ .
\end{equation}
This means in turn that the FRG equation (\ref{frg2}) is fully integrable,
a feature not immediately obvious if one does not know that it
originates from a self-consistent equation (an observation
not made in Ref.\ \cite{BalentsDSFisher1993}). To better understand this integrability
property let us show that (\ref{frg2}) can be transformed
into a {\it linear} equation. Let us first take a derivative
of (\ref{frg2}) and express it in terms of the new function $y(x)$ (\ref{new}) 
\begin{equation}
 - m \partial_m y = (\epsilon - 2 \zeta) y + 2 \zeta x y'
- y' ( y -  y_0 ) +  T_m \frac{y'}{ 1 - \frac{y'_0}{\epsilon}} 
\label{y}
\ ,
\end{equation}
where we denote $y'_0=y'(0)$. Converting this into an equation for
the inverse function $x(y)$ one finds:
\begin{equation}
m \partial_m x = (\epsilon - 2 \zeta) y x' + 2 \zeta x - ( y - y_0
) + \frac{ T_m \epsilon x'_0}{\epsilon x'_0 - 1} \label{linear}
\end{equation}
with $x'_0=x'(y_0)$ \footnote{Note the misprint in formula (11) of
Ref.\ \cite{LeDoussalWiese2001} corrected here.}.  We have used that $m
\partial_m y(x) = - y'(x) m \partial_m x(y(x))$ and have canceled a
factor of $\frac{1}{x'(y)}$ on both sides. (The validity near $x=0$
beyond the Larkin length is reexamined below).

One recovers now that the general solution of this linear equation is
(\ref{self}) since it is the sum of the general solution of the
homogeneous part
\begin{equation}\label{lf44}
 x =  m^{2 \zeta} \phi[ y m^{\epsilon - 2 \zeta } ] 
\ ,
\end{equation}
where $\phi$ is an arbitrary function, and of a particular solution
\begin{equation}\label{lf45}
 x = \frac{1}{\epsilon} (y - y_0) - \tilde{T}_m \ .
\end{equation}
The $y$ dependence obviously satisfies (\ref{linear}) and for the
constant part to work we use:
\begin{eqnarray}
 - m \partial_m y_0 &=& (\epsilon - 2 \zeta) y_0 + T_m \frac{y'_0}{ 1
- \frac{y'_0}{\epsilon}}
\label{dy0} \\
- m \partial_m \tilde{T}_m &=& - 2 \zeta \tilde{T}_m + T_m \ .
\end{eqnarray}
The first line comes from evaluating (\ref{y}) at $x=0$ and assuming
analyticity, i.e.\ that $\lim_{x \to 0} y'(x) (y(x) - y_0) = 0$, and
equality which will not work beyond the Larkin length ($m < m_c$), as
found below.

Now that we have clarified the connections between the two approaches
(self-consistent equation and FRG) we can try to find solutions valid
in the small mass limit. To analyze the solutions of the large-$N$ FRG
equation (\ref{rgscaled}), two approaches are legitimate,
corresponding to different points of view. The first, natural in mean
field, is exact integration. But then one discovers that the solution
becomes non-analytic upon reaching the Larkin mass.  It thus raises
the non-trivial question on how to continue this solution beyond the
Larkin length.  Before doing so, we will first examine a second point
of view, more familiar from standard RG arguments.

\subsection{The FRG point of view: Search for fixed points}
\label{frgapproach}

The standard RG approach amounts to construct and compute the
$\beta$-function of the theory, and then search for a fixed point
(function) which describes the large scale physics. Usually, finding
the basin of attraction of the fixed point, or relating arbitrary
initial conditions to the final approach of the fixed point is an
unmanageably difficult task. It is fortunately also besides the goal
of the RG which is to compute universal large scale physics
independently of the irrelevant details of the bare model. Here,
however, because of the large-$N$ limit, we can integrate the RG flow
exactly and in principle ``solve'' any bare model. Let us temporarily
ignore this integrability feature and focus on finding the zeroes of
the $\beta$-function.

The $\beta$-function was derived previously within an $\epsilon$
expansion and non-analytic fixed points were found to one loop
\cite{DSFisher1986,BalentsDSFisher1993,GiamarchiLeDoussal1995} and
also to two loops
\cite{ChauveLeDoussalWiese2000a,LeDoussalWieseChauve2003,ChauveLeDoussal2001}.
In the latter case additional ``anomalous'' terms are present in the
$\beta$-function for the non-analytic theory to be renormalizable and
a meaningful fixed point to exist. Viewing the right hand side of
(\ref{frg2}) as the large-$N$ limit of the true $\beta$-function, let
us follow the same strategy and ask whether we can find non-trivial
fixed points.

Let us study $T=0$ and use the equivalent linear form of the
FRG equation. We want to find the solutions $y(x)$ of
\begin{equation}
 (\epsilon - 2 \zeta) y x' + 2 \zeta x - ( y - y_0) = 0\ .
\end{equation}
$y_0$ is a fixed number (we want to impose $y_0=y(0)$), since we are
looking for a fixed point function. Keeping $y_0$ arbitrary, one first
tries a linear solution $x=a y + b$ which yields $a=1/\epsilon$ and
$b=-y_0/(2 \zeta)$. Writing $x(y) = (y/\epsilon) - y_0/(2 \zeta) +
\phi(y)$ one finds a homogeneous equation for $\phi$ and thus
\begin{equation}
x(y) = \frac{y}{\epsilon} - \frac{y_0}{2 \zeta} + \alpha y^{- \frac{2
\zeta}{\epsilon - 2 \zeta} } \ .
\end{equation}
Imposing now $y_0=y(0)$, i.e.\ $x(y_0)=0$, fixes the value of $\alpha$
and one finds the {\it family of zero temperature fixed point}
functions, parameterized by $\zeta$:
\begin{equation}
x = x^*(y)= \frac{y}{\epsilon} - \frac{y_0}{2 \zeta} + \frac{ \epsilon
- 2 \zeta}{ 2 \zeta \epsilon} y_0^{ \frac{\epsilon}{\epsilon - 2
\zeta} } y^{- \frac{2 \zeta}{\epsilon - 2 \zeta} }
\label{fixedpoints}
\ .
\end{equation}
Since $x>0$, $y_0>0$ one must have $\frac{2 \zeta}{\epsilon - 2
\zeta}>0$ and thus
\begin{equation}\label{lf46}
 0 < \zeta < \frac{\epsilon}{2}\ .
\end{equation}
The case $\zeta = \frac{\epsilon}{2}$ corresponds to a Larkin random
force model. For the same reason, we must exclude the branch $y>y_0$
and thus $x^*(y)$ is given by the unique solution of
(\ref{fixedpoints}) with $x>0$ and $0 \leq y \leq y_0$. Finally, for
$\zeta =0$ we find the fixed point:
\begin{equation}
x = x^*(y)  = \frac{1}{\epsilon} (y - y_0 - y_0 \ln(y/y_0) )
\ .
\end{equation}
An important observation is that all of these fixed points exhibit
automatically the expected cusp. Indeed one finds that $x'(y_0)=0$,
i.e.\ $x(y)$ in (\ref{fixedpoints}) vanishes and has a minimum at
$y=y_0$:
\begin{equation}
x^*(y) = \frac{1}{2(\epsilon - 2 \zeta) y_0} (y-y_0)^2 + O((y-y_0)^3)\ .
\end{equation}
This gives
\begin{equation}
\tilde b'(x) = \tilde b'(0) + A \sqrt{x} + O(x) \ ,
\label{bp}
\end{equation}
with $A=\sqrt{2(\epsilon - 2 \zeta)|\tilde b'(0)|}$, 
implying that the second derivative diverges as $x \to 0^+$ 
\begin{equation}
\tilde b''(x) \sim \frac{A}{2 \sqrt{x}} + O(x^{0})
\label{bpp}
\ .
\end{equation}
Recalling that $y=-b'(x)$ we see that all fixed points with $\zeta>0$
correspond to a power-law long-range correlator $b(x)$, while
$\zeta=0$ corresponds to a gaussian short range disorder.  If we
follow the standard RG arguments, we can now sort the models
(\ref{correlator}) into these universality classes. Since for the bare
model
\begin{equation}\label{lf47}
 B'(z) \sim z^{- \gamma}\ ,
\end{equation}
and since the decay of $R(u)$ in (\ref{correlator}) at large
$u$ can be argued to be identical for $B$ and $\tilde{B}$
(for LR fixed points) we find
\begin{equation}\label{zetag}
 \zeta = \zeta(\gamma) = \frac{4-d}{2(1+\gamma)}
\end{equation}
or $\zeta=0$ for short range correlations. These values are valid to
dominant order in $1/N$. In \cite{BalentsDSFisher1993} the effect of
the $O(1/N)$ terms in the 1-loop FRG equation was studied, i.e.\ the
corrections of $\zeta$ were estimated to order $O(\epsilon)$ and at
zero temperature. For SR disorder it was found that the result of the
GVM (i.e.\ Flory) is corrected by terms $a_N$ exponentially small in
$N$, i.e.\  $\zeta_{\mathrm{SR}}= \zeta(\gamma=\frac{N}{2}+1) + a_N
\epsilon + O(\epsilon^2)$.  For LR disorder with $\gamma >
\gamma^*(N)$ the result (\ref{zetag}) was found to be uncorrected to
$O(\epsilon)$. (The crossover SR to LR occurs at $\gamma^*$ such that
$\zeta(\gamma^*) = \zeta_{\mathrm{SR}}$).  One can in fact argue that
(\ref{zetag}) is always exact in the LR case (see e.g.~discussion in
Ref.~\cite{LeDoussalWieseChauve2003}).

Several important remarks are in order. First we have found the fixed
points of the inverted linear form (\ref{linear}) of the FRG equation.
A valid question is whether this is equivalent to finding the fixed
points of the initial form of the $\beta$-function
(\ref{frg2}). Second we have found fixed points {\it assuming} that $m
\partial_m y_0 = 0$. Since this is {\it different} from what has been found
previously in (\ref{dy0}) at $T=0$, one can ask whether these result
are compatible.

These two questions have a common answer. Examining more closely what
has really been done in this Section, we note that it is equivalent to declaring both
(\ref{frg2}) and (\ref{linear}) valid for any $x>0$ and interpreting
everywhere $y_0=y(0^+)$ in (\ref{linear}) and, equivalently $b'(0)$ as
$b'(0^+)$ defined by continuity as $x \to 0^+$.  This is legitimate
since the transformation from (\ref{frg2}) to (\ref{linear}) is
certainly valid for $x>0$ and we note that this answer the second
question above since Eq.\ (\ref{y}), i.e.\ the derivative of (\ref{frg2}),
evaluated at $x \to 0^+$ yields:
\begin{equation} \label{cusp}
 - m \partial_m y(0^+) = (\epsilon - 2 \zeta) y(0^+) - \lim_{x \to 0^+}
y'(x) (y(x) - y'(0^+))  
\ ,
\end{equation}
which works both in the regime $m>m_c$ where the solution is analytic
$y(0^+)=y(0)$ and in the fixed point regime $m \to 0$ when the cusp
has developed and the last term in (\ref{cusp}) has a non-zero limit
according to (\ref{bp},\ref{bpp}).

We expect these fixed points to be the physically correct solutions at
small $m$.  We now investigate whether we can confirm this by
providing the solution at infinite $N$, for arbitrary mass $m$, i.e
continue our solution (\ref{self}) beyond the Larkin length.

\subsection{Full solution beyond the Larkin mass}
\label{fullsolumc}
We now show that one can connect the two regimes, i.e.\ the regime for
$m > m_c$ where an analytic solution exists to the asymptotic one, for
$m \to 0$, studied in the previous Section. This can be done here
because of the full integrability of the infinite-$N$ limit and
provides a rare and non-trivial insight into what happens around the
Larkin scale.

It is instructive to start our analysis with the specific power law
models with LR correlations (\ref{corrlr}), together with the case of
SR correlations (\ref{corrsr}), in the form of a Gaussian. The solution
for an arbitrary bare potential $B(z)$ is more subtle, and will be
given in Section \ref{sec:solarbdocor}, and appendix \ref{app:convergence}.

For the power law correlators the inverse function in (\ref{self}) is:
\begin{equation}
 - B'(z) = \frac{g}{(a^2 + z)^\gamma} \quad \Leftrightarrow \quad z =
\Phi(y) = \Big(\frac{y}{g}\Big)^{-1/\gamma} - a^2\ .
\end{equation}
For gaussian correlations it is:
\begin{equation}
 - B'(z) = g e^{-z} 
 \quad \Leftrightarrow \quad z = \Phi(y) = \ln(g/y) 
\ .
\end{equation}
We can now insert this result into the general solution (\ref{self2})
of the self-consistent equation. $\zeta$ is arbitrary, but the
convenient choice (to later obtain a fixed point) is
$\zeta=\zeta(\gamma)$ such that the $m$ dependence of the first term
drops. Let us define
\begin{equation}
\tilde{g} = 4 A_d g \ .
\end{equation}
We then obtain, for power law models:
\begin{eqnarray}
x &=& \left(\frac y {\tilde{g}}\right)^{- 1/\gamma} +
\frac{1}{\epsilon} (y - y_0)
- m^{2 \zeta} a^2  - \tilde{T}_m  \qquad\quad  \label{solu1}  \\
 -b'(0) &=& y_0 = \tilde{g} (m^{2 \zeta} a^2 + \tilde{T}_m)^{-\gamma}
\label{solu2} \ ,
\end{eqnarray}
since we want $y(0)=y_0$ i.e.\ $x(y_0)=0$. This solution is valid for
$m>m_c$ and the value of $b'(0)$ is the DR result (\ref{dr}). For
short range disorder the solution for $m>m_c$ is
\begin{eqnarray}
  x  &=& \ln(\tilde{g} m^{-\epsilon}/y) + {\epsilon}^{{-1}} (y - y_0)
- \tilde{T}_m  \label{solu1exp} \qquad \\
 -b'(0)&=& y_0 = \tilde{g} m^{-\epsilon} \rme^{- \tilde{T}_m} \ ,
\label{solu2exp}
\end{eqnarray}
having set $\zeta=0$ in that case. We recall that $y(x) = - b'(x)$. 
Note that the bare disorder is recovered for $m\to\infty$. We
have kept temperature, but here we discuss only the case where
\begin{equation}\label{lf48}
\theta = \theta(\gamma) = d - 2 + 2 \zeta(\gamma) > 0 \ ,
\end{equation}
i.e.\ $2 < d < 4$, or $ d < 2$ with $\gamma < \gamma_c(d) = 2/(2-d)$.
In that case $\tilde{T}_m$ decreases as $m$ decreases, and, as mentioned
above the role of temperature is minor.

\begin{figure}[t] \centerline{ \fig{0.5\textwidth}{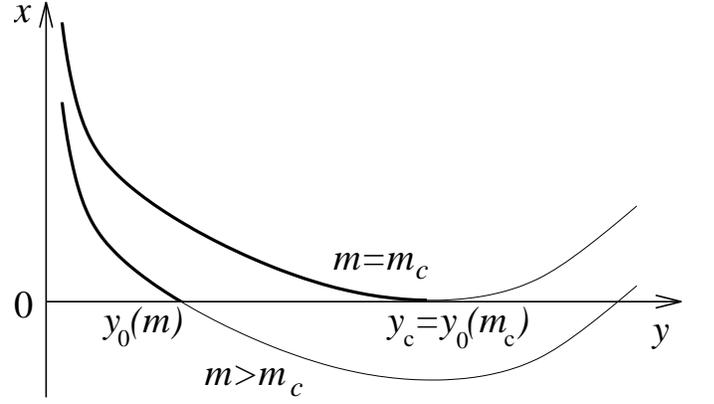} }
\caption{The function $x(y)$ given by (\ref{solu1}) or
(\ref{solu1exp}). The physical branch is the one with $y < y_0$.}
\label{fig2}
\end{figure}

Let us plot the r.h.s of (\ref{solu1}), (\ref{solu1exp}) on
Fig.~\ref{fig2}.  The curve $x(y)$ has the indicated shape in all
cases. It cuts the axis $x=0$ at $y=y_0$ and has a minimum $x'(y_c)=0$
at $y=y_{c}$ with
\begin{equation}\label{lf49}
 y_c = \tilde{g}^{1/(1+\gamma)} (\epsilon/\gamma)^{\gamma/(1+\gamma)}\ ,
\end{equation}
{\em independent} of $m$, and $y_c=\epsilon$ for SR disorder. For
$m>m_c$ the minimum occurs at negative $x$ and the slope at
$y=y_0<y_{c}$ is non-zero, indicating an analytic solution
$y(x)=-b'(x)$. For large $m$ only the first term on the r.h.s. of
(\ref{self2}) contributes and one recovers essentially the bare
disorder $B$. Decreasing $m$ simply amounts to translate the curve
upward along positive $x$, and $y_0$ increases as the curve $x(y)$
cuts the axis $x=0$ closer to the minimum. It reaches it at the Larkin
mass, solution of $y_0=y_c$, i.e.
\begin{equation}\label{deftc}
 m_c^{2 \zeta} a^2 + \tilde{T}_{m_c} = (\tilde{g}
\gamma/\epsilon)^{1/(1+\gamma)} =: \tilde{T}_c \ .
\end{equation}
For SR disorder $y_0=y_c=\epsilon$ gives
$m_c^\epsilon=\tilde{g}/\epsilon$.  Exactly as $m \to m_c^{+}$ the
solution acquires a cusp and one finds:
\begin{equation}\label{lf50}
 b'(x) - b'(0) \approx \sqrt{-2 (\epsilon - 2 \zeta) b'(0) x}
\end{equation}
i.e.\ the same result as (\ref{bp}).

Although it is a priori not obvious how to follow this
solution for $m < m_c$, the following remarkable property indicates
how to proceed. If we compute the $\beta$-function, i.e.\ the r.h.s.\
of (\ref{linear}) using (\ref{solu1}) at $m=m_c$ and
$\zeta=\zeta(\gamma)$ we find that it {\it exactly
vanishes}. Similarly the $\beta$-function for $b'(x)$ also {\it
exactly vanishes} for all $x>0$ provided we use also (\ref{cusp}), i.e
all $b'(0)$ are defined as $b'(0^+)$.  Thus at $m=m_c$ the function
{\it has already reached its fixed point form} $x=x^*(y)$, and {\it
freezes} for $m < m_c$. For the disorder correlators studied here,
$b(x)$ evolves according to (\ref{solu1}) or (\ref{solu1exp}) until
$m_c$ where it reaches its fixed point $b=b^*(x)$, and does not evolve
for $m<m_c$. In particular $y_0=-b'(0^+)$ freezes at $m_c$ and one has
$-m \partial_m y_0 = 0$ for $m<m_c$, exactly as was discussed in the
previous Section.

The solution for $m <m_c$ is thus:
\begin{eqnarray}
 x &=& (y/\tilde{g})^{- 1/\gamma} + {\epsilon}^{{-1}} (y - y_0) -
\tilde{T}_c \quad m < m_c \nonumber\\
 -b'(0^+)&=& y_0 = \tilde{g} \tilde{T}_c^{-\gamma}
\ ,
\end{eqnarray}
where the parameter $\tilde{T}_c$ is defined in (\ref{deftc}), thus it
exactly identifies with the zero temperature fixed point
(\ref{fixedpoints}) with $\zeta=\zeta(\gamma)$, as can be
explicitly verified. This is easily understood a posteriori, 
since the same functions appear and in both cases we have
two conditions to fix the two undetermined amplitudes
$x(y_0)=x'(y_0)=0$. It does however heavily rely on the exact power
law form of the model, so it is not immediately obvious how it will
extend to an {\it arbitrary} bare model $B(z)$. One clearly cannot
expect in the general case that convergence to the fixed point will be
completed within a finite scale.  The solution to this puzzle is
given below.

Similarly the solution for the Gaussian SR disorder correlator for $m
\leq m_c$ is given by setting $m=m_c$ in (\ref{solu1exp}),
(\ref{solu2exp}) with $y_0=\epsilon$ (which determines $m_c$).

The result of this section thus provides unambiguously a solution
beyond the Larkin scale which connects with the zero temperature fixed
point. It justifies the previous Section and the value obtained for
$\zeta$. We found that for power law and gaussian models the freezing
mechanism apparent in (\ref{cusp}) leads to:
\begin{eqnarray}
 - m \partial_m y(0^+) &=& (\epsilon - 2 \zeta) y(0^+) \quad , \quad
 m> m_c  \qquad  \\
 - m \partial_m y(0^+) &=& 0 ~~~~~~~~~~~~~~~~~~~~~~ \quad , \quad m <
m_c\ .
\end{eqnarray}
The fixed point is reached at $m=m_c$.

\subsection{Role of temperature} In the case where disorder is
relevant i.e.\ for $\theta(\gamma)>0$ (i.e.\ $2 < d < 4$; $d<2$ for
$\gamma < \gamma_c = 2/(2-d)$) we found in the previous Section that
temperature plays only a minor role since the convergence to the non
analytic zero temperature fixed point occurs on a finite (Larkin) RG
scale. Whether it should be called a zero temperature fixed point can
also be debated since it is reached when $T_m=T_{m_c}$. A proper
definition of the renormalized temperature may then include the
denominator in (\ref{frg2}).

Let us now examine the marginal case $\theta(\gamma) = 0$,
$\gamma=\gamma_c(d)$ and $d=2$ for SR disorder. We give here
the main results, further details are examined in the
Appendix \ref{app:finiteT}

The analytic solution is given by (\ref{solu1}) and $y_0$ given by
(\ref{solu2}), where here $\tilde T_m = 4 A_d T/(\epsilon (2-d) )$ does not
flow as $m$ is lowered. Let us examine the second derivative,
\begin{eqnarray}
 \frac{1}{\tilde{b}''(0)} &=& - x'(y_0) = \frac{1}{\gamma y_0}
\left(\frac{\tilde{g}}{y_0}\right)^{\!\!\frac{1}{\gamma}} -
\frac{1}{\epsilon}\nonumber  \\ 
& =& \frac{1}{\epsilon} \left[  \left(\frac{T}{T_c} m^\theta + \frac{a^2 m^{2
\zeta}}{\tilde T_c}\right)^{1+\gamma} -1\right] \label{div}
\ ,
\end{eqnarray}
which is a rescaled version of (\ref{b2}). 
The first line in (\ref{div}) holds more generally (in the infinite UV cutoff
limit) and to obtain the second we have set $\zeta=\zeta(\gamma)$,
$\theta=\theta(\gamma)$ and assumed $d<2$. Setting now $\gamma=\gamma_c$,
i.e.\ $\theta = 0$, we find 
that there is a transition at a temperature $T=T_c$ defined by
\begin{equation}\label{lf52}
 T_c = \frac{\epsilon (2-d)}{4 A_d} \tilde{T}_c = 
\frac{\epsilon (2-d)}{4 A_d} (\tilde{g} \gamma/\epsilon)^{1/(1+\gamma)}\ ,
\end{equation}
such that for $T > T_c$ the solution is analytic for all $m$ down to $m=0$, 
given by (\ref{solu1}) and $\tilde{b}''(0)$ remains
finite and given by (\ref{div}). This is a line of analytic fixed points which
terminates at $T_c$. For $T < T_c$ the solution freezes
as in the previous Section, and becomes
non-analytic at and below the Larkin mass
\begin{equation}\label{lf53}
a^2 m_c^{2 \zeta} = \tilde T_c \left(1 - \frac{T}{T_c}\right)
\ .
\end{equation}

The case $d=0$, $\gamma=1$ corresponds to the logarithmically
correlated disorder $B(z) = - g \ln(a^2 + z)$. It has been studied 
for finite $N$ in
\cite{CarpentierLeDoussal2001} where it was shown that there is a
transition {\it for any $N$} at $T_c = \sqrt{g}$ ($g=\sigma/N$ in the
notations of Ref.\  \cite{CarpentierLeDoussal2001}). The above result
is in agreement with this value for $T_c$. There, for $N=1,2$ there is
also a line of fixed points for $T>T_c$ with a continuously varying
dynamical exponent (and also one for $T<T_c$ with a different
dynamical exponent and some form of RSB). Since the dynamical
exponent is perturbatively related to $\tilde{b}''(0)$,
obtained above for infinite $N$, it would be particularly interesting 
to study the $1/N$ corrections
in this case. 

Let us now examine the case of SR disorder (\ref{corrsr}) in
$d=2$. More details are given in the Appendix \ref{app:finiteT}.  One
has $\tilde T_m = 2 T I_1 = (T/\pi) \ln (\Lambda/m)$. The analytic
solution (\ref{solu1exp}), (\ref{solu2exp}) becomes
\begin{eqnarray}
 x  &=& \ln(y_0/y) + {\epsilon}^{{-1}} (y - y_0) \label{solu0} \\
 y_0 &=& \tilde{g} m^{T/\pi - \epsilon} \Lambda^{-T/\pi}
\end{eqnarray}
with $\epsilon=2$. Thus there is a transition at $T=T_c=2 \pi$. For $T
< T_c$, we find $y_0$ to increase as $m$ decreases and reach $y_0 =
\epsilon$ at the Larkin mass. For $m< m_c$ the solution remains frozen
to (\ref{solu0}) with $y_0=\epsilon$. For $T>T_c$, we find that $y_0$
flows to zero and disorder is irrelevant. The physics is the same as
the one contained in the variational method for the periodic model in
$d=2$ \cite{GiamarchiLeDoussal1995} which exhibits a (so-called
marginal) 1-step RSB solution.

The case $\gamma > \gamma_c(d)$, ($d < 2$) is discussed in Appendix
\ref{app:sr}. Although an analytic solution exists as $m \to 0$ and
disorder is formally irrelevant, there are some freezing phenomena at
intermediate $m$. It corresponds to the case where MP find, in
addition to a RS solution, a 1-step RSB solution which is so called
non-marginal (different in nature from the one step solutions obtained
in the case $\theta=0$).

\section{Comparison between the RSB and the FRG approach} 
\label{sec:ComparisonRSBFRG}
In this Section we compare the FRG approach at large $N$ with the GVM
using RSB. Since the two methods study the same model in the same
limit (large $N$) a precise connection should exist.

We start by comparing the two methods at the level of the results of
the calculations. We first perform the comparison for power law
models.  Then we generalize the FRG solution to arbitrary bare
disorder correlator. Based on these results, we address the deeper
connections between the two methods, and emphasize what we learn from
them about the physical consequences.

\subsection{Zero momentum correlation function from the FRG}
Our main result up to now is a non-trivial
solution for the renormalized disorder correlator
$\tilde{B}(z)$ as a function of the scale parameter
$m$, i.e.\ the effective action for the zero momentum
mode. Since this function is once differentiable,
i.e.\ $\tilde{B}(z) = \tilde{B}(0) + \tilde{B}'(0) z + O(z^{3/2}) $, 
we can extract from its first derivative the 2-point correlation
function at zero momentum (see Section \ref{orderparam}):
\begin{eqnarray}
\langle v_a(q) \cdot v_b(q') \rangle &=& \frac{1}{N} \langle u_a(q) 
\cdot u_b(q') \rangle \nonumber \\
& =& G_{ab}(q) (2 \pi)^d \delta^d(q+q')  \\
 G_{ab}(q=0) &=& \frac{T}{m^2} \delta_{ab} - 2
 \frac{\tilde{B}'(0)}{m^4}  \nonumber  \\
& =& \frac{T}{m^2} \delta_{ab} - \frac{b_m'(0)}{2 A_d} m^{-d - 2
\zeta}\ ,
\end{eqnarray}
where in the last equation we have used the definition (\ref{lf43}) for the
rescaled function $b$, and added the index $m$ to recall its
dependence on the mass.

In the case $\theta >0$, for the power-law models (\ref{corrlr}), we
thus find, using $b_m'(0)=-y_0$ from (\ref{solu2})
\begin{eqnarray}
 G_{a \neq b}(q=0) &=& 2 g (m_c^{2 \zeta} a^2 +
\tilde{T}_{m_c})^{-\gamma} m^{-d - 2 \zeta} \nonumber \\ 
&=& \frac{1}{2 A_d} \tilde g^{\frac{1}{1+\gamma}}
\left(\frac{\epsilon}{\gamma}\right)^{\frac{\gamma}{1+\gamma}} m^{-d -
2 \zeta}\ , \label{resfrg}
\end{eqnarray}
and, for $m > m_c$ the DR result (\ref{lf41}), (\ref{dr}) (where $m_c$
is determined by (\ref{deftc})) and, we recall, $\zeta= \epsilon
/(2(1+\gamma))$.

\subsection{Explicit full RSB solution at large $N$} 
\label{mpfull}

Let us recall the RSB solution at large $N$ and resolve carefully the
MP saddle-point equation in presence of a mass.  We only assume that
there is indeed full RSB, to be checked a posteriori. Let us first
reexpress the general solution, valid for an arbitrary $B$, in a
rather compact form.

\begin{figure}[t] \centerline{\Fig{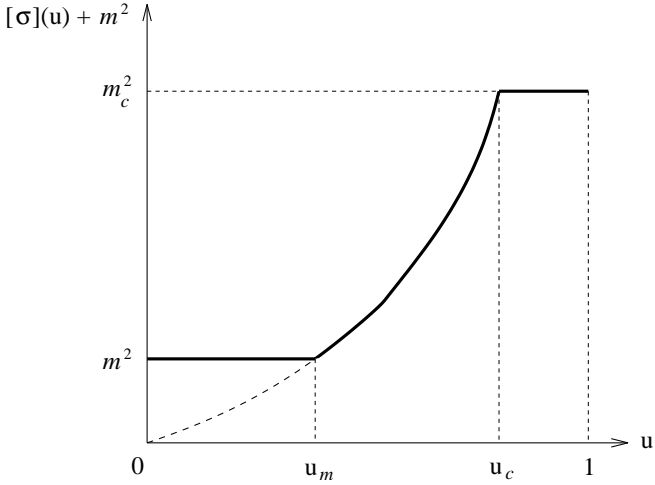}} \caption{Full RSB
solution for the function $[\sigma ]({\sf {\sf u}})+ m^2$ and a finite
mass $m$. $\sigma({\sf u})$ has identical behavior, with two plateaus
at values $\sigma({\sf u}=0)$ and $\sigma({\sf u}={\sf u}_c)$. In both
cases, upon increasing the mass only, ${\sf u}_m$ varies (increases)
and the lower plateau moves up, the rest of the function being
unchanged, see Eqs.~(\ref{rgrsb1}) f.\ in the text. The dashed line is
the zero mass solution.  The lower breakpoint ${\sf u}_m$ reaches the
upper one ${\sf u}_c$ at the Larkin mass $m=m_c$ above which the
solution becomes RS.  The FRG gives exactly the lower plateau value
for $\sigma({\sf u}=0)=\sigma({\sf u}={\sf u}_m)$ and its
$m$-dependence. From it, the full RSB solution can be reconstructed,
see Section \ref{reconstruct}.} \label{MPfunction}
\end{figure}

In the RSB method one first parameterizes the correlation matrix as
$G_{a b}(k)=G(k,{\sf u})$ and the self-energy matrix $T G_{ab}^{-1}(k)
- (k^2 + m^2) \delta_{ab} = \sigma_{ab} = \sigma({\sf u})$, in terms
of the overlap $0<{\sf u}<1$ between (distinct) replicas $a$ and $b$
(and denote $\tilde{G}=G_{aa}$).  The saddle point equations then read
\begin{eqnarray}
\sigma({\sf u}) &=& - \frac{2}{T} B'( \tilde{\chi}({\sf u}) ) 
\label{sadsig} \\
\tilde{\chi}({\sf u}) &=& 2 \int_k (\tilde{G}(k) - G(k,{\sf u}) ) \nonumber \\
& = & \tilde{\chi}({\sf u}^c) + \int_{\sf u}^{{\sf u}^c} \rmd {\sf w}
\int_k \frac{2 T \sigma'({\sf w})}
{(k^2 + m^2 + [\sigma]({\sf w}))^2}\qquad \label{sadsig2} \\ 
 \tilde{\chi}({\sf u}^c) &=& 2 T \int_k \frac{1}{k^2 + m^2 + \Sigma_c}
\label{sadsig3}
\end{eqnarray}
with 
\begin{equation}\label{sigmasimgabracket}
[\sigma]({\sf u})={\sf u} \sigma({\sf u}) - \int_0^{\sf u} \rmd
{\sf w}\, \sigma({\sf w})
\end{equation}
and $\Sigma^c = [\sigma]({\sf u} \geq {\sf
u}_c)$. The last two equations are the RSB-matrix inversion formulae;
$\sigma({\sf u})$ is assumed to be continuous. 
Taking a derivative of (\ref{sadsig}) w.r.t.~$\sf u$ gives 
\begin{equation}\label{yyyy}
 \sigma ' ({\sf u}) = \sigma ' ({\sf u})\, 4 B''( \tilde{\chi}({\sf u})
) \int_k \frac{1}{(k^2 + m^2 + [\sigma]({\sf u}))^2}\ .
\end{equation}
This equation admits two solutions: Either $\sigma({\sf u})$ is
constant, or satisfies the {\it marginality condition}
\begin{equation} \label{marginality}
 1 = 4 B''( \tilde{\chi}({\sf u}) ) \int_k \frac{1}{(k^2 + m^2 + [\sigma]({\sf u}))^2} \ .
\end{equation}
We thus look for a solution of the full RSB equations (see
Fig.~\ref{MPfunction}) with a non-trivial function $\sigma({\sf u})$
for ${\sf u}_m < {\sf u} < {\sf u}_c$ joined by two plateaus
\begin{eqnarray}
 \sigma({\sf u}) &=& \sigma({\sf u}_c) \ , ~\,\qquad {\sf u} \geq {\sf u}_c \\
 \sigma({\sf u}) &=& \sigma(u_m) \ , \qquad {\sf u} \leq {\sf u}_m\ .
\end{eqnarray}
Similar forms are valid for $G(k,{\sf u})$ and $\tilde{\chi}({\sf
u})$.  The breakpoint ${\sf u}_c$ is related to the physics at the
Larkin scale $m_{c}$, which, at weak disorder, can be much smaller
than the UV scale $\Lambda$, while ${\sf u}_m$ depends on the IR
cutoff $m$. (\ref{marginality}) also yields by continuity a closed
equation which determines $\Sigma_c$
\begin{equation}
1 = 4 B''\left( 2 T \int_k \frac{1}{(k^2 +  m^2 + \Sigma^c)} \right) 
\int_k \frac{1}{(k^2 + m^2 + \Sigma^c)^2}\ ,
\label{scsc}
\end{equation}
as well as
\begin{equation} \label{lower}
 1 = 4 I_2 B''(\tilde{\chi}({\sf u}_m)) \ ,
\end{equation}
since $[\sigma]({\sf u})=0$ for ${\sf u} \leq {\sf u}_m$. To solve these
equations one firsts determines the function $[\sigma]({\sf u})$ (see below),
then finds ${\sf u}_c$ and ${\sf u}_m$.

One can already note at this stage that (\ref{scsc}) is exactly the
condition (\ref{lf39}) which determines the Larkin mass $m_c$,
equivalent to the vanishing of the replicon:
\begin{equation}
\Sigma_c = m_c^2 - m^2  \qquad \mbox{for}\ m < m_c
\end{equation}
and $\Sigma_c=0$ (no RSB) for $m > m_c$.

To find $[\sigma]({\sf u})$ for arbitrary $B$ and cutoff, one notes
\cite{GiamarchiLeDoussal1995} that with the help of
(\ref{marginality}) and (\ref{sadsig}) $\sigma$ can be expressed as a
function of $[\sigma]$ as
\begin{equation} \label{sigmausol}
\sigma({\sf u}) = - \frac{2}{T} B'\left( (B'')^{-1}\bigg ( \frac{1}{4
\int_k \frac{1}{(k^2 + m^2 + [\sigma]({\sf u}))^2}} \bigg)\right) \ ,
\end{equation}
where $(B'')^{-1}$ is the inverse function of $B''$. Then one notes
that $\sf u$ as a function of $\sigma$ of $ [\sigma]$ is from
(\ref{sigmasimgabracket}) simply $1/{\sf u} =
\rmd\sigma/\rmd[\sigma]$. This yields immediately, using the chain
rule:
\begin{eqnarray} \label{usolu}
 {\sf u} &=& - 4 T \,\frac{ \big[\int_k \frac{1}{(k^2 +
m^2 + [\sigma]({\sf u}))^2}\big]^3}{\int_k
\frac{1}{(k^2 + m^2 + [\sigma]({\sf u}))^3}}
 \nonumber \\
&&\times B'''\left( (B'')^{-1}\bigg ( \frac{1}{4 \int_k \frac{1}{(k^2
+ m^2 + [\sigma]({\sf u}))^2}} \bigg)\right) \ .\qquad 
\end{eqnarray}
Upon inversion one obtains the exact function $[\sigma]({\sf u})$, and
inserting into (\ref{sigmausol}) $\sigma({\sf u})$. More precisely, we
see that the sum $[\sigma]({\sf u}) + m^2$ is a $m$-independent
function of ${\sf u}$, and thus from (\ref{sigmausol}) $\sigma({\sf
u})$ is also $m$-independent. Then one solves the self-consistent
equation (\ref{scsc}) for $\Sigma_c$, and finally obtains ${\sf u}_c$
from the above. The result can be written using (\ref{scsc}) in the
simple form
\begin{equation}
{\sf u}_c = - 4 T \frac{\big[\int_k \frac{1}{(k^2 + m^2_c)^2}\big]^3}{\int_k
\frac{1}{(k^2 + m^2_c)^3}} ~~ B'''\left( (B'')^{-1}\bigg ( \frac{1}{4
\int_k \frac{1}{(k^2 + m^2_c)^2}} \bigg )\right)\ .
\end{equation}
Thus ${\sf u}_c$ depends only on the Larkin mass and is independent of
$m$ (See appendix \ref{app:mprg} for another derivation and a
discussion of this useful property).  Similarly one obtains:
\begin{eqnarray}
{\sf u}_m &=& - 4 T \frac{I_2^3}{I_3} B'''\left( (B'')^{-1}\bigg ( \frac{1}{4
I_2} \bigg)\right)  \nonumber \\
 I_3 &=& \int_k \frac{1}{(k^2 + m^2)^3}\ .
\end{eqnarray}
Let us apply these considerations to the power law model
(\ref{corrlr}). For this model the Larkin mass is determined by
(\ref{deftc}). Next one has:
\begin{equation}\label{lf87}
 B'''\left( (B'')^{-1}( y )\right) = - \frac{\tilde{g} \gamma
(1+\gamma)}{4 A_d} \left( \frac{4 A_d y}{\gamma \tilde{g}} \right)^{\frac{2 +
\gamma}{1 + \gamma}} \ .
\end{equation}
In the limit of infinite UV cutoff $\Lambda$ limit, using $I_2=
(m^2)^{-\epsilon/2} A_d/\epsilon$ and $I_3= (m^2)^{-1 -\epsilon/2}
A_d/4$ we obtain from (\ref{usolu})
\begin{eqnarray}
 m^2 + [\sigma]({\sf u}) &=& (\tilde A {\sf u})^{2/\theta} \\
 \tilde A &=& \frac{1}{4 T A_d (1+ \gamma)} \epsilon^{\frac{1 + 2
\gamma}{1 + \gamma}}
(\gamma \tilde g)^{\frac{1}{1 + \gamma}}\qquad  \\
 {\sf u}_m &=& m^{\theta}/\tilde A \\
 {\sf u}_c &=& m_c^{\theta}/\tilde A 
\end{eqnarray}
with $\theta= d-2 + 2 \zeta$, $\zeta=\epsilon/(2(1+ \gamma))$. Using
(\ref{sigmausol}) one finds the $m$-independent result
\begin{equation}\label{lf88}
 \sigma({\sf u}) = \frac{2}{2 - \theta} \tilde A^{2/\theta} {\sf
u}^{-1 + \frac{2}{\theta}} \ .
\end{equation}
In particular one has the value of the lower plateau (see
Fig.\ \ref{MPfunction})
\begin{equation}
\sigma_m(0) = \sigma({\sf u}_m) 
= \frac{2}{2 - \theta} \frac{m^2}{{\sf u}_m}  = \frac{2}{2 - \theta} \tilde A m^{2 - \theta}
\end{equation}
Let us already note the relation ${\sf u}_m \partial_{m^2} \sigma_m(0)
=1$ which will be demonstrated to hold more generally below.

\subsection{Correlation function in MP solution compared to FRG} The
inversion formula yielding the diagonal correlation from the RSB
solution is
\begin{equation}\label{diagcorr}
 G_{aa}(q=0) = \frac{T}{m^2} \left[1 + \frac{\sigma_m(0)}{m^2} +
\int_{{\sf u}_m}^1 \frac{\rmd {\sf u}}{{\sf u}^2} \frac{[\sigma]({\sf
u})}{m^2 + [\sigma]({\sf u})} \right]
\end{equation}
and is a sum of contributions from all overlaps $0 \leq {\sf u} \leq 1$.
In particular the contribution from states with {\it zero overlap},
i.e.\ the most distant states, is:
\begin{equation}\label{lf89}
 G_{aa}(q=0)|_{{\sf u}=0} = G(q,{\sf u}=0) = \frac{T \sigma_m(0)}{m^4}\ .
\end{equation}
We can now compare with the FRG. One has, using $\theta=d-2 + 2
\zeta$, $2 \zeta= \epsilon (1+\gamma)$:
\begin{eqnarray}
 G_{aa}(q=0)|_{{\sf u}=0} &=& \frac{T \sigma_m(0)}{m^4}
= \frac{2}{2 - \theta} T \tilde A m^{- 2 - \theta} \nn
\\
& =& \frac{2 (1+ \gamma)}{\epsilon \gamma} \frac{\epsilon^{\frac{1 + 2
\gamma}{1 + \gamma}}}{4 A_d (1+ \gamma)}
(\gamma \tilde g)^{\frac{1}{1 + \gamma}} m^{-d-2\zeta }\nonumber  \\
&=& G_{a \neq b}^{\mathrm{FRG}}(q=0) \ ,
\end{eqnarray}
as given by (\ref{resfrg}). Thus, for this power-law model, we found
that the FRG gives exactly and only the contribution from the most
distant states (the lower plateau in the RSB solution). Before
discussing the reasons and consequences, let us show that this feature
is much more general than power law models, and holds in any case
where full RSB holds.

\subsection{Solution of the FRG equation for arbitrary 
disorder correlator $B$}
\label{sec:solarbdocor}
In Section \ref{fullsolumc} we found how to continue the solution of
the FRG equation beyond the Larkin scale. It involved freezing of the
$m$ dependence of $y_0=-b'(0)$ at $m=m_c$ and worked only for two
special forms of disorder correlators, which happened to be already
fixed point forms. It is important to find the solution for a more
general form of the bare correlator $B(z)$, and this is what we
achieve here.

Let us examine whether we can find a solution for any $m$ of the FRG
equation (\ref{linear}) in inverted variables
\begin{equation} \label{frgeqlin}
 m \partial_m x_m(y) = (\epsilon - 2 \zeta) y x_m'(y) + 2 \zeta x_m(y)
- y + y_0 \ ,
\end{equation}
which correspond to a more general function $B(z)$.  We take special
care here to indicate that $x_m(y)$ is an $m$ dependent function of
$y$ (we note $x_m'(y) = \partial_y x_m(y)$ and we recall that
$y_m(x)=- b'_m(x)$).  The idea is to play with the $m$ dependence
of $y_0=y_0(m)$ since this is really all the freedom we have. Let
us restrict our analysis for simplicity to $T=0$, the generalization being
straightforward. The definition of $y_{0}(m)$ is given implicitly by
\begin{equation}\label{lf90}
 x_m(y_0(m)) = 0
\end{equation}
for all $m$. The total derivative thus vanishes:
\begin{eqnarray}
  m \frac{\rmd}{\rmd m} ( x_m(y_0(m)) )  &=& m \partial_m x_m(y_0) + x_m'(y_0) m \partial_m y_0 \nonumber  \\
 &=& 0 \ .
\end{eqnarray}
Together with (\ref{frgeqlin}) at $y=y_{0}$, it yields (recall that
$x_{m} (y_{0})=0$):
\begin{equation}\label{lf91}
 ( m \partial_m y_0 + (\epsilon - 2 \zeta) y_0)  x_m'(y_0) = 0
\ .
\end{equation}
There are only two possible solutions:
\begin{eqnarray}
  m \partial_m y_0 + (\epsilon - 2 \zeta) y_0 &=& 0  \\
 x'(y_0) &=& 0 \ .
\end{eqnarray}
The first holds before the Larkin scale and the second, which implies
a non-analytic $b(y)$, beyond.  We now want to find the solution
beyond the Larkin scale, i.e.\ assuming that $x_m'(y_0) = 0$, together
with $x(y_0(m)) = 0$, which of course implies $m \partial_m x_m(y_0) =
0$.

Equation~(\ref{frgeqlin}) with $y_0=y_0(m)$ is trivially
separable and admits the general solution
\begin{eqnarray} \label{6.78}
 x_m(y) &=& m^{2 \zeta} \Phi\left(\frac{y}{4 A_d m^{2 \zeta -
 \epsilon}}\right) + \frac{1}{\epsilon} y \nn \\
&& - m^{2 \zeta} \int_m^\infty \frac{\rmd m'}{m'} y_0(m') m'^{- 2
\zeta} \ ,
\end{eqnarray}
where for now $y_0(m)$ is arbitrary and so is the function
$\Phi(y)$. (It will be identified below with $(- B')^{-1}$ as in
Section \ref{detailed}).  The first condition one must impose is the
definition $x_m(y_0(m)) = 0$, i.e.\ 
\begin{eqnarray}
 0 &=& m^{2 \zeta} \Phi\left(\frac{y_0(m)}{4 A_d m^{2 \zeta -
\epsilon}}\right) + \frac{1}{\epsilon} y_0(m) \nn \\
&& - m^{2 \zeta}
\int_m^\infty \frac{\rmd m'}{m'} y_0(m') m'^{- 2 \zeta} \label{cond1}\ ,
\end{eqnarray}
which should be valid both for $m>m_c$ and $m<m_c$.  Taking $m
\frac{\rmd}{\rmd m}$ of (\ref{cond1}) yields, using (\ref{cond1})
again
\begin{equation}\label{master}
 \left[\frac{1}{\epsilon} + \frac{m^\epsilon}{4 A_d}
\Phi'\left(\frac{y_0(m)}{4 A_d m^{2 \zeta - \epsilon}}\right) \right]
\left[ m \partial_m y_0 + (\epsilon - 2 \zeta)y_{0} \right] =0 \ .
\end{equation}
In order to satisfy this equation, at least one of the factors must
vanish. The regime $m<m_{c}$ corresponds to the first, the regime
$m>m_{c}$ to the second factor being zero.

For $m>m_c$ one has $\left[ m \partial_m y_0(m) + (\epsilon - 2
\zeta)y_{0} (m)\right]=0$ leading to
\begin{equation}
 y_0(m) = A m^{2 \zeta - \epsilon}
\end{equation}
and the above solution becomes:
\begin{equation}\label{lf92}
x = m^{2 \zeta} \Phi\left(\frac{y}{4 A_d m^{2 \zeta -
\epsilon}}\right) + \frac{1}{\epsilon} (y - y_0) \ .
\end{equation}
This can clearly be identified with the analytic solution of the
self-consistent equation (\ref{self}) found before in Section
\ref{detailed}, and thus implies that $\Phi$ is the reciprocal
function of $-B'$. Eq.~(\ref{cond1}) is trivially satisfied by
\begin{equation}\label{6.80}
\Phi\left(\frac{y_0}{4 A_d m^{2 \zeta - \epsilon}}\right) = 0\ .
\end{equation}
Applying $-B'$ to (\ref{6.80}) fixes $A$ to be
\begin{equation}
A =- 4 A_{d} B' (0)\ ,
\end{equation}
and one recovers the dimensional reduction result.

The interesting new information is obtained for $m<m_c$. Then the
first factor in (\ref{master}) vanishes, i.e.\
\begin{eqnarray} \label{cuspcond}
  0 = \frac{1}{\epsilon} + \frac{m^\epsilon}{4 A_d}
\Phi'\left(\frac{y_0(m)}{4 A_d m^{2 \zeta - \epsilon}}\right)
\label{cond2} \ .
\end{eqnarray} 
Deriving (\ref{6.78}) w.r.t.\ $y$ one sees that (\ref{cuspcond})
correctly implies
\begin{equation}\label{cond3}
x'(y_0) = 0\ ,
\end{equation}
thus the solution for $b' (x)$ has a cusp.  (\ref{cond2}) determines
the function $y_0(m)$ for $m<m_c$. Note that if the power law in the
correlator holds only asymptotically, $y_0(m)$ will nicely converge to
a constant (for the right choice of $\zeta$) due to the asymptotic
power law tail, but may vary arbitrarily according to the irrelevant
corrections to power law. This is studied in more details in appendix
\ref{app:convergence}. 

It is convenient to rewrite the final result, i.e.\ Eqs.\
(\ref{cond1}), (\ref{cuspcond}) in the form:
\begin{eqnarray}
 b'_m(0) &=& 4 A_d m^{2 \zeta - \epsilon} B'( \tilde \chi_m(0) )
\label{rep3} \\
 4 I_2 &=& \frac{4 A_d}{\epsilon} m^{-\epsilon} = \frac{1}{B''(\tilde
 \chi_m(0))} \label{rep2} \\
 \tilde \chi_m(0) &=& \frac{b'_{m}(0)}{\epsilon} m^{- 2 \zeta} -
\int_m^\infty \frac{\rmd m'}{m'} b'_{m'}(0) m'^{- 2 \zeta}\ , \qquad
\label{inversionchi}
\end{eqnarray}
where we use the notation $\tilde \chi_m(0) \equiv
\tilde \chi^{\mathrm{FRG}}_m(0)$. The connection with the RSB solution
becomes obvious in this form. Comparing with (\ref{lower}), the
equation (\ref{rep2}) of the FRG solution identifies with the
marginality condition at ${\sf u}={\sf u}_m$, the lower plateau of the
RSB solution, see Fig.~\ref{MPfunction}. It allows to determine
$\tilde \chi_m(0)$;  the two other equations are self-consistently
obeyed and give $b'_m(0)$.  Comparing with (\ref{sadsig}) at ${\sf
u}={\sf u}_m$ yields the identification
\begin{eqnarray}
 \tilde \chi_m({\sf u}=0) &=& \tilde \chi({\sf u}={\sf u}_m) = \tilde
 \chi^{\mathrm{FRG}}_m(0)  \\
 \frac{T \sigma_m(0)}{m^4} &=& \frac{T \sigma({\sf u}_m)}{m^4} =
\frac{- b'_m(0)}{2 A_d} m^{-d + 2 \zeta}\ ,
\end{eqnarray}
and thus we obtain:
\begin{equation}
G_{ab}^{\mathrm{FRG}}(q) = G^{\mathrm{GVM}}(q,{\sf u}=0) =
\tilde{G}^{\mathrm{GVM}}(q)|_{{\sf u}=0} \ .
\end{equation}
It thus holds for an arbitrary disorder correlator, provided a
solution to Eqs.\ (\ref{rep3}), (\ref{rep2}) exists, i.e.\ for the
class of functions $B(u)$ which yield full RSB (also called continuous
RSB) within the MP approach. Of course, equations (\ref{rep3}),
(\ref{rep2}) were derived {\it without any assumption about replica
symmetry breaking}.

Extension to $T>0$ is obvious. Adding the last term of (\ref{linear})
and following the same steps as above, one finds:
\begin{equation}
\left[ m \partial_m y_0 + (\epsilon - 2 \zeta) y_0 + \frac{\epsilon
T_m}{\epsilon x_m'(y_0)-1} \right] x_m'(y_0) = 0 \ .
\end{equation}
Vanishing of the first factor yields the finite $T$ analytic solution
studied in the previous Section (equivalent to the RS solution of
MP). Continuation beyond the Larkin mass implies $x_m'(y_0)=0$, in
which case the additional temperature term in (\ref{linear}) vanishes
and one is back to the $T=0$ equations (\ref{rep2}), (\ref{rep3}):
Thus only the value of the Larkin mass depends on temperature,
everything else is independent of $T$.

\subsection{Full RSB solution from the FRG result} \label{reconstruct}
In the previous Section we have shown that the FRG yields
$\sigma_m({\sf u}=0)$ (via $b'_m(0)$) i.e.\ the value of the RSB
function only at ${\sf u}=0$.  In fact, as we now discuss, by varying
the mass one can scan the whole function $\sigma({\sf u})$ of MP for
any ${\sf u}$, and thus the FRG yields the same information as
contained in the function $\sigma({\sf u})$. Remarkably, we can obtain
an explicit expression for $\sigma({\sf u})$, even though the argument
of this function, the ``overlap'' is not obviously related to any
quantity in the FRG. Furthermore, we can also compute the full
correlation-function of Mezard Parisi, if one knows only $\sigma_m(0)$
for all $m$, which is given by the FRG.

Thus from now on we assume that we know only $\sigma_m({\sf u}=0)$ as
a function of $m$ through the FRG, together with some general
properties of the MP solution. As we have already found in Section
\ref{mpfull}, and is shown more directly in Appendix \ref{app:mprg},
the GVM saddle point equations, upon assuming full continuous RSB,
satisfy the two ``RG equations''
\begin{eqnarray} \label{rgrsb1}
\partial_m \sigma_m({\sf u}) &=& 0 \\
 \partial_m ( [\sigma_m]({\sf u}) + m^2 ) &=& 0 \label{rgrsb2}
\ ,
\end{eqnarray}
valid for any ${\sf u}$ such that $\sigma'({\sf u}) \neq 0$. One can
thus relate the solution $\sigma_m({\sf u})$ at finite $m$ to the
solution at zero mass $\sigma_0({\sf u})$.

Note that Eqs.\ (\ref{rgrsb1}) and (\ref{rgrsb2}) have been
hypothesized by Parisi and Toulouse for the SK-model
\cite{ParisiToulouse1980}. However, it has been shown that there they
are only approximately satisfied, see e.g.~\footnote{ A.\ Crisanti,
T.\ Rizzo, and T.\ Temesvar, {\em On the Parisi-Toulouse hypothesis
for the spin glass phase in mean-field theory}, cond-mat/{\bf
0302538}.}\footnote{We thank J.~Kurchan for pointing this out to us.}.

Analysis of these equations shows that, up to the breakpoint, one has:
\begin{eqnarray}
 \sigma_m(u)&=&\sigma_0(u) \qquad~~ u > u_m \\
 \left[\sigma_m\right](u) + m^2 &=& [\sigma_0](u) \qquad u > u_m \\
 \sigma_m(u)&=&\sigma_m(0) \qquad~ u < u_m \\
\left[\sigma_m\right](u) &=& 0 \qquad~~~~~~~~~\, u < u_m \ .
\end{eqnarray}
${\sf u}_m$ is thus uniquely defined from the solution at zero
mass by
\begin{eqnarray}
 \sigma_m(0) &=& \sigma_0({\sf u}_m) \label{sig1} \\
\left[\sigma_0\right]({\sf u}_m) &=& m^2 \label{sig2} \ .
\end{eqnarray}
Indeed one has, taking derivatives of (\ref{sig1}) and (\ref{sig2})
w.r.t.\ $m^2$:
\begin{eqnarray}
\partial_{m^2} \sigma_m(0) &=&  \sigma_0'({\sf u}_m) \partial_{m^2} 
{\sf u}_m \\
 {\sf u}_m \sigma_0'({\sf u}_m) \partial_{m^2} {\sf u}_m &=& 1\ ,
\end{eqnarray}
where here we introduce for convenience $\partial_{m^2} = \frac{1}{2
m} \partial_{m}$. These two equations give
\begin{equation}\label{umu}
{\sf u}_m = \frac{1}{\partial_{m^2} \sigma_m(0)} \ .
\end{equation}
One thus finds that the function $\sigma_0({\sf u})$ is implicitly
given by
\begin{eqnarray}
\sigma_0\left({\sf u}=\frac{1}{\partial_{m^2} \sigma_m(0)}\right) =
\sigma_m(0) \ .
\end{eqnarray}
Since $\sigma_m(0)$ can be extracted from the FRG, we see that {\it we
can  obtain the full function $\sigma({\sf u})$ from the FRG}.

One notes that the upper breakpoint ${\sf u}^c_m={\sf u}_c$ is
independent of $m$. As shown in Section (\ref{mpfull}), ${\sf u}_m$
increases upon increase of $m$, and reaches ${\sf u}_c$ at the Larkin
mass, i.e.\ for ${\sf u}_{m_c}={\sf u}_c$.

Let us show how one can recast the correlation function of MP, given
in (\ref{diagcorr}) at zero momentum, entirely using FRG data.
\begin{eqnarray}
 \frac{G_{aa}(q=0)}T& =& \frac{1}{m^2} + \frac{\sigma_m(0)}{m^4} \nonumber \\
&& + \int_{{\sf u}_m}^{{\sf u}^c} \frac{\rmd{\sf u}}{{\sf u}^2}
\left(\frac{1}{m^2} - \frac{1}{m^2 +
[\sigma_m]({\sf u})}\right) \nn \\
&& + \int_{{\sf u}^c}^{1} \frac{\rmd{\sf u}}{{\sf u}^2}
\left(\frac{1}{m^2} - \frac{1}{m^2 + \Sigma^c }\right) \ .
\end{eqnarray}
Using our previous results gives:
\begin{eqnarray}
\frac{G_{aa}(q=0)}T &=&  \frac{1}{m^2} + \frac{\sigma_m(0)}{m^4} 
\\
&& + \int_{{\sf u}_m}^{{\sf u}_c} \frac{\rmd{\sf u}}{{\sf u}^2}
\left(\frac{1}{m^2} - \frac{1}{[\sigma_0]({\sf u})}\right)\nonumber \\
 &&+ \left( \frac{1}{{\sf u}_c} -1\right) \left(\frac{1}{m^2} -
\frac{1}{m^2_c}\right) \ . \nn
\end{eqnarray}
Shifting from the variable ${\sf u}$ to the variable $\mu$ defined by
${\sf u}_{\mu}={\sf u}$, one finds that the correlation function can
be expressed {\em entirely} from the knowledge of $\sigma_m(0)$. To
see this, note that
\begin{equation}
\frac{\rmd {\sf u}}{{\sf u}^{2}} = - \rmd \left(\frac{1}{{\sf u}}
\right) = - \rmd (\partial _{m^2} \sigma _{m} (0)) = - \partial
_{m^2}^{2} \sigma _{m } (0) \rmd (m^2) \ .
\end{equation}
This gives
\begin{eqnarray}
\frac{G_{aa}(q=0)}T  &=& \frac{1}{m^2} + \frac{\sigma_m(0)}{m^4} \nonumber \\
&& + \int_{m^{2}}^{m_c^{2}} \rmd \mu^{2} \left( \frac{1}{m^2}-\frac{1}{\mu^2}
\right){\partial_{\mu^2}^2 \sigma_\mu(0)} \nn \\
&& + \left( \frac{1}{u_c} -1\right)
\left(\frac{1}{m^2} - \frac{1}{m^2_c}\right) \ ,
\end{eqnarray}
where we have used that $[\sigma_0](u_\mu) = \mu^2$. After an
integration by part and using again (\ref{umu}) at $u=u_c$ one finds
the remarkably simple formula
\begin{eqnarray}
\frac{G_{aa}(q=0)}T &=& \frac{\sigma_m(0)}{m^4} +
\int_{m^{2}}^{m_c^{2}} \frac{\rmd \mu^{2}}{\mu^4} \partial_{\mu^2}
\sigma_\mu(0) +
\frac{1}{m^2_c}  \nn \\
 &=& \frac{\sigma_m(0)}{m^4} + \int_{m^{2}}^{m_c^{2}} 
\frac{\rmd \sigma_\mu(0)}{\mu^4} +
\frac{1}{m^2_c} \ ,
\end{eqnarray}
valid for $m \leq m_c$.  Recalling the relation between $\sigma_m(0)$
and $\tilde B'(0)$ we see that the MP result is a simple average of
the correlations corresponding to masses between $m$ and $m_c$. Note
that one can derive a similar formula for $G(k,u)$ obtained by MP as a
function of $\sigma_\mu(0)$.

In summary, although strictly speaking our FRG result gives only the
contribution of distant states to the 2-point correlation function, it
does allow to obtain the whole MP result, although we do not yet have
a derivation within the framework of the FRG alone.  One should also
note that the formula (\ref{inversionchi}) is in a sense equivalent to
the inversion formula of hierarchical matrices which relates $\tilde
\chi({\sf u}) = 2 \int_k \tilde{G} - G(k,{\sf u})$ to the self energy
$\sigma({\sf u})$. This raises the question of whether the FRG
equations ``know'' about ultra-metric matrix inversion.  These result
hold for continuous RSB and the case of non-marginal RSB (when the
marginality condition is not obeyed) is discussed in Appendix
\ref{app:sr}.

\subsection{Discussion: Explicit versus spontaneous replica symmetry
breaking} Let us examine what has been achieved and how it compares
with other works.

We are interested in the behavior of the effective action of the
replicated field theory for $N$ large.  Let us focus here on the
uniform configuration, for which $\Gamma(u) = L^d \tilde \Gamma(u)$,
where we denote $\tilde \Gamma(u)$ the effective action per unit
volume. $T \tilde \Gamma(u)$ then represents the free energy per unit
volume, depending on a set $u_a$, $a=1,\dots,n$ in presence of
external sources $J_a$ which impose field averages $ \langle u_a(x)
\rangle =u_a$. The usual free energy ${\cal F}$ is recovered for
$u=0$, ${\cal F}=T \Gamma(u=0)$.

The saddle point method allows to write in the limit of infinite $N$:
\begin{equation}\label{superpos}
\rme^{- \Gamma(u) } = \rme^{- L^d \tilde \Gamma(u) } \approx \sum_\pi
\rme^{- N L^d \hat \Gamma_\pi(v=u/\sqrt{N}) } \ ,
\end{equation}
where $\hat \Gamma(v)$ satisfies the saddle point equations
(\ref{lf33}),(\ref{freeenergy}), and (\ref{saddle}), and the
$\sum_\pi$ denotes a sum over saddle points whenever more than one
solution exists \footnote{For a uniform mode one easily sees that upon
Legendre transform the superposition (\ref{lf25}) yields
(\ref{superpos}).}.  To be accurate the saddle-point method computes
\begin{eqnarray}
\lim_{N \to \infty} \frac{1}{N L^d} \Gamma(u=v \sqrt{N})\ ,
\end{eqnarray}
where the limit is taken {\em at fixed $v$}.

As mentioned in Section \ref{sec:sp}, the saddle point equations
(\ref{saddle}) contain {\it both} the FRG and the GVM. They depend on
the set of $v_a$, and when expanded in cumulants, taking {\it all}
$v_{ab} \equiv v_a - v_b \neq 0$, they lead to the FRG equations.
This approach clearly consists in imposing an {\it explicit} breaking
of replica symmetry. Also we expect that in that case a single saddle
point exists. This is indicated by the fact that the quantity
$\tilde{B}''(v^2)^{-1} = B''(\chi_v)^{-1} - 4 I_2$ plays the role of a
replicon eigenvalue and remains frozen and positive for $v >0$.

By contrast, MP studied the case where all $v_{ab}=0$ and found {\it
spontaneous} replica symmetry breaking, i.e.\ multiple saddle points,
differing by permutations $\pi$ of replicas.

We can now make contact between the two approaches and understand why
we have obtained via the FRG the correlations of MP corresponding to
the distant states.  Let us focus on the mode $q=0$, and define the
center of mass variable $\tilde{u} :=\frac{1}{L^d} u(x)$ (i.e.\ without
rescaling in $N$) and consider:
\begin{equation} \label{zJ}
 {\cal Z} (J) = \int \rmd \tilde{u}_1 \ldots \rmd \tilde{u}_n\,
\overline{P_V(\tilde{u}_1) \dots P_V(\tilde{u}_n)} \,\rme^{ - L^d
\sum_a J_a \tilde{u}_a}\ .
\end{equation}
$P_V(\tilde{u})$ is the probability distribution of $\tilde{u}$ in a
given disorder configuration:
\begin{equation} 
P_V(\tilde{u}) = \int {\cal D} [u] \delta\left( \tilde{u} -
\frac{1}{L^d} \int_x u(x) \right) \rme^{- {\cal H}_V[u]/T}\ .
\end{equation}
In the present paper we have computed (\ref{zJ}), scaling $J_a \sim
\sqrt{N}$, and taking all $J_a$ different in order to impose all
$v_{ab} \neq 0 \sim O(1)$.  Because of this scaling with $N$ we
obtained a {\it different} saddle point than MP (shifted by $j$, see
(\ref{spW})), and since all $J_a$ are different, this saddle point has
explicit RSB.  According to Section \ref{sec:calculgamma} this gives
$\Gamma[u]$ when $u$ scales as $\sqrt{N}$, i.e.\ we determined the
averaged probability (\ref{prob1}), (\ref{prob2}) for fixed
$w=\tilde{u}/\sqrt{N}$.

On the other hand MP found that:
\begin{equation}
\label{mpproba} \overline{P_V(\tilde u_1) \dots P_V(\tilde u_n)} =
\sum_{\pi} \rme^{- \frac{1}{2} L^d (G_{\pi})_{ab}(q=0) \tilde u_a
\cdot \tilde u_b }\ .
\end{equation}
One should in principle be able to recover the 2-point correlation
function (\ref{diagcorr}) obtained by MP adding small sources $J_a$ as
in (\ref{zJ}) and taking derivatives at $J_a=0$.  Clearly, to
reproduce the MP result, these should be taken as $J_a = O(N^{0}) \to
0$ and not $O(\sqrt{N})$ so as to maintain the unperturbed MP saddle
point. For instance the diagonal 2-point correlation function is
obtained using $J_{1}=J$, all others $J_{a}=0$, and differentiating
twice w.r.t $J$. (The off diagonal one involves $(J,-J,0,\dots
)$). Equivalently it should be obtainable from the effective action
for $\tilde u_{ab} = \tilde u_a - \tilde u_b =O (N^{0})$.  Thus
scaling $\tilde u_{ab} \sim \sqrt{N}$ and $J \sim \sqrt{N}$ as was
done here selects the distant states in the MP solution.  The fact
that we obtain exactly the MP result for these states shows that there
is no intermediate scaling regime.

We emphasize that our primary aim here is not to recover the MP
result, but to understand what exactly the FRG predicts, in view of
getting a better understanding of FRG within e.g. the
$\epsilon$-expansion. Extension of the FRG beyond the Larkin scale
requires giving a meaning to the limit $u\to 0^+$.  We find here that
what the FRG actually computes (from $b'(0^+)$) is a second moment of
$w$ in presence of a small extra external source $\sqrt{N} j_a$ such
that all $v_{ab} \neq 0$, i.e.\ an average, such that when there are
several states, the different replicas are chosen in maximally
separated states (${\sf u}=0$). 

Note that the quantity computed by the FRG specifies the system's
preparation, while such a procedure still has to be worked out for the
MP solution. In presence of a broken symmetry this is an important
issue, and the FRG gives a natural solution. 

It would be interesting to understand further the limit $v_{ab} \to 0$
coming from our solution, which one can call the crossover from FRG to
RSB. It is clearly non-trivial. For instance, one question is what we
get if we take a source $J = (j \sqrt{N} , -j \sqrt{N},0,\dots ,0)$ so
that we still have spontaneous RSB in $n-2$ copies, or if we divide
the replicas in two groups of $n/2$ each, $J = (j \sqrt{N},..j
\sqrt{N}, -j \sqrt{N},0,..-j \sqrt{N})$ so that RSB persists within
each packet.

Another important issue is what happens at large but finite $N$. For
any $N$, if one parameterizes the 2-replica part of the effective
action using $\tilde R(u_{ab}) = N \tilde{B}_N(u_{ab}^2/N)$, one can
write the 2-point correlation function as
\begin{eqnarray}
G_{ab}(q=0) = - 2 \tilde{B}_N'(0) m^{-4}
\end{eqnarray}
for $a \neq b$.  We have determined the function $\tilde B_N(x)$ for
$x = O(N^0)$, i.e.\ $u_{ab} =O (\sqrt{N})$. To obtain
$\tilde{B}_N'(0)$ for finite $N$ however, one needs a priori to know
$\tilde R(u_{ab})$ for $u_{ab} \sim O(N^0)$, i.e.\ $\tilde B_N(x)$ for
$x = O(1/N)$. The two could be the same, or there could be a boundary
layer of size $ 1/N$. A priori the knowledge of this requires
including the $1/N$ corrections in the FRG equation (as is examined in
\cite{LeDoussalWiesePREPg}). This may help to better understand the
connection of this regime to RSB. This is important since there are
cases (e.g.\ for $d=0$, $\theta <0$) where we know that Paris-type RSB
cannot survive at finite $N$.

\subsection{Interpretation: Comparison with BBM approach} A previous
study \cite{BalentsBouchaudMezard1996} aimed at connecting the RSB
solution to the FRG. The authors defined, for each configuration of
the disorder, an ``effective random potential'' $U_V(\phi_0)$ for a
given mode (e.g.\ the zero mode). Starting from the MP solution
(\ref{mpproba}), they computed the second cumulant of $U_V$ and showed
that it exhibits a non-analyticity, reminiscent of the one found in
the FRG. A parallel was drawn with a $d=0$, $N=1$ toy model where
$U_V$ satisfies an exact RG equation of the Burgers-KPZ type with
random initial conditions. Shocks, well known to develop in this
equation, provide an appealing physical picture for the singularities
in the energy landscape responsible for the non-analyticity in the FRG
beyond the Larkin length $\xi_{\mathrm{LO}}$.

Comparison between this study and the present one shows several
important differences, with interesting physical consequences. The
scaling in $L$, in $N$, and the definition of the ``renormalized''
disorder are different. As here, the authors of Ref.\
\cite{BalentsBouchaudMezard1996} focus on the zero mode, but with a
different scaling with system size: They define $\phi_0:=L^{- d/2}
\int_x u(x)= L^{d/2} \tilde{u}$, such that fluctuations of $\phi_0$
remain of order one. Other quantities are:
\begin{eqnarray}
 U_V(\phi_0) &=& -T \ln P_V(\phi_0)  \\
 \overline{ U_V(\phi_0) U_V(\phi_0') } &=& R_{\mathrm{BBM}}( \phi_0 - \phi_0' )
= \tilde{B}_{\mathrm{BBM}}( (\phi_0 - \phi_0')^2 )\nonumber\ . \\
\end{eqnarray}
This definition means that the mode $\phi_0$ sees an equivalent $d=0$
toy model, with random potential $U_V(\phi_0)$.  Comparing with
(\ref{prob2}) and (\ref{lf21}), we see that since the rescaling in $L$
is different there can be no relation between $R_{\mathrm{BBM}}''(0)$
or $\tilde{B}_{\mathrm{BBM}}'(0)$ and the two point correlation
function, neither the one of MP, nor the one obtained here in the
FRG. To obtain the 2-point correlation of MP one would still have to
{\it solve} the toy model defined by $U_V(\phi_0)$, i.e.\ compute:
\begin{equation}\label{toysolve}
\int \rmd \phi_0^1 \dots \rmd \phi_0^n\, \phi_0^a \phi_0^b\, \overline{
\rme^{- \sum_a U_V(\phi_0^a)/T} }\ .
\end{equation}
This task is difficult, since it requires not only the second
cumulant, but also higher ones (not computed in
\cite{BalentsBouchaudMezard1996}). More importantly, it requires the
large argument behavior $\phi_0 - \phi_0'$ of $R_{\mathrm{BBM}}(y)$,
not obtained in \cite{BalentsBouchaudMezard1996}, were attention was
focused on the Larkin scale (see below).  Thus the information
contained in $R_{\mathrm{BBM}}$ is physically interesting but not
obviously related to large scale correlations. It is in a sense
(e.g. for the $d=0$ case discussed in
\cite{BalentsBouchaudMezard1996}) closer in spirit to
Wilson-Polchinski type RG \cite{Polchinski1984} versus an RG based on
the effective action (see
\cite{ChauveLeDoussal2001}\footnote{G. Schehr, P. Le Doussal, {\em
Exact multilocal renormalization on the effective action: Application
to the random sine Gordon model, statics and non-equilibrium dynamics},
cond-mat/{\bf 0304486}.}).

Reexpressed in the variables of the present work, the result of
\cite{BalentsBouchaudMezard1996} reads:
\begin{equation}
 \tilde{B}(\tilde u^2) = B'(0) \tilde u^2 + c
\left(\frac{L}{\xi_{\mathrm{LO}}}\right)^{d/2} \tilde u^3\ ,
\end{equation} 
where $c$ is a constant.  Because of the different rescaling, the
non-analytic term has a coefficient growing with the system size,
which expresses again that it is not an effective action. However,
since the $\tilde u^2$ term is simply the bare disorder, and the
non-analytic term involves only the Larkin scale $\xi_{\mathrm{LO}}$,
it seems that this carries information for and only for the physics
below and around the Larkin length, and does not contain any
information about large scale behavior. Thus, despite exhibiting
shock behavior at the Larkin scale, we think it has little to do with
the FRG as a perturbative method to obtain large scale behavior. Not
surprisingly then, $R_{\mathrm{BBM}}$ is non-perturbative in
$\epsilon=4-d$, contrarily to the one obtained in standard FRG, which
is of order $\epsilon $.

Another important difference with the present approach is the scaling
with $N$. The approach of \cite{BalentsBouchaudMezard1996} used the
unperturbed MP saddle point and thus, as was extensively discussed in
the previous Section, it focuses on $\tilde{u}_{ab} = O(N^0)$ while we
focus on $\tilde{u}_{ab} = O(N)$, (i.e.\ $v_{ab}^2 \sim 1/N$ there and
$v_{ab}^2 \sim 1$ here) . Further work is needed to connect these
regimes. On the other hand it seems that the thermal boundary layer
can be found within this approach \footnote{We thank Leon Balents for
help in clarifying these issues.}.

\section{Discussion and conclusion}
\label{sec:Discussion and conclusion}
In this paper we have studied the FRG at large $N$.  From an exact
saddle point calculation of the replicated effective action at large
$N$ we have derived the exact renormalization group equation, valid in
any dimension $d$ for infinite $N$, for the field theory of pinning.
It is expressed as the $\beta$-function for the second cumulant of the
disorder correlator, and is exact as the second cumulant satisfies
a remarkably simple closed equation.  To order $O(\epsilon)$ it agrees
with the one derived by Balents and Fisher.

This result teaches us a lot about how the FRG works and helps put the
FRG approach to the $\epsilon$ expansion on more solid grounds. Since
here the FRG flow equation is formally equivalent to a self-consistent
saddle point equation, it is fully integrable, i.e.\ one can follow
the RG flow from any initial condition. It is thus possible to examine
in detail what happens around the Larkin length and how the disorder
correlator develops the non-analyticity. Let us emphasize that this is
the first time that the emergence of non-analytic behaviour in the FRG
is proven rigourously, beyond perturbative calculations. Indeed, the
1-loop FRG is insufficient per se to provide such a proof since the
runaway of $R''''(0)$ could very well be argued to be the analog of
the famous Landau ghost, i.e.\ a flow to a strong coupling fixed point
without the need, or better the possibility, for renormalization
within a non-analytic functional space. Here we demonstrate that this
is not the case, at least for infinite $N$.

If we had restricted the analysis to the self-consistent equation, the
continuation beyond the Larkin scale would have seemed quite
problematic. Remarkably, the FRG equation, equivalent below the Larkin
scale, provides an unambiguous way to continue the flow equation
beyond the Larkin scale. Even more remarkably, its solution reproduces
exactly the small overlap result of the full RSB solution of MP, a
non-trivial result which, within MP, cannot be obtained without
constructing the full RSB solution. The mechanism for this seems to be
that the FRG solution in that case naturally satisfies the so-called
marginality condition.  In fact, it turns out to be equivalent to it,
and we were able to find a formula yielding the complete RSB solution
for all overlaps. This is striking since we did not make any
assumption about Parisi RSB. We avoided the issue altogether by using
a method where RSB is not spontaneous, but explicit.

Given that the validity of the Parisi Ansatz, e.g.\ for the SK model,
has not yet been proven (despite recent progress \cite{Guerra2003}),
it is interesting to find a method which does {\it not} rely on
RSB. In fact there may well be deeper connections to be unveiled
between the Parisi algebra of ultrametric matrices and the type of
singular differential equations arising in the FRG. Another example
where a RSB solution can equivalently be obtained via an RG type
equation is the Derrida Spohn solution of the DP on the Cayley tree
\cite{DerridaSpohn1988}. This has inspired a similar solution for a
model with translationally invariant disorder in
\cite{CarpentierLeDoussal2001}.

We have thus shown agreement with the main results of the full and the
marginal one-step RSB solutions of Mezard-Parisi. This is also
interesting since it has been widely debated \cite{NewmanStein2002}
whether the RSB method captures the physics: Our results raise no
doubt for infinite $N$. 

More puzzling is the situation for SR disorder. There MP find both a
stable replica symmetric solution and a 1-step solution where
minimization over the breakpoint has to be enforced (marginality
condition violated). For gaussian disorder both solutions of MP have
$\zeta=(2-d)/2$.  Similarly the FRG naturally finds the finite
temperature RS solution with $\zeta=(2-d)/2$ (and one fixed point
solution at $T=0$ with $\zeta=0$).  A non-analytic solution also seems
to exist in the FRG, and work is in progress to analyze it further and
elucidate whether it is related to the non-marginal 1-step solution of
MP.  Let us note that in $d=0$ we essentially know that (apart from
$\gamma=1$) RSB does not hold at finite $N$ (the phase transition
predicted by the GVM in that case must go away at finite $N$, $T>0$).
So there is little doubt that the correct branch at finite $N$ is the
RS one, as also given by the FRG. For the DP problem with $d=1$ on the
other hand, it is not yet clear whether both branches (a $T=0$ fixed
point starting from $\zeta=0$ or a finite $T$ non-analytic solution
with $\zeta=(2-d)/2$) can coexist.  One scenario is that they would
cross over at some lower value of $N=N_c$ yielding the upper critical
dimension of KPZ. The calculation of the FRG $\beta$-function to next
order in $1/N$ should shed light on this question, and is thus of high
interest. It is presented in \cite{LeDoussalWiesePREPg}. Our method
thus provides a unique candidate for a field theory of the
strong-coupling phase.

To summarize, the present method is promising in solving mean field
models, by using explicit rather than spontaneous RSB. It would be of
interest to investigate whether other models like the SK-model could
be solved via the same route. More importantly, it may open an
alternative road to tackle disordered systems from a different
direction than expanding around RSB saddle points, a task which still
has to be accomplished. Of course, in the end, the same difficulties
may be in store. They could hide in the subtelties due to the
non-analytic behaviour of the $\beta$-function at large $N$.  However
we are optimistic, since we have understood the infinite-$N$ limit, at
least in the full-RSB case.  Also, a solution has been found for $N=1$
to two \cite{ChauveLeDoussalWiese2000a,LeDoussalWieseChauve2002,%
LeDoussalWieseChauve2003} and three loop order
\cite{LeDoussalWiesePREPb}, and for finite $N$ at 2-loop order.

Let us close by indicating that many extensions of this work are
possible and some in progress.  One example is the random field
problem, still under intense debate
\cite{BrezinDeDominicis1998,BrezinDeDominicis2001}, for which we have
also computed \cite{LeDoussalWiesePREPf} the effective action at large
$N$, and at 2-loop order.  Finally, the same method applies to the
dynamics, classical or quantum: it has been shown \cite{BookYoung}
that the mode-coupling approximation in glasses
\cite{CugliandoloKurchan1993} identifies with (non-marginal) mean
field (large $N$) dynamics, exhibiting aging solutions. However this
picture leaves out thermally activated processes, and our $1/N$ method
may be promnising there too.

We gratefully acknowledge discussions with L.\ Balents, E.\ Br\'ezin,
J.P.\ Bouchaud, L.\ Cugliandolo, Y.Y.\ Goldschmidt, J.\ Kurchan and
M.\ M\'ezard.

K.J.W.\ is supported by the Deutsche Forschungsgemeinschaft through
Heisenberg grant Wi 1932/1-1, and in part by the National Science
Foundation under grant PHY99-07949.

\appendix 
\def\theequation{\thesection.\arabic{equation}}

\section{Variational formulation} \label{sec:var} Let us extend the
variational method of Ref.\ \cite{MezardParisi1991} to the case where
the average of the field $u_a(x)$ is fixed to a non-zero value denoted
here $\overline{u}_a(x) = \sqrt{N} \overline{v}_a(x)$. One defines the
variational Hamiltonian and free energy: 
\begin{eqnarray}
 {\cal H}_{\mathrm{var}}[u] &=& \frac{1}{2} \int_{xy} \left[u_a(x) -
\overline{u}_a(x) \right] \nonumber \\
&&\qquad \times (G_{\mathrm{var}})_{xa,yb}
\left[u_a(y) - \overline{u}_b(y) \right] \nonumber \\
 {\cal F}_{\mathrm{var}}[G_{\mathrm{var}},\overline{v}] &=& - T \ln \int
{\cal D}[u] \exp\left( - {\cal H}_{\mathrm{var}}[u]/T\right)
\nonumber \\
&& + \left< {\cal H} - {\cal H}_{\mathrm{var}} \right>_{{\cal
H}_{\mathrm{var}}} \ ,
\end{eqnarray}
which satisfies (for $n$ positive integer) the usual bound ${\cal F} =
- T \ln {\cal Z} \leq {\cal F}_{\mathrm{var}}$. Here ${\cal H} = N
{\cal S}[u,0]$ defined in the text, is the replicated Hamiltonian.
Comparing with (\ref{var}) one finds that
\begin{equation}\label{lf71}
 {\cal F}_{\mathrm{var}}[G_{\mathrm{var}},\overline{v}]/T =
\Gamma_0[G_{\mathrm{var}},\overline{v}, \hat{U}]\ ,
\end{equation} 
where the last argument indicates that for $N$ finite $U(\chi)$ should
be replaced by $\hat U(\chi)$; in the infinite-$N$ limit 
$\hat{U} = U$. Restricting to a bare model with only a second cumulant
one finds (omitting the bars on $v$):
\begin{align}
& \hat{U}(vv(x),(G_v)_{xx}) = - \frac{1}{2 T^2} 
\sum_{ab} \hat{B}(v_{ab}(x)^2, (\tilde G_v)_{xx}^{ab}) \nonumber \\
& \hat{B}(v^2, z) = \int \frac{\rmd^N w}{(2 \pi)^{N/2}}\, \rme^{- w^2/2}
B(v^2 + 2 \frac{v \cdot w}{\sqrt{N}}
\sqrt{z} + \frac{w^2}{N} z)  \nonumber \\
& \tilde G^{ab} = G^{aa} + G^{bb} - G^{ab} - G^{ba}\ .
\end{align}
In general, $\hat{B} (s,z)$ is a function of two variables, which
becomes a function of the sum $\hat{B}(s,z) \to B(s+z)$ only as $N \to
\infty$, since in that limit $\langle vw\rangle =0$ and $\langle
w^{2}\rangle=N$, without fluctuations, in the gaussian measure
$\sim\rme^{-w^{2}/2}$.  
%

\section{Effective action in non-uniform background: general
formulation} \label{sec:appB} In some applications bilocal terms may
already be present in the starting action.  Let us thus give a more
general and compact result, which also includes that case. It is
derived by a simple extension of the methods of Section
\ref{sec:calculgamma}.

Let us consider a $N$-component field $\phi_x$, with components
$\phi^i_x$, $i=1,\dots ,N$, which can carry other indices, 
coordinates, or be a set of fields, etc\dots. A general $O(N)$-invariant
form for the action functional is
\begin{eqnarray}
 {\cal S}[\phi] &=& \frac{1}{2} \phi :\! C^{-1}\! : \phi +
N S_{\mathrm{int}}[\psi] \\
 \psi_{xy} &=& \frac{1}{N} \phi_x \cdot \phi_y \label{psi} \ ,
\end{eqnarray}
where $S_{\mathrm{int}}[\psi]$ is a functional of the {\it bilocal
field} $\psi_{xy}$ (which is also a bi-index matrix if the field
$\phi$ carries other indices, etc.).  Then the effective action
associated to ${\cal S}$ can be written as
\begin{equation}\label{lf72}
 \Gamma[\phi] = \frac{1}{2} \phi :\! C^{-1}\! : \phi + N
\Gamma^0[\psi] + \Gamma^1[\psi] + \dots \ ,
\end{equation}
where $\Gamma^0[\psi]$ is a functional of the bilocal field
$\psi_{xy}$ in (\ref{psi}) and satisfies the self-consistent equation:
\begin{eqnarray}
 \frac{\delta \Gamma^0}{\delta \psi_{xy}}[\psi] &=&
\frac{\delta S_{\mathrm{int}}[\chi]}{\delta \chi_{xy}} \\
 \chi_{xy} &=& \psi_{xy} + G[\psi]_{xy}  \\
 G[\psi]_{xy} &=& \left( C^{-1} + 2 \frac{\delta
\Gamma^0[\psi]}{\delta \psi} \right)^{-1}_{xy} \ 
\end{eqnarray}
$\chi_{xy}$ is another (set of) bilocal fields.

\section{Calculation of higher cumulants} \label{sec:highercum}
In this Appendix we compute the third and fourth renormalized
cumulants of the disorder. One uses the parameterization:
\begin{eqnarray}
 \tilde{U}(vv) &=& - \frac{1}{2 T^2} \sum_{ab} \tilde{B}(v^2_{ab}) -
\frac{1}{6 T^3} \sum_{abc} \tilde{S}(v^2_{ab},v^2_{bc},v^2_{ac}) \nonumber \\
&& - \frac{1}{24 T^4} \sum_{abcd}
\tilde{Q}(v^2_{ab},v^2_{bc},v^2_{cd}, v^2_{ad},v^2_{ac},v^2_{bd}) \ +
\dots\nonumber \\
\end{eqnarray} 
We need the matrix $M_{ab} = (- 2 T \partial_\chi
\tilde{U}(\chi))_{ab}|_{\chi = vv}$ up to the fourth cumulant:
\begin{eqnarray}
 M_{ab} &=& \frac{2}{T} (\delta_{ab} \sum_c \tilde{B}'_{ac} -
\tilde{B}'_{ab}) \nonumber \\
&& + \frac{2}{T^2} [ \delta_{ab} \sum_{c g}
\tilde{S}'_{1,acg} - \sum_g \tilde{S}'_{1,abg} ] \nonumber \\
&& + \frac{1}{T^3} [ \delta_{ab} \sum_{c g h} \tilde{Q}'_{1,acgh} -
\sum_{gh} \tilde{Q}'_{1,abgh} ] \ .\qquad 
\end{eqnarray}
The equality of (\ref{chipower1}) (pushed to the fourth cumulant,
i.e.\ the above formula) and of (\ref{power1}) using (\ref{simpli}),
implies:
\begin{eqnarray}
 \tilde B_{ab} &=& B'(\tilde{\chi}^{(0)}_{ab}) \\
 \frac{1}{T} \sum_g \tilde{S}'_{1,abg} &=&
B''(\tilde{\chi}^{(0)}_{ab}) \tilde{\chi}^{(1)}_{ab} \\
\frac{1}{2 T^2} \sum_{gh} \tilde{Q}'_{1,abgh} &=&
B''(\tilde{\chi}^{(0)}_{ab}) \tilde{\chi}^{(2)}_{ab} 
+ \frac{1}{2} B'''(\tilde{\chi}^{(0)}_{ab})
(\tilde{\chi}^{(1)}_{ab})^2 \ .\qquad  \label{eqcum2}
\end{eqnarray}
Thus we need compute the terms with one and two free replica sums,
$\tilde{\chi}^{(1)}_{ab}$ and $\tilde{\chi}^{(2)}_{ab}$. Because of
(\ref{p}) it means that we need to compute $\chi^{(1)}_{ab}$,
$\chi^{(1)}_{a}$, $\chi^{(2)}_{ab}$ and $\chi^{(2)}_{a}$.  To compute
them we use the definitions (\ref{chidef}), (\ref{defoff}). We thus
need powers of the matrix $M$, but only terms with zero, one or two
replica sums. The expression of $(M^2)_{ab}$ given in (\ref{power2})
is thus sufficient, and we also need:\begin{widetext}
\begin{equation}
 (M^3)_{ab} = \frac{8}{T^3} \sum\limits_{ef} \bigg[ \tilde{B}'_{ae}
\tilde{B}'_{be} ( \tilde{B}'_{af} + \tilde{B}'_{bf} + \tilde{B}'_{ef}) - \tilde{B}'_{ae} \tilde{B}'_{bf} \tilde{B}'_{ef} - \tilde{B}'_{ab}
(\tilde{B}'_{af} \tilde{B}'_{ae} + \tilde{B}'_{bf} \tilde{B}'_{be} +
\tilde{B}'_{ae} \tilde{B}'_{bf} \bigg]
\end{equation}
dropping all terms with three or more sums. One then finds:
\begin{eqnarray}
 \chi^{(1)}_{a} &=& 2 I_2 \sum_c \tilde{B}'_{ac} \\
 \chi^{(1)}_{ab} &=& 
- \frac{2}{T} I_2  \sum_g \tilde{S}'_{1,abg} +  \frac{4}{T} I_3 (
- \tilde{B}'_{ab} \sum_f (\tilde{B}'_{af} + \tilde{B}'_{bf})
+ \sum_c \tilde{B}'_{ac} \tilde{B}'_{cb} ) \\
 \chi^{(2)}_{a} &=& \frac{2}{T} I_2  \sum_{cg} \tilde{S}'_{1,acg} 
+ \frac{4}{T} I_3 \sum_{ef} \tilde{B}'_{ae} \tilde{B}'_{af} 
\\
 \chi^{(2)}_{ab} &=& - \frac{1}{T^2} I_2 \sum_{gh} \tilde{Q}'_{1,abgh}
+ \frac{4}{T^2} I_3
\!\left[ - \tilde{B}'_{ab} \sum_{gh} ( \tilde{S}'_{bgh} {+}
\tilde{S}'_{agh} ) - \sum_{eh} (\tilde{B}'_{ae} \tilde{S}'_{abh} +
\tilde{B}'_{be} \tilde{S}'_{abh})
+ \sum_{h c} (\tilde{B}'_{ac} \tilde{S}'_{cbh} {+} \tilde{B}'_{bc}
\tilde{S}'_{cah})\right]\!\! \nonumber \\
&&+ \frac{8}{T^2} I_4 \sum_{ef} ( \tilde{B}'_{ae} \tilde{B}'_{be} (
\tilde{B}'_{af} + \tilde{B}'_{bf} + \tilde{B}'_{ef}) - \tilde{B}'_{ae}
\tilde{B}'_{bf} \tilde{B}'_{ef} - \tilde{B}'_{ab} (\tilde{B}'_{af}
\tilde{B}'_{ae} + \tilde{B}'_{bf} \tilde{B}'_{be} + \tilde{B}'_{ae}
\tilde{B}'_{bf}) \ .
\end{eqnarray}
which yields $\tilde{\chi}^{(1)}_{ab}$ and $\tilde{\chi}^{(2)}_{ab}$
using (\ref{p}).

\subsection{Third cumulant}
To obtain the third cumulant we now insert
$\tilde{\chi}^{(1)}_{ab}$ in (\ref{eqcum2}). One can rewrite
$B''(\tilde{\chi}_{ab}^0)$, indeed taking the derivative of
(\ref{saddlepointforB}) with respect to $v_{ab}^2$ shows that:
\begin{equation}
 B''(\tilde{\chi}_{ab}^0) = \frac{\tilde{B}''_{ab}}{1 + 4 I_2
\tilde{B}''_{ab}} \ .
\end{equation}
This becomes, regrouping the terms in $\tilde{S}'_{1abg}$ and 
dividing by the common denominator $1/(1 + 4 I_2 \tilde{B}_{ab}'')$
we obtain:
\begin{eqnarray}
&& \!\! \!\! \!\!\sum_g  \tilde{S}'_{1abg} + 4 I_2 \tilde{B}''_{ab}
(\frac{1}{2} \tilde{S}'_{1,aag} + \frac{1}{2} \tilde{S}'_{1,bbg})\nonumber \\
&& = 2 T I_2 \tilde{B}''_{ab} \sum_c (\tilde{B}'_{ac} +
\tilde{B}'_{bc} ) + 8 I_3 \tilde{B}''_{ab} \sum_g \Big[
\tilde{B}'_{ab} (\tilde{B}'_{ag} + \tilde{B}'_{bg}) - \tilde{B}'_{aa}
\tilde{B}'_{ag} - \tilde{B}'_{bb} \tilde{B}'_{bg} - \tilde{B}'_{ag}
\tilde{B}'_{gb} + \frac{1}{2} \tilde{B}'_{ag} \tilde{B}'_{ga} +
\frac{1}{2} \tilde{B}'_{bg} \tilde{B}'_{gb} \Big] \ .\nonumber \\
&&
\end{eqnarray}
This first yields:
\begin{equation}
 \tilde{S}'_{1aab} = \frac{ 4 T I_2 \tilde{B}''(0) }{1 + 4 I_2
\tilde{B}''(0)} \tilde{B}'_{ab} \ .
\end{equation}
Inserting this back yields:
\begin{eqnarray}
\tilde{S}'_{1abc}  &=& 
 \frac{ 2 T I_2 \tilde{B}''_{ab} }{1 + 4 I_2 \tilde{B}''(0)}
(\tilde{B}'_{ac} + \tilde{B}'_{bc} ) 
+ 8 I_3 \tilde{B}''_{ab}
\left[  (\tilde{B}'_{ab} - \tilde{B}'(0)) (\tilde{B}'_{ac} + \tilde{B}'_{bc} )
- \tilde{B}'_{ac} \tilde{B}'_{bc} 
+ \frac{1}{2} (\tilde{B}'_{ac})^2 + \frac{1}{2} (\tilde{B}'_{bc} )^2
\right]\ .
\end{eqnarray}
In terms of functions it gives:
\begin{eqnarray}
 \tilde S'_1(x,y,z) &=& 
 \frac{ 2 T I_2 }{1 + 4 I_2 \tilde{B}''(0)} \tilde{B}''(x)  
(\tilde{B}'(y) + \tilde{B}'(z) )  \nonumber \\
&& + 8 I_3 \tilde{B}''(x) 
\left[ (\tilde{B}'(x) - \tilde{B}'(0)) (\tilde{B}'(y) + \tilde{B}'(z) )  - \tilde{B}'(y) \tilde{B}'(z) 
+ \frac{1}{2} (\tilde{B}'(y))^2 + \frac{1}{2} (\tilde{B}'(z) )^2  \right]\ .
\end{eqnarray}
Integrating once, this yields the simple expression for the third cumulant
given in the text. Note that, up to terms which vanish at $n=0$, it can 
be expressed only in terms of the function $\tilde{B}'(x) - \tilde{B}'(0)$.

\subsection{Fourth cumulant}
\begin{figure}
\fig{0.5\textwidth}{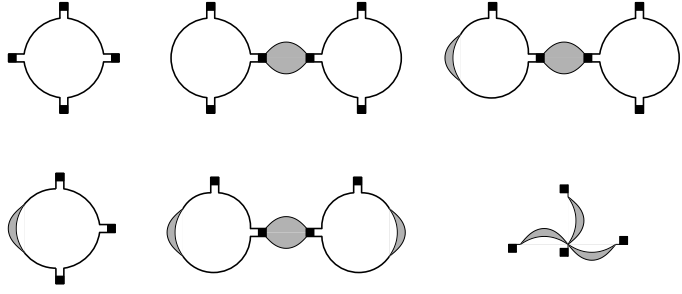} \caption{Graphical representation
of the fourth cumulant. The notation is explained in
\protect\cite{LeDoussalWiesePREPg}.  Each diagram corresponds to a
square bracket in Eq.~(\ref{final4th}), in the same order.}
\label{f:4thcumulant}
\end{figure}
From (\ref{eqcum2}) one has:
\begin{eqnarray}
 \sum_{e} \tilde{S}'_{1,abe} &=& \sum_{e} \frac{ \tilde{B}''_{ab} }
{ 1 + 4 I_2 \tilde{B}''_{ab}} Y_{abe} \\
 \frac{1}{2} \sum_{ef} Q'_{1abef} &=& \sum_{ef} \Big[ \frac{
\tilde{B}''_{ab} }{ 1 + 4 I_2 \tilde{B}''_{ab}} Z_{abef} + \frac{
\tilde{B}'''_{ab} }{ 2 (1 + 4 I_2 \tilde{B}''_{ab})^3} Y_{abe} Y_{abf}
\Big] \ ,\qquad
\end{eqnarray}
where we have used
$B'''(\tilde{\chi}^{(0)}_{ab})=\tilde{B}'''_{ab}/((1 + 4 I_2
\tilde{B}''_{ab})^3$ obtained by further differentiation of
(\ref{saddlepointforB}) with respect to $v_{ab}^2$. We also  define
\begin{eqnarray} 
 Y_{abe} = T \tilde \chi^{(1)}_{ab} &=& 4 I_2 \Big[ - \frac{1}{2}
(\tilde{S}'_{1,aae} +
\tilde{S}'_{1,bbe})  + \tilde{S}'_{1,abe}  \Big]  + 2 I_2 T
(\tilde{B}'_{ae} + \tilde{B}'_{be})  \nn \\
&& + 8 I_3 \Big[ - \frac{1}{2} (\tilde{B}'_{aa} + \tilde{B}'_{bb})
(\tilde{B}'_{ae} + \tilde{B}'_{be}) + \frac{1}{2} (\tilde{B}'_{ae}
\tilde{B}'_{ae} + \tilde{B}'_{be} \tilde{B}'_{be}) + \tilde{B}'_{ab}
(\tilde{B}'_{ae} + \tilde{B}'_{be}) - \tilde{B}'_{ae} \tilde{B}'_{be}
\Big]\ ,\qquad 
\end{eqnarray}
and
\begin{eqnarray} 
 Z_{abef} &=& T^2 \tilde \chi^{(2)}_{ab} = 2 I_2 \Big[
\tilde{Q}'_{1,abef} - \frac{1}{2} \tilde{Q}'_{1,aaef} - \frac{1}{2}
\tilde{Q}'_{1,bbef} \Big] + 2 I_2 T (\tilde{S}'_{1,aef} +
\tilde{S}'_{1,bef}) + 4 I_3 T ( \tilde{B}'_{ae} \tilde{B}'_{af} +
\tilde{B}'_{be} \tilde{B}'_{bf}) \nn \\
&& + 8 I_3 \Big[ - (\tilde{B}'_{aa} \tilde{S}'_{1,aef} + \tilde{B}'_{bb}
\tilde{S}'_{1,bef}) - (\tilde{B}'_{ae} \tilde{S}'_{1,aaf} +
\tilde{B}'_{be} \tilde{S}'_{1,bbf}) + (\tilde{B}'_{af}
\tilde{S}'_{1,fae} + \tilde{B}'_{bf}
\tilde{S}'_{1,fbe}) \nn \\ 
&&\qquad\quad + \tilde{B}'_{ab} ( \tilde{S}'_{1,bef} +
\tilde{S}'_{1,aef} ) + (\tilde{B}'_{ae} \tilde{S}'_{1,abf} +
\tilde{B}'_{be} \tilde{S}'_{1,abf}) - (\tilde{B}'_{af}
\tilde{S}'_{1,fbe} +
\tilde{B}'_{bf} \tilde{S}'_{1,fae})\Big] \nn \\
&& + 8 I_4 \Big[ \tilde{B}'_{ae} \tilde{B}'_{ae} ( 2 \tilde{B}'_{af} +
\tilde{B}'_{ef}) + \tilde{B}'_{be} \tilde{B}'_{be} (2 \tilde{B}'_{bf}
+ \tilde{B}'_{ef}) - (\tilde{B}'_{ae} \tilde{B}'_{af} +
\tilde{B}'_{be} \tilde{B}'_{bf}) \tilde{B}'_{ef} - 3 \tilde{B}'_{aa}
(\tilde{B}'_{af} \tilde{B}'_{ae} +
 \tilde{B}'_{bf} \tilde{B}'_{be}) \nn \\
&& \qquad \quad - 2 \tilde{B}'_{ae} \tilde{B}'_{be} ( \tilde{B}'_{af} +
\tilde{B}'_{bf} + \tilde{B}'_{ef}) + 2 \tilde{B}'_{ae} \tilde{B}'_{bf}
\tilde{B}'_{ef} + 2 \tilde{B}'_{ab} (\tilde{B}'_{af} \tilde{B}'_{ae} +
\tilde{B}'_{bf} \tilde{B}'_{be} + \tilde{B}'_{ae} \tilde{B}'_{bf})
\Big] \nn \ ,
\end{eqnarray}
which should be further symmetrized over $e,f$ for later use
(indicated by $\text{sym}_{ef}$ below).

The fourth cumulant equation yields, regrouping and simplifying the
denominators (and using also the third cumulant equation):
\begin{eqnarray} 
&& \frac{1}{2} \tilde Q'_{1abef} + 2 I_2 \tilde{B}''_{ab} (
\frac{1}{2} \tilde Q'_{1aaef} + \frac{1}{2} \tilde Q'_{1bbef}) =
\frac{1}{2} \tilde{B}'''_{ab}
\frac{\tilde{S}'_{1,abe}}{\tilde{B}''_{ab}}
\frac{\tilde{S}'_{1,abf}}{\tilde{B}''_{ab}} + 2 I_2 T \text{sym}_{ef}
[ \tilde{B}''_{ab} (\tilde{S}'_{1,aef} + \tilde{S}'_{1,bef}) ] \nn \\
&& + 4 I_3 T \tilde{B}''_{ab} ( \tilde{B}'_{ae} \tilde{B}'_{af} +
\tilde{B}'_{be} \tilde{B}'_{bf}) + 8 I_3 \tilde{B}''_{ab}
\text{sym}_{ef} \Big[ - (\tilde{B}'_{aa} \tilde{S}'_{1,aef} +
\tilde{B}'_{bb} \tilde{S}'_{1,bef}) - (\tilde{B}'_{ae}
\tilde{S}'_{1,aaf} + \tilde{B}'_{be} \tilde{S}'_{1,bbf}) \nn \\ && +
(\tilde{B}'_{af} \tilde{S}'_{1,fae} + \tilde{B}'_{bf}
\tilde{S}'_{1,fbe}) + \tilde{B}'_{ab} ( \tilde{S}'_{1,bef} +
\tilde{S}'_{1,aef} ) + (\tilde{B}'_{ae} \tilde{S}'_{1,abf} +
\tilde{B}'_{be} \tilde{S}'_{1,abf}) - (\tilde{B}'_{af}
\tilde{S}'_{1,fbe} +
\tilde{B}'_{bf} \tilde{S}'_{1,fae}) \Big] \nn \\
&& + 8 I_4 \tilde{B}''_{ab} \text{sym}_{ef} \Big[ \tilde{B}'_{ae}
\tilde{B}'_{ae} ( 2 \tilde{B}'_{af} + \tilde{B}'_{ef}) +
\tilde{B}'_{be} \tilde{B}'_{be} ( 2 \tilde{B}'_{bf} + \tilde{B}'_{ef})
 - (\tilde{B}'_{ae} \tilde{B}'_{af} + \tilde{B}'_{be}
\tilde{B}'_{bf}) \tilde{B}'_{ef}
-  3 \tilde{B}'_{aa} (\tilde{B}'_{af} \tilde{B}'_{ae} +
\tilde{B}'_{bf} \tilde{B}'_{be}) \nn   \\
&& - 2 \tilde{B}'_{ae} \tilde{B}'_{be} ( \tilde{B}'_{af} +
\tilde{B}'_{bf} + \tilde{B}'_{ef}) + 2 \tilde{B}'_{ae} \tilde{B}'_{bf}
\tilde{B}'_{ef} + 2 \tilde{B}'_{ab} (\tilde{B}'_{af} \tilde{B}'_{ae} +
\tilde{B}'_{bf} \tilde{B}'_{be} + \tilde{B}'_{ae} \tilde{B}'_{bf})
\Big]  .
\end{eqnarray}
Setting $a=b$ and solving one finds:
\begin{equation} 
 \frac{1}{2} \tilde Q'_{aaef} = \frac{4 I_2 T \tilde B''(0)}{1 + 4 I_2
\tilde B''(0)} \text{sym}_{ef} \tilde S'_{1,aef} + \left[ \frac{8 T^2
I_2^2 \tilde B'''(0)}{(1 + 4 I_2 \tilde B''(0))^3} + \frac{8 I_3
\tilde B''(0) T}{1 + 4 I_2 \tilde B''(0)} \right] \tilde B'_{ae}
\tilde B'_{af}
\end{equation}
This gives the final result for the fourth cumulant:
\begin{eqnarray}\label{lf94}
Q_{abcd} &=& \mbox{Sym}_{abcd}\bigg \{\nonumber \\
48\!\!&\Big[&\! -4\tilde B'_0\tilde B'_{ab}\tilde B'_{ac}\tilde
B'_{ad}+4(\tilde B'_{ab})^2\tilde B'_{ac}\tilde B'_{ad}+2(\tilde
B'_{ab})^2\tilde B'_{ad}\tilde B'_{bc}-4\tilde B'_{ab}\tilde
B'_{ac}\tilde B'_{ad}\tilde B'_{bc}+\tilde B'_{ab}\tilde B'_{ad}\tilde
B'_{bc}\tilde B'_{cd}\Big] I_{4}\nonumber \\
+ 192\!\!&\Big[&\! 4(\tilde B'_0)^2\tilde B'_{ab}\tilde B'_{cd}{\tilde
B''}_{ac}-4\tilde B'_0(\tilde B'_{ab})^2\tilde B'_{cd}{\tilde
B''}_{ac}-8\tilde B'_0\tilde B'_{ab}\tilde B'_{ac}\tilde
B'_{cd}{\tilde B''}_{ac}+4(\tilde B'_{ab})^2\tilde B'_{ac}\tilde
B'_{cd}{\tilde B''}_{ac}+4\tilde B'_{ab}(\tilde B'_{ac})^2\tilde
B'_{cd}{\tilde B''}_{ac}\nonumber \\
&& +8\tilde B'_0\tilde B'_{ab}\tilde B'_{bc}\tilde B'_{cd}{\tilde
B''}_{ac}-8\tilde B'_0\tilde B'_{ac}\tilde B'_{bc}\tilde
B'_{cd}{\tilde B''}_{ac}-8\tilde B'_{ab}\tilde B'_{ac}\tilde
B'_{bc}\tilde B'_{cd}{\tilde B''}_{ac}+4(\tilde B'_{ac})^2\tilde
B'_{bc}\tilde B'_{cd}{\tilde B''}_{ac}+4(\tilde B'_{ab})^2\tilde
B'_{ac}\tilde
B'_{ad}{\tilde B''}_{ad}\nonumber \\
&& -4(\tilde B'_{ab})^2\tilde B'_{ac}\tilde B'_{cd}{\tilde
B''}_{ad}+2\tilde B'_{ab}\tilde B'_{ac}\tilde B'_{bd}\tilde
B'_{cd}{\tilde B''}_{ad}+(\tilde B'_{ab})^2(\tilde B'_{cd})^2{\tilde
B''}_{ad}+4(\tilde B'_0)^2\tilde B'_{ac}\tilde B'_{cd}{\tilde
B''}_{bc}-4\tilde B'_0(\tilde B'_{ac})^2\tilde B'_{cd}{\tilde
B''}_{bc}\nonumber \\
&& +(\tilde B'_{ad})^2(\tilde B'_{cd})^2{\tilde B''}_{bd}\Big] 
I_{3}^{2}\nonumber \\
+ 192 T \!\!&\Big[&\!  2\tilde B'_{ab}\tilde B'_{ac}\tilde B'_{ad}\tilde
B''_{ab}+2\tilde B'_{ab}\tilde B'_{ad}\tilde B'_{bc}\tilde
B''_{ab}-2\tilde B'_0\tilde B'_{ab}\tilde B'_{ad}\tilde
B''_{ac}+(\tilde B'_{ab})^2\tilde B'_{ad}\tilde B''_{ac}-2\tilde
B'_{ab}\tilde B'_{ad}\tilde B'_{bc}\tilde B''_{ac}-2\tilde B'_0\tilde
B'_{ab}\tilde B'_{cd}\tilde B''_{ac}\nonumber \\
&& +(\tilde B'_{ab})^2\tilde B'_{cd}\tilde B''_{ac} -2\tilde
B'_{ab}\tilde B'_{ac}\tilde B'_{ad}\tilde B''_{0}\Big]\frac{
I_{2}I_{3}}{1+4 I_{2}\tilde B''_{0} }\nonumber \\
+ 32 T \!\!&\Big[\!& \tilde B'_{ab}\tilde B'_{ac}\tilde B'_{ad}\Big]
\frac{I_{3}}{1+4 I_{2}\tilde B''_{0}} \nonumber \\ 
+ 48 T^{2} \!\!&\Big[&\! \tilde B'_{ab}\tilde B'_{ad}\tilde
B''_{ac}+\tilde B'_{ab}\tilde B'_{cd}\tilde B''_{ac} \Big]
\frac{I_{2}^{2}}{(1+4 I_{2}\tilde B''_{0})^{2}}\nonumber \\
-128 T^{2}  \!\!  &\Big[ &\! \tilde B'_{ac}\tilde B'_{bc}\tilde
B'_{cd}\tilde B'''_{0}\Big] \frac{I_2^3}{(1+4 I_{2}\tilde
B''_{0})^{3}} \bigg  \}  
\label{final4th}
\end{eqnarray}
\end{widetext}
where $\mathrm{Sym}_{abcd}$ denotes $1/24$ times the sum of all 24 permutations
of the indices $a,b,c,d$, and we note $\tilde B'_{0}=\tilde B'(0)$,
$\tilde B''_{0}=\tilde B''(0)$ and $\tilde B'''_{0}=\tilde B'''(0)$.
Note that all terms containing $\tilde B'(0)$ can be eliminated by the
redefinition $\tilde B'_{\mathrm{new}}(x) = \tilde B'(x) -\tilde B'(0)$.

\section{Cumulant expansion for non-local effective action}
\label{scnonloc}

The cumulant expansion can be generalized to study the effective
action for non-uniform configurations.  The functional $\tilde
U[v\cdot v]$ is a functional of the field $v_{ab}^2(x)$ and can be
expanded as:
\begin{equation}\label{functcumulexp}
\tilde{U}[v v]  =  \frac{- 1}{2 T^2} \sum_{ab}
\tilde{B}[v_{ab}^2] - \frac{1}{3! T^3} \sum_{abc}
\tilde{S}[v_{ab}^2,v_{bc}^2,v_{ca}^2] + \dots  
\ .
\end{equation}
The self-consistent equation (\ref{scfunctional}) then yields, by a
similar expansion in numbers of sums:
\begin{align}\label{lf108}
& \frac{\delta \tilde B[v\cdot v]}{\delta (v_a(x) \cdot v_b(x)) }
=B'\bigg (v_{ab}^2(x) + 2 T I_1 
 \nonumber \\
& \qquad \quad~ + 4 \int_y C_{xy}^2 \Big[\frac{\delta \tilde B[v\cdot
v]}{\delta (v_a(y) \cdot v_b(y)) } - \frac{\delta \tilde B[v\cdot
v]}{\delta (v_a(y) \cdot v_a(y)) }\Big]\bigg)
\end{align}

\section{Analysis for all $m$ and short range case} 
\label{app:sr}
\begin{figure*}
\fig{\textwidth}{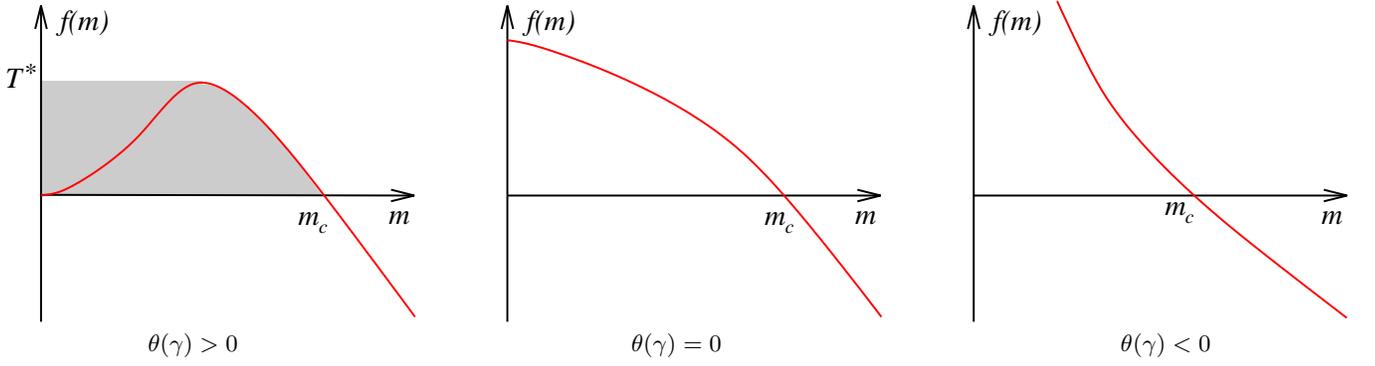}
\centerline{\qquad\qquad\qquad $\theta(\gamma) >0$\hfill  $\theta(\gamma) =0$
\hfill  $\theta(\gamma) < 0$\qquad\qquad\qquad  }
\caption{The function $f (m)$ defined in Eq.\ (\protect\ref{E.2}).}
\label{f:supercurve}
\end{figure*}
To refine the analysis and study the behavior for all values of $m$,
let us rewrite from (\ref{secder0}) the condition that $\tilde{B}$
remains analytic for all $m$. It reads:
\begin{align}\label{E.1}
&T > f(m)  \qquad \text{for all}  \quad m \\
& f(m) = \frac{(B'')^{-1}\left(\frac{1}{4 I_2}\right)}{2 I_1} =
\frac{(4 g \gamma I_2)^{\frac{1}{1+ \gamma}} - a^2}{2 I_1}\ ,
\label{E.2}
\end{align}
where we have also inserted the value of the inverse function, with
$B''((B'')^{-1}(x))=x$, for the power law models. This condition is
equivalent to the vanishing of the replicon, i.e.\ it is the line
where the RS solution of the GVM becomes unstable to RSB. 

One can then plot (see Fig.\ \ref{f:supercurve}) the function $f(m)$
for the three cases defined in the text, LR $\theta(\gamma) >0$,
marginal $\theta(\gamma) =0$ and SR $\theta(\gamma) < 0$ where
$\theta(\gamma)=d-2 + \frac{4-d}{1+ \gamma}$.  The LR and marginal
cases, which correspond to continuous and one step marginal RSB
solutions, have been discussed in the text and there the FRG gives
back exactly $\sigma_m(0)$ of the MP solution. We defer the detailed
study of the SR case to further work, and give here a few general
remarks.

First one notices on Fig.\ \ref{f:supercurve} that solving the FRG
equation decreasing the mass from infinity one first has the analytic
solution which coincides with the RS one. For $T>T^*=\max_m f(m)$ it
remains analytic down to $m=0$. For $T<T^*$ a cusp arises when the
left branch of the line is reached. Thus despite the reentrance of the
analytic solution at small $m$, freezing of the FRG solution has
already occurred and it is clearly important to understand how to
extend the FRG in the shaded region. On the other hand, a 1-step
solution of the MP saddle point equations exists, obtained by varying
the free energy w.r.t.\ $u_c$. Its precise boundary depends on the
model, but it is generally contained within the shaded region (for
details see
\cite{MezardParisi1991,Goldschmidt1993,Engel1993,CugliandoloLeDoussal1996b}).
An intriguing property of the GVM is the simultaneous existence,
within the rightmost portion of the shaded region near the axis $m=0$,
of two locally stable solutions, one RS and the other 1-step
RSB. Thus, although the line in Fig.\ \ref{f:supercurve} is the locus
of a continuous transition from RS to RSB, in this rightmost portion
of the $m-T$ diagram, the 1-step non-marginal RSB solution appears
discontinuously, before the line is reached as $T$ is lowered. Work is
in progress to make contact between FRG and RSB in this SR case, and
in particular to understand whether there are also two branches of
solutions of the FRG equation in that region.

\section{Finite-temperature fixed points}
\label{app:finiteT} 
From the RG point of view it is interesting to search for finite
temperature fixed points (FP) of the FRG equation, especially in view
of future extensions to finite $N$ (since we know at least in some
cases, that these persist at finite $N$). It is convenient to use Eq.\
(\ref{linear}). These FPs exist only in the marginal case $\theta=0$,
i.e.\ $\zeta=(2-d)/2$ for $d<2$ or $\zeta=0$ for $d=2$, so that $T_m=
4 A_d T/\epsilon$ does not flow. This is the case studied here.

\subsection{$d < 2$} Following the same steps as in Section
\ref{frgapproach}, the general solution of the fixed-point condition $m
\partial_m x(y)=0$ in (\ref{linear}) for a fixed value of $\zeta >0$,
imposing $x(y_0)=0$ is
\begin{eqnarray} \label{fp1}
x^*(y) &=& \frac{y}{\epsilon} - \frac{y_0}{2 \zeta} + \frac{\epsilon -
2 \zeta}{2 \zeta \epsilon} \, y_0 \left(\frac{y_0}{y}\right)^{\frac{2
\zeta}{\epsilon - 2 \zeta}} \nonumber \\
&& + \frac{T_m \epsilon x'_0}{2 \zeta(\epsilon x'_0 -1)} \left[
\left(\frac{y_0}{y}\right)^{\frac{2 \zeta}{\epsilon - 2 \zeta}}
-1\right]\ .
\end{eqnarray}
Taking a derivative at $y=y_0$, we obtain a self-consistency condition
for $x'_0$. One solution is $x'_0 =0$, the ``zero-temperature'' fixed
point discussed in Section \ref{frgapproach}. The other one is
\begin{equation}\label{lf97}
- \epsilon x'_0 = \frac{\epsilon T_m}{y_0 (\epsilon - 2 \zeta)} - 1\ ,
\end{equation}
with the condition that it must be positive. Reinserting in
(\ref{fp1}) we obtain the final form for the finite-temperature fixed
point:
\begin{eqnarray}\label{lf95}
x^*(y) &=& \frac{y}{\epsilon} - \frac{y_0}{2 \zeta} + \frac{\epsilon -
2 \zeta}{2 \zeta \epsilon}\, y_0 \left(\frac{y_0}{y}\right)^{\!\frac{2
\zeta}{\epsilon
- 2 \zeta}} \nonumber \\
&& + \frac{1}{2 \zeta} \left[ T_m - \frac{y_0 (\epsilon - 2
\zeta)}{\epsilon}\right] \left[\left(\frac{y_0}{y}\right)^{\!\frac{2
\zeta}{\epsilon - 2 \zeta}} -1\right]\qquad \nonumber  \\
&=& \frac{y-y_{0}}{\epsilon} + \frac{T_{m}}{2\zeta}
\left[\left(\frac{y_{0}}{y} \right)^{\!\frac{2 \zeta}{\epsilon - 2
\zeta}}-1 \right]
\end{eqnarray}
The term in the second line of (\ref{lf95}) was not present in the
``zero-temperature'' fixed-point solution (\ref{fixedpoints}). Note
that it does not vanish at $T=0$ but at the higher temperature $T=T_c$
such that $T_m = \frac{y_0 (\epsilon - 2 \zeta)}{\epsilon}$. At this
point, also $x'_0$ vanishes and the solution $b'(x) = - y(x)$ becomes
non-analytic. The fixed-point analysis alone does neither fix the value of
$y_0$, nor $T_c$.

However we can now explicitly check that this fixed-point solution
identifies with the analytic solution (\ref{solu1}), (\ref{solu2}) when
setting $m \to 0$, using $\tilde{T}_m = T_m/(2-d)$ and
$\zeta=1/\gamma=(2-d)/2$.  This identification works only for $T >
T_c$, and since $y_0$ is now fixed by (\ref{solu2}), we can compute
$T_c$ and find that it is given by (\ref{lf52}).  Below $T_c$ the
solution freezes at $m=m_c$ at the zero temperature fixed point.

\subsection{$d = 2$}
Let us now solve the fixed point condition $m \partial_m x(y)=0$ in
(\ref{linear}) for $\zeta = 0$, imposing $x(y_0)=0$. One finds:
\begin{eqnarray} \label{fp1d2}
x^*(y) = \frac{y - y_0}{\epsilon} - \left(\frac{y_0}{\epsilon} +
\frac{T_m x'_0}{\epsilon x'_0 -1} \right) \ln(y/y_0)\ .
\end{eqnarray}
Determining $x'_0$ again one finds, in addition to the solution
$x'_0 =0$
\begin{equation}\label{lf98}
- \epsilon x'_0 = \frac{T_m}{y_0} - 1\ .
\end{equation}
Reinserting one finds finally the finite-temperature fixed points
\begin{equation}\label{lf99}
x^*(y) = \frac{y - y_0}{\epsilon} - \frac{T}{\epsilon \pi} \ln(y/y_0)\ ,
\end{equation}
with $\epsilon=2$ and $T_m = T/\pi$.  There is thus a line of fixed
points with $\zeta=0$ in $d=2$, parameterized by temperature, $y_0$
being again undetermined.

To compare with the solution of the flow equation, we obviously need
to consider a broader class of SR models with $B'(z)= - g
\exp(-z/a^2)$. The solution is then:
\begin{eqnarray}
 x  &=& a^2 \ln(y_0/y) + {\epsilon}^{{-1}} (y - y_0) \label{solu0b} \\
 y_0 &=& \tilde{g} m^{T/(\pi a^2) - \epsilon} \Lambda^{-T/(\pi a^2) }
\end{eqnarray}
with $\epsilon=2$. For small disorder $\tilde{g}$, and $T > T_c
=\epsilon \pi a^2 =2 \pi a^2$, $y_0(m)$ flows to zero as $m \to 0$ and
the solution remains analytic. For $T < T_c$ the solution develops a
cusp when $y_0$ reaches $y_0 = y_0(m_c) = \epsilon a^2$, i.e.\ at the
Larkin mass:
\begin{equation}\label{lf100}
m_c = (\tilde{g}/a^2 \epsilon)^\frac{1}{2 (1- T/T_c)}
\Lambda^\frac{T}{T_c - T}\ .
\end{equation}
Thus only for $T=T_c$ the solution reaches for $m \to 0$ an analytic
finite-$T$ fixed point associated with $\zeta=0$, of the form
(\ref{lf99}).  Thus in $d=2$ the line of finite-temperature fixed
points with $\zeta=0$ corresponds to the line of critical fixed points
as the parameter $a$ is varied.

\section{RG formulation of the RSB solution} \label{app:mprg} In this
appendix, we derive simple RG equations for the MP solution in the
full RSB case. This gives a more direct derivation of the
key-equations (\ref{rgrsb1}) and (\ref{rgrsb2}).  We start from (see
(\ref{sadsig}) and (\ref{sadsig2}), equivalent to (5.4) of MP):
\begin{equation}\label{MP(5.4)}
\sigma(u) = - \frac{2}{T} B'\left( 2 T \int_k (\tilde{G}(k) - G(k,u) )
\right) \ .
\end{equation}
Taking a derivative with respect to $u$ yields
\begin{equation}\label{lf67}
1 = 4 B''\left( 2 T \int_k  \tilde{G}(k) - G(k,u)  \right) \int_k
\frac{1}{(k^2 + \mu + [\sigma](u))^2}
\end{equation}
for all $u$ such that $\sigma'(u) \neq 0$.  Using that due to (see
(\ref{sadsig2}), equivalent to (5.2) of MP):
\begin{eqnarray}\label{MP(5.2)}
\tilde G (k)-G (k,u) &=& \frac{1}{u (k^{2}+{m^2} +[\sigma ] (u))}\nonumber \\
&&- \int_{u}^{1}\frac{\rmd v}{v^2} \frac{1}{k^{2}+{m^2} +[\sigma ]
(v)}
\end{eqnarray}
we have
\begin{equation}\label{6.15}
\partial_u ( \tilde{G}(k) - G(k,u) ) = - \frac{\sigma'(u)}{(k^2 + {m^2}
+ [\sigma](u))^2}  
\ .
\end{equation}
On the other hand, we can take a derivative of (\ref{MP(5.4)}) with
respect to ${m^2}$:
\begin{eqnarray}\label{lf68}
\partial_{m^2} \sigma(u) &=& - 4 B''\left( 2 T \int_k (\tilde{G}(k) -
G(k,u) )   \right) \nonumber \\
&&\times \int_{k} \partial_{m^2} ( \tilde{G}(k) - G(k,u) ) \ .
\end{eqnarray}
Eliminating $B''(\dots )$ one finds:
\begin{eqnarray}\label{lf69}
&&\partial_{m^2} \sigma(u) \int_k \frac{1}{(k^2 + {m^2} +
[\sigma](u))^2} \nonumber \\
&&\qquad = -\int_{k} \partial_{m^2} ( \tilde{G}(k) - G(k,u) ) \ .\qquad 
\end{eqnarray}
Taking another derivative with respect to $u$ after  using 
(\ref{6.15}) gives
\begin{equation}
\partial _{u }\! \int_{k} \frac{\partial _{{m^2} }\sigma (u)}{(k^{2}+{m^2}
+[\sigma ] (u))^{2}}= \partial_{{m^2}}\! \int_{k} \frac{\sigma '
(u)}{(k^{2}+{m^2} +[\sigma ] (u))^{2}}
\ .
\end{equation}
Noting that the derivatives of the numerator cancel, we get
\begin{equation}
 \int_{k} \frac{u \sigma ' (u)\partial _{{m^2} }\sigma
(u)}{(k^{2}+{m^2} +[\sigma ] (u))^{3}}=
\int_{k} \frac{\sigma ' (u) (1+\partial _{{m^2} }[\sigma ] (u))
}{(k^{2}+{m^2} +[\sigma ] (u))^{3}} 
\ .
\end{equation}
Since for all $u$, $\int_{k}{(k^{2}+{m^2} +[\sigma ] (u))^{-3}}\neq 0$
and by assumption $\sigma ' (u)\neq 0$, RSB reveals its universality
in the simple relation:
\begin{equation}\label{lf70}
u \partial_{m^2} \sigma(u) = 1 + \partial_{m^2} [\sigma](u)
\ ,
\end{equation}
which, upon another derivation, yields the two ``RG equations'':
\begin{eqnarray}\label{lf10b}
{m^2} \frac{\rmd }{\rmd {m^2} } \left({m^2} +[\sigma ] (u) \right) &=&0 \\
{m^2} \frac{\rmd }{\rmd {m^2} } \sigma (u)&=&0 \ .
\end{eqnarray}

\section{Convergence to the fixed point}
\label{app:convergence}
\begin{figure}[b] \centerline{\fig{0.8\columnwidth}{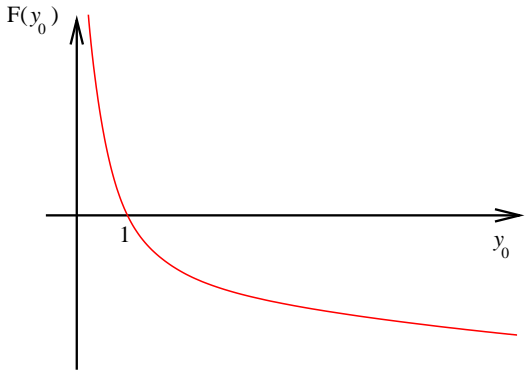}}
\caption{The function (\protect\ref{Fy4}), describing the approach to
the fixed point in presence of an additional bare power law tail.}
\label{nonleadingcurve}
\end{figure}
Since we have found the solution of the FRG equation for arbitrary
disorder correlations, it is instructive to study the convergence to
the FRG fixed point in the case where the inital disorder is not of
the simple form (\ref{corrlr}) on an explicit example. We start from a
disorder correlator which is the superposition of two power laws:
\begin{equation}\label{Fy1}
- B' (z) = \frac{g}{(a^2 + z)^{\gamma }} + C' \frac{g}{(a^2 + z)^{\alpha }}
\end{equation}
with $\alpha >\gamma $, s.t.\ for large $z$ the first term dominates
and fixes the exponent $\zeta =\frac{\E}{2 (1+\gamma )}$.  
We will determine the inverse function $\Phi (x)$ only
for $C'$ small. One finds
\begin{equation}\label{Fy2}
-\frac{\E}{4A_{d}}\,\Phi'\! \left(\frac{y}{4 Ad} \right) =
y^{-\frac{1}{\gamma }-1} + C y^{-\frac{1}{\delta }-1}\ ,
\end{equation}
with $C' \sim C$, and we have defined
\begin{equation}
\frac{1}{\delta} = \frac{1 - \alpha + \gamma}{\gamma} \ .
\end{equation}
We have chosen $\tilde{g} = (\gamma/\epsilon)^{\gamma}$ to simplify 
all prefactors.

Inserting into (\ref{cuspcond}), we obtain:
\begin{equation}\label{Fy3}
(y_{0}m^{\E-2\zeta})^{-\frac{1}{\gamma }-1}+C
(y_{0}m^{\E-2\zeta})^{-\frac{1}{\delta }-1} = m^{-\E} \ .
\end{equation}
By multiplying with $m^{\E}$ and $y_{0}^{\frac{\delta +1}{\delta }}$,
we obtain the equivalent formula
\begin{equation}\label{Fy4}
{\mathrm{F}} (y_{0}):= y_{0}^{\frac{\gamma -\delta }{\gamma \delta
}}-y_0 ^{\frac{\delta +1}{\delta }} = -C m^{\frac{\delta -\gamma
}{(1+\gamma )\delta }\epsilon } \ .
\end{equation}
The left-hand side is plotted on figure \ref{nonleadingcurve}; note
that always the first exponent is negative, and the second and third
are positive. Thus the solution for $m= 0$ or $C=0$ is simply
$y_{0}=1$.  For non-vanishing $C$ and $m$, the solution can be
obtained graphically as the intersection of $ {\mathrm{F}} (y_{0})$
with $-C m^{\frac{\delta -\gamma }{(1+\gamma )\delta }\epsilon
}$. Note that there is a solution for any $C$.

For $m\to 0$, the approach to $y_0=1$ is obtained by linearizing
$\mathrm{F} (y_{0})$, with the result
\begin{equation}
y_{0}-1 \approx \frac{\gamma C}{1+\gamma }m^{\frac{\delta -\gamma
}{(1+\gamma )\delta }\epsilon } \ .
\end{equation}

\section{Pure $O(N)$ models, non-analytic effective action} In this
Section we recall the corresponding result for the effective action of
the generic pure $O(N)$ model at large $N$. One mechanism by which the
effective action may becomes non-analytic is given on standard example
of $\phi^4$.

\subsection{Self-consistent equation} The generic $O(N)$ model in
dimension $d$ is defined by the action:
\begin{equation}
S = \frac{1}{T} \int_x \left[ \frac{1}{2} \left(\nabla u(x)\right)^2 +
\frac{1}{2} m^2 u(x)^2 + N V\Big(\frac{u(x)^2}{N}\Big) \right]
\end{equation}
Here $m$ is used as a parameter, the bare mass being $m_b^2 = m^2 + 2 V'(0)$.
For a uniform mode one has $\Gamma[u]= L^{d} \tilde{\Gamma}[v]$ in terms of
the rescaled field $v=u/\sqrt{N}$. One defines:
\begin{equation}
\tilde{\Gamma}[v] = \frac{1}{2} m^2 v^2 + \tilde V(v^2) = \tilde W(v^2)\ .
\end{equation}
Similarly, one defines 
\begin{equation}\label{lf95a}
W(z) = \frac{1}{2} m^2 z + V(z)\ ,
\end{equation}
whether absorbing or not the mass into the (bare or renormalized)
potential.  Again, for $m=\infty$, one has $\tilde V=V$.  The same
method as in Section (\ref{sec:calculgamma}) yields the saddle point
equations for infinite $N$:
\begin{eqnarray}
 \tilde V'(v^2) &=& V'(v^2 + G(v)) \qquad \\
 G(v) &=& \int_q \frac{T}{q^2 + m^2 + 2 V'(v^2 + G(v)) } \ .
\end{eqnarray}
More details, a graphical derivation, and the $1/N$ expansion are
given in \cite{LeDoussalWiesePREPg}.  A condition for the stability of
the theory is that:
\begin{equation}\label{lf109}
 2 \tilde W'(v^2) := m^2 + 2 \tilde V'(v^2) \geq 0 \quad \text{for
all}\ v^2 \ .
\end{equation}

\subsection{Solution and FRG equation} Let us start from the form
\begin{equation}\label{lf81}
 \tilde{W}'(x) = W'\left(x + T \int_q^{\Lambda}
\frac{1}{q^2 + 2 \tilde{W}'(x)} \right) \ ,
\end{equation}
The self-consistent solution of this equation
is formally:
\begin{eqnarray}
 x &=& (W')^{-1}\left(y \right)
- T \int_q^{\Lambda} \frac{1}{q^2 + 2 y} \\
 y &=& \tilde{W}'(x) \ .
\end{eqnarray}
and $y \geq 0$. 

Let us write the associated FRG equation. One has
\begin{eqnarray}
 - m \partial_m \tilde{W}'(x) &=& - m^2 - 2 T W''\left(x + T
\int_q^{\Lambda} \frac{1}{q^2 + 2 \tilde{W}'(x)} \right) \nonumber \\
&&\times \left(- m \partial_m \tilde{W}'(x)\right)
\int_q^{\Lambda} \frac{1}{(q^2 + 2 \tilde{W}'(x))^2}\nonumber \\  \\
 \tilde{W}''(x) &=& W''\left(x + T \int_q^{\Lambda} \frac{1}{q^2 + 2
\tilde{W}'(x)} \right) \nonumber \\
&&\times \left( 1 - 2 T \tilde{W}''(x) \int_q^{\Lambda} \frac{1}{(q^2
+ 2 \tilde{W}'(x))^2} \right) \ . \nonumber \\
\end{eqnarray}
Thus:
\begin{equation}\label{lf101}
 - m \partial_m \tilde{W}'(x) = 2 T m^2 \tilde{W}''(x)
\int_q^{\Lambda} \frac{1}{(q^2 + 2 \tilde{W}'(x))^2} - m^2 \ .
\end{equation}
For $d<4$ taking $\Lambda$ to infinity, this becomes
\begin{equation}\label{lf102}
 - m \partial_m \tilde{W}'(x) = 2 T \frac{A_d}{\epsilon} m^2
\tilde{W}''(x) (2 \tilde{W}'(x))^{- \epsilon/2} - m^2 \ .
\end{equation}

\subsection{$\phi^4$-theory and non-analytic behavior}
 For the
$\phi^4$ theory $V(x) = \frac{g}{4} (x-1)^2$ this reads:
\begin{eqnarray}
 \tilde{W}'(x) &=& \frac{1}{2} m^2 + \frac{g}{2} (x - 1) + \frac{g
T}{2}
\int_q^{\Lambda} \frac{1}{q^2 + 2 \tilde{W}'(x)} \qquad \quad \\
 x(y) &=& \frac{2}{g} y + 1 - \frac{m^2}{g} - T \int_q^{\Lambda}
\frac{1}{q^2 + 2 y} \label{phi4}
\end{eqnarray} 
for $y \geq y_0$ with $x(y_0)=0$.  At $T=0$, as $m$ is decreased,
there is a transition in any $d$ when $m_b^2 = m^2 - g$ vanishes. For
$d>2$, the transition persists for $T<T_c$, and occurs when $m^2 - g +
g T \int_q^{\Lambda} \frac{1}{q^2}$ vanishes, with the standard
result:
\begin{equation}\label{lf82}
T_c \int_q^{\Lambda} \frac{1}{q^2} = 1  \ ,
\end{equation}
which depends strongly on the UV-cutoff $\Lambda$. $y_0$ vanishes at
the transition, and in the ordered phase the effective action has a
non-analytic form. In addition of the branch (\ref{phi4}) for $y \geq
0$, $x>x_c = 1 - \frac{m^2}{g} - T \int_q^{\Lambda} \frac{1}{q^2}$,
the function $x(y)$ has a branch $y=0$ for $0 < x < x_c$, where $x_c$
is the order parameter.

Exactly at $T=T_c$ we should recover that the effective action
exhibits the standard power-law non-analyticity $\Gamma[M] = M^{1+
\delta}$. Indeed, from the self-consistent solution (\ref{phi4}),
subtracting the same with $x=x_c$ and $y=0$ one gets
\begin{equation}\label{lf107}
 x (y) = \frac{2}{g} y + (2 y)^{(d-2)/2} T_c \int_k^{\Lambda/2y}
\frac{1}{k^{2}} \frac{1}{k^2 + 1}
\end{equation}
for $2<d<4$. This corresponds to $\delta=(2+d)/(2-d)$. It can be
recovered by solving the FRG equation.  One can look for fixed-point
solutions of the form
\begin{figure}[t] \fig{0.8\columnwidth}{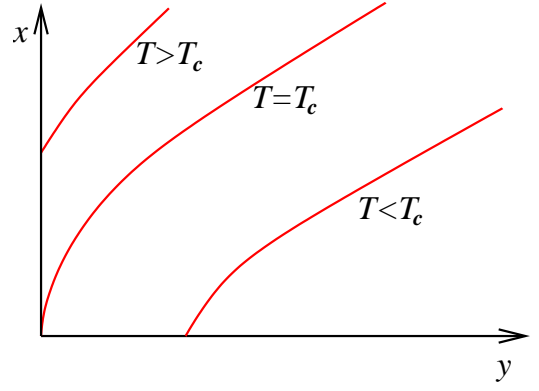} \caption{Relation
between $x$ and $y$ in scalar field theory: above, at, and below the
critical temperature.}  \label{f:phi4phase}
\end{figure}
\begin{equation}\label{lf103}
 \tilde{W}'(x) = m^\alpha f(m^\beta x) \ .
\end{equation}
If one wants the two first terms to dominate and to scale in the same
way, one needs $\alpha \le 2$ and $\beta = \alpha \epsilon/2 - 2$. For
all three terms to scale the same way one needs $\alpha=2$,
$\beta=2-d$. Inserting (\ref{lf103}) into (\ref{lf102}) yields
\begin{equation}\label{lf104}
 1 - \alpha f(z) - \beta z f'(z) = 2 T \frac{A_d}{\epsilon} f'(z) (2
f(z))^{-\epsilon/2} \ .
\end{equation}
Again this can be transformed into a linear RG equation for $z(f)$:
\begin{equation}\label{lf105}
 (1 - \alpha f) z' (f) - \beta z (f) = 2 T \frac{A_d}{\epsilon} (2
f)^{-\epsilon/2} \ .
\end{equation}
The solutions of the above equation with  $\alpha=2$,
$\beta=2-d$ are:
\begin{equation}\label{lf106}
 z (f) = - 2 T \frac{A_d}{\epsilon} (2 f - 1)^{(d-2)/2} \int_g^f
t^{-\epsilon/2} (2 t - 1)^{-d/2}
\end{equation}
A particular solution is
\begin{eqnarray}
 z (f) &=& (2 f -1)^{(d-2)/2} \\ 
\tilde{W}'(x) &=& \frac{m^{2}}{2} + x^{2/(d-2)}\ .
\end{eqnarray}
In the limit of zero mass this yields $\tilde{V}'(x) = x^{2/(d-2)}$.

One can also pursue the RG approach in the ordered phase, as is done
usually in the form of a non-linear sigma model, and deal with a
non-analytic effective action.

Although the mechanism for the disordered systems studied in the main
text seems to be different from $\phi^4$-models, it raises the
question of the meaning of the non-analyticity in the disordered
problem. Is it the signature that we are dealing with a glass phase,
where a symmetry has been broken? We know that for infinite $N$, this
is also accompanied by RSB, but this does not have to be so in
general, i.e.\ the cusp can arise without RSB, just from localization
(single ground state dominance).


\begin{widetext}
\tableofcontents

\bigskip 

\end{widetext}

\end{document}